\documentclass[runningheads]{llncs}
\usepackage{graphicx}
\usepackage{subfig}
\usepackage{mathtools}
\usepackage[ruled]{algorithm2e}
\newcommand{\ourAlgorithm}{Kadabra}

\begin{document}
\title{\ourAlgorithm: Adapting Kademlia for the Decentralized Web}
%
%
\author{Yunqi Zhang\and
Shaileshh Bojja Venkatakrishnan}


\institute{The Ohio State University\\
\email{\{zhang.8678, bojjavenkatakrishnan.2\}@osu.edu}}

\maketitle              
\begin{abstract}

Blockchains have become the catalyst for a growing movement to create a more decentralized Internet.
A fundamental operation of applications in a decentralized Internet is data storage and retrieval. 
As today's blockchains are limited in their storage functionalities, in recent years a number of peer-to-peer data storage networks have emerged based on the Kademlia distributed hash table protocol. 
However, existing Kademlia implementations are not efficient enough to support fast data storage and retrieval operations necessary for (decentralized) Web applications. 
In this paper, we present \ourAlgorithm, a decentralized protocol for computing the routing table entries in Kademlia to accelerate lookups. 
\ourAlgorithm~is motivated by the multi-armed bandit problem, and can automatically adapt to heterogeneity and dynamism in the network. 
Experimental results show \ourAlgorithm~achieving between 15--50\% lower lookup latencies compared to state-of-the-art baselines. 

\end{abstract}
\section{Introduction}

Decentralized peer-to-peer applications (dapps) fueled by successes in blockchain technology are rapidly emerging as secure, transparent and open alternatives to conventional centralized applications.
Today dapps have been developed for a wide gamut of application areas spanning payments, decentralized finance, social networking, healthcare, gaming etc., and have millions of users and generate billions on dollars in trade~\cite{dappusage}.  
These developments are part of a growing movement to create a more ``decentralized Web'',  in which no single administrative entity (e.g., a corporation or government) has complete control over important web functionalities (e.g., name resolution, content hosting, etc.) thereby providing greater power to application end users~\cite{trautwein2022design,alabdulwahhab2018web}.

A fundamental operation in dapps is secure, reliable data storage and retrieval.    
Over the past two decades, the cloud (e.g., Google, Facebook, Amazon) together with content delivery networks (CDNs; e.g., Akamai, CloudFlare) have been largely responsible for storing and serving data for Internet applications.
Infrastructure in the cloud or a CDN is typically owned by a single provider, making these storage methods unsuitable for dapps.  
Instead dapps---especially those built over a blockchain (e.g., ERC 721 tokens in Ethereum)---directly resort to using the blockchain for storing application data. 
However, mainstream blockchains are notorious for their poor scalability which limits the range of applications that can be deployed on them. 
In particular, realizing a decentralized Web that supports sub-second HTTP lookups at scale is infeasible with today's blockchain technology.  

To fill this void, a number of recent efforts have designed decentralized peer-to-peer (p2p) data storage networks---such as IPFS~\cite{benet2014ipfs}, Swarm~\cite{team2021swarm,tron2020book,ethswarm}, Hypercore protocol~\cite{hypercore}, Safe network~\cite{safenetwork} and Storj~\cite{storj2018storj}---which are seeing rapid mainstream adoption. 
E.g., the IPFS network has more than 3 million client requests per week with hundreds of thousands of storage nodes worldwide as part of the network~\cite{trautwein2022design}.  
In these networks, each unique piece of data is stored over a vast network of servers (nodes) with each server responsible for storing only a small portion of the overall stored data unlike blockchains. 
The networks are also characterized by their permissionless and open nature, wherein any individual server may join and participate in the network freely.  
By providing appropriate monetary incentives (e.g., persistent  storage in IPFS can be incentivized using Filecoin~\cite{ipfsfilecoin,filecoin}) for storing and serving data, the networks encourage new servers to join which in turn increases the net storage capacities of these systems.  

A key challenge in the p2p storage networks outlined above is how to efficiently locate where a desired piece of data is stored in the network.  
Unlike cloud storage, there is no central database that maintains information on the set of files hosted by each server at any moment. 
Instead, p2p storage networks rely on a distributed hash table (DHT) protocol for storage and retrieval by content addressing data. 
While tens of DHT constructions have been proposed in the past, in recent years the Kademlia DHT~\cite{maymounkov2002kademlia} has emerged as the de facto protocol and has been widely adopted by practitioners. 
For instance, IPFS, Swarm, Hypercore protocol, Safe network and Storj are all based on Kademlia. 
To push or pull a data block from the network, the hash of the data block (i.e., its content address) is used to either recursively or iteratively route a query through the DHT nodes until a node responsible for storing the data block is found.  
  
For latency-sensitive content lookup applications, such as the Web where a delay of even a few milliseconds in downloading webpage objects can lead to users abandoning the website~\cite{webloadtime}, it is imperative that the latency of routing a query through Kademlia is as low as possible. 
Each Kademlia node maintains a routing table, which contains IP address references to other Kademlia nodes in the network. 
The sequence of nodes queried while performing a lookup is dictated by the choice of routing tables at the nodes. 
Today's Kademlia implementations choose the routing tables completely agnostic of where the nodes are located in the network.  
As a result, a query in Kademlia may take a route that criss-crosses continents before arriving at a target node costing significant delay. 
Moreover, the open and permissionless aspects makes the network inherently heterogeneous:  nodes can differ considerably in their compute, memory and network capabilities which creates differences in how fast nodes respond to queries; 
data blocks published over the network vary in their popularity, with demand for some data far  exceeding others; 
the network is also highly dynamic due to peer churn and potentially evolving user demand for data (e.g., a news webpage that is popular today may not be popular tomorrow). 
Designing routing tables in Kademlia that are tuned to the various heterogeneities and dynamism in the network to minimize content lookup delays is therefore a highly nontrivial task. 

Prior works have extensively investigated how to design location-aware routing tables in Kademlia.  
For example, the proximity neighbor selection (PNS)~\cite{castro2002exploiting} advocates choosing routing table peers that are geographically close to a node (more precisely, peers having a low round-trip-time (RTT) ping delay to the node), and proximity routing (PR)~\cite{castro2002exploiting} favors relaying a query to a matching peer with the lowest RTT in the routing table. 
While these location-aware variants have been shown to exhibit latency performance strictly superior to the original Kademlia protocol~\cite{maymounkov2002kademlia}, they are not adaptive to the heterogeneities in the network.
PNS is also prone to Sybil attacks which diminishes its practical utility~\cite{pecori2016s}---an adversary controlling a large number of fake Kademlia nodes at a location can cause a nearby node's routing table to be completely filled with adversarial IP addresses. 
Real world Kademlia implementations in libp2p~\cite{libp2pkad}, IPFS and other file sharing networks therefore have resorted to maintaining the peer routing tables largely per the original Kademlia protocol. 
S/Kademlia~\cite{baumgart2007s} is a particularly popular implementation which uses public-key cryptography for authentication and proof-of-work puzzles to avoid Sybil attacks.  

We propose \ourAlgorithm, a decentralized, adaptive algorithm for selecting routing table entries in Kademlia to minimize object lookup times (to push or get content) while being robust against Sybil attacks.    
\ourAlgorithm~is motivated by the (combinatorial) multi-armed bandit (MAB) problem~\cite{slivkins2019introduction,bubeck2012regret}, with each Kademlia node acting as an independent MAB player and the node's routing table configurations being the arms of the bandit problem. 
By balancing exploring new routing table configurations with exploiting known configurations that have resulted in fast lookup speeds in the past, a node is able to adaptively discover an efficient routing table that provides fast lookups. 
Importantly, the discovered routing table configuration at a node is optimized precisely to the pattern of lookups specific to the node. 
Our proposed algorithm is fully decentralized, relying only on local timestamp measurements for feedback at each node (time between when a query was sent and its corresponding response received) and does not require any cooperation between nodes. 
To protect against Sybil attacks, \ourAlgorithm~relies on a novel exploration strategy that explicitly avoids including nodes that have a low RTT to a node within the node's routing table with the RTT threshold specified as a security parameter. 
At the same time, \ourAlgorithm's exploration strategy also avoids selecting nodes very far from a node. 
To accelerate discovery of an efficient routing table configuration, \ourAlgorithm~decomposes the problem into parallel independent MAB instances at each node, with each instance responsible for optimizing peer entries of a single $k$-bucket.  
In summary, the contributions of this paper are:  
\begin{enumerate}
\item We consider the problem of efficient routing table design in Kademlia and formulate it as an instance of the multi-armed bandit problem.
Using data-driven techniques for optimizing lookup speeds in structured p2p networks has not been proposed before, to our best knowledge.  
\item We propose \ourAlgorithm, a fully decentralized and non-cooperative algorithm for learning the routing table entries to accelerate lookups. \ourAlgorithm~is adaptive to both the traffic demand patterns of the users and the heterogeneities in the network. 
\item We validate \ourAlgorithm~through simulations under various network and traffic settings. In each case, we observe \ourAlgorithm~to consistently outperform baselines by between 15--50\% in latency.  
\end{enumerate} 

\section{Background}

\subsection{Kademlia}

\begin{figure}[!tbp]
  \centering
  \subfloat[]{\includegraphics[width=0.37\textwidth]{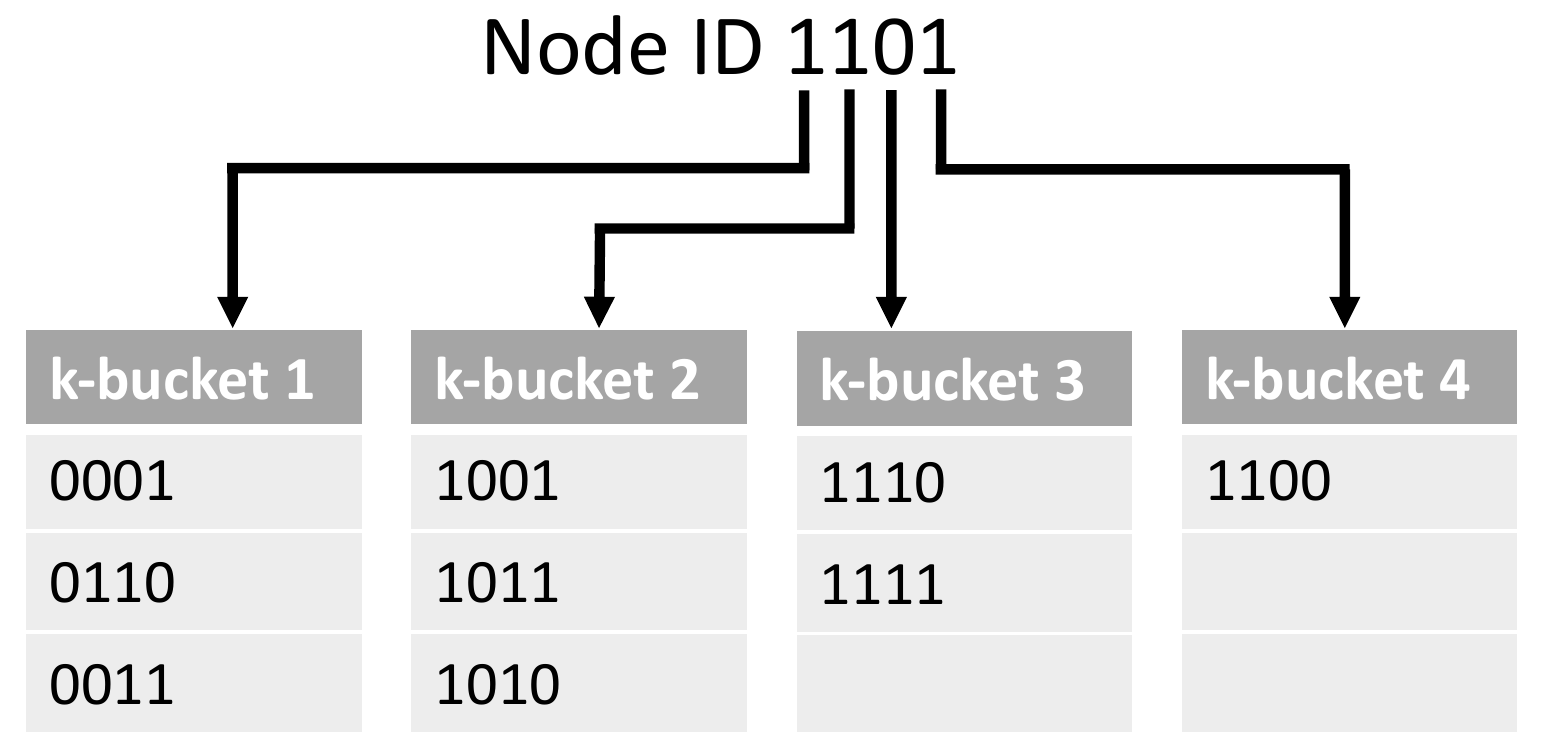}\label{fig:exbuckets}}
  \hfill
  \subfloat[]{\includegraphics[width=0.6\textwidth]{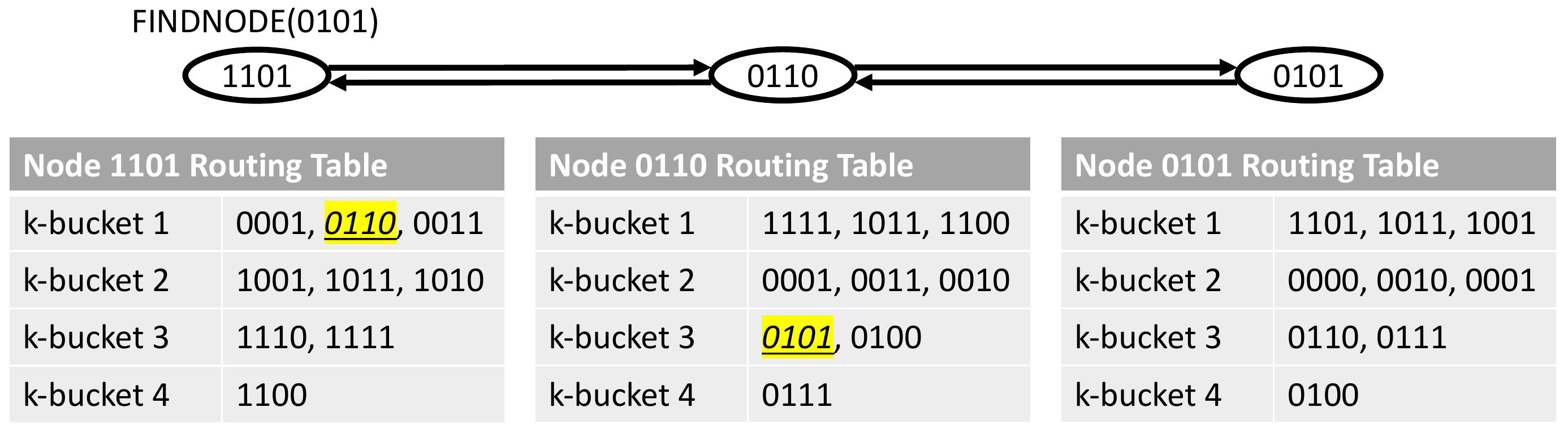}\label{fig:expathsnotree}}
  \caption{(a) Example of $k$-buckets at a node in a network with 4-bit node IDs. (b) Example of a recursive routing path taken to lookup key 0101. A yellow-highlighted node ID is a peer to which the query is forwarded for the next hop.}
\end{figure}

{\bf Overview.}
Kademlia is arguably the most popular protocol for realizing a structured p2p system on the Internet today. 
In a Kademlia network, each node is assigned a unique binary node ID from a high-dimensional space (e.g., 20 byte node IDs are common). 
When the network size is large, it is difficult for a node to know the node ID of every single node in the network. 
A node may have knowledge of node IDs of only a small number (such as logarithmic in network size) of other nodes. 
The most basic operation supported by Kademlia is {\em key-based routing} (KBR) wherein given a key from the node ID space as input to a node, the protocol determines a routing path from the node to a target node whose ID is the `closest' to the input key. 
Closeness between a key and a node ID in Kademlia is measured by taking the bitwise-XOR between the two binary strings, and converting the resultant string as a base-10 integer. 
The basic KBR primitive can be used to realize higher-order functions such as a distributed hash table (DHT).  
In a DHT, a (key, value) store is distributed across nodes in the network.  
A (key, value) pair is stored at a node whose node ID is the closest to the key according to the XOR distance metric. 
To protect against node failures, in practice a copy of the (key, value) is also stored at a small number (e.g., 20) of sibling nodes that are nodes whose IDs are closest to the initial storing node.  
To store a (key, value) in the network a node invokes a \textsc{Store}(key, value) remote procedure call (RPC), and to fetch a value corresponding to a key the node calls a \textsc{FindValue}(key) RPC~\cite{maymounkov2002kademlia,baumgart2007s}. 
KBR is implemented as a \textsc{FindNode}(key) RPC, which returns the Kademlia node having the closest ID to key. 

\smallskip 
\noindent 
{\bf Routing.}
Each Kademlia node maintains a routing table containing node ID, IP address and port information of other peers using which \textsc{Store}, \textsc{FindValue} or \textsc{FindNode} queries are routed to their appropriate target nodes. 
For node IDs that are $n$ bits long, the routing table at each node comprises of $n$ $k$-buckets, where each $k$-bucket contains information about $k$ peers.  
The IDs of peers in the $i$-th $k$-bucket of a node's routing table share the first $i-1$ bits with the node's ID, while differing in the $i$-th bit (Fig.~\ref{fig:exbuckets}). 
For a network with $m$ nodes, it can be shown that on average only the first $\log(m)$ $k$-buckets can be filled with peer entries while the remaining $k$-buckets are empty due to lack of peers satisfying the prefix constraints. 

Queries in Kademlia are routed either recursively or iteratively across nodes. 
In a recursive lookup, a query is relayed sequentially from one node to the next until a target node is found. 
The response from the target node is then relayed back on the reverse path to the query initiator. 
In an iterative lookup, a query initiating node itself sequentially contacts nodes until a target node is found, and receives a response directly from the target node.  
We focus primarily on recursive routing in this work (Fig.~\ref{fig:expathsnotree}). 

When a query for key $x$ is received at a node $v$, the node searches for a peer $v'$ within its routing table with an ID that is closest to $x$. 
If the distance between the IDs $x$ and $v'$ is less than the distance between $x$ and $v$, then $v$ forwards the query to node $v'$. 
Later when $v$ receives a response to the query from $v'$, $v$ relays the response back to the node from whom it received the query. 
If the distance between $x$ and $v$ is less than than distance between $x$ and $v'$, the node $v$ issues an appropriate response for the query to the node from whom $v$ received the query. 
To avoid lookup failures, a query initiator issues its query along $\alpha$ (e.g., $\alpha = 3$) independent paths. 
This basic lookup process described above is fundamental to implementing the \textsc{Store}, \textsc{FindValue} and \textsc{FindNode} functions. 
We point the reader to prior papers~\cite{maymounkov2002kademlia,baumgart2007s} for more details on the lookup process. 

\subsection{Lookup Latency and Node Geography}
\label{s:lookupLatency}

A Kademlia node may include any peer it has knowledge of within its $k$-buckets, provided the peer satisfies the required ID prefix conditions for the $k$-bucket. 
Nodes get to know of new peers over the course of receiving queries and responses from other nodes in the network. 
As node IDs are assigned to nodes typically in a way that is completely independent of where the nodes are located in the world, in today's Kademlia it is likely that the peers within a $k$-bucket belong to diverse geographical regions around the world without any useful structure.  
E.g., a recent study~\cite{trautwein2022design} measuring performance on the IPFS network reports a 90-th percentile content storing latency of 112s with 88\% of it attributed to DHT routing latency. 
For retrieving content, the reported 90-th percentile delay is 4.3s which is more than $4\times$ the latency of an equivalent HTTPS lookup.
Similar observations have been made on other Kademlia systems in the past as well~\cite{crosby2007analysis}.   

There has been an extensive amount of work on reducing lookup latencies in DHTs by taking the physical location of nodes on the underlay~\cite{hildrum2002distributed,jain2003study,karger2002finding,plaxton1997accessing,ratnasamy2002topologically,zhao2002locality}.  
For instance, Kaune et al.~\cite{kaune2008embracing} propose an algorithm that takes the ISPs of nodes into consideration, and also uses network coordinates for reducing latencies. 
Jimenez et al.~\cite{jimenez2011sub} tune the number of parallel lookup queries sent or bucket size to achieve speedup. 
Chen et al.~\cite{chen2010design} minimize the mismatch between Kademlia's logical network and the underlying physical topology through a landmark binning algorithm and RTT detection. 
Gummadi et al.~\cite{gummadi2003impact} do a systematic comparison of proximity routing and proximity neighbor selection on different DHT protocols. 
The algorithms proposed in these and other prior works are hand-crafted designs, which are not tuned to the various heterogeneities in the network.  
Moreover, security in these proposed methods has not been discussed as a first-order concern. 
Indeed, today's DHT implementations have not adopted these proposals into their systems.

\subsection{Security in Kademlia}

A Kademlia node is susceptible to various attacks, especially in permissionless settings. 
We consider the following attacks in this work. 

\smallskip 
\noindent 
{\bf Eclipse and Sybil attacks.}
In an Eclipse attack, an attacker blocks one or more victim nodes from connecting to other nodes in the network by filling the victim nodes' routing table with malicious nodes. 
In a Sybil attack, the attacker creates many fake nodes with false identities to spam the network, which may eventually undermine the reputation of the network. 
Today's Kademlia implementations circumvent these attacks using ideas largely inspired from  S/Kademlia~\cite{baumgart2007s}.
In S/Kademlia, the network uses a supervised signature issued by a trustworthy certificate authority or a proof-of-work puzzle signature to restrict users' ability to freely generate new nodes. 

\smallskip 
\noindent 
{\bf Adversarial routing.}
In Kademlia, 
a malicious node within an honest node's routing table may route messages from the honest node to a group of malicious nodes. This attack is called adversarial routing, and it may cause delays and/or make the queries unable to find their target keys. 
To alleviate adversarial routing, S/Kademlia makes nodes use multiple disjoint paths to lookup contents at a cost of increased network overhead. 



\smallskip 
\noindent 
{\bf Churn attack.}
Attackers can also enter and exit the network constantly to induce churns to destabilize the network. Kademlia networks handle these kind of attacks by favoring long-lived nodes \cite{maymounkov2002kademlia,koutrouli2012taxonomy}.

\section{System Model}
\label{s:system model}

We consider a Kademlia network over a set of nodes $V$ with each node $v \in V$ having a unique IP address and a node ID from ID space $\{0, 1\}^n$. 
Each node maintains $n$ $k$-buckets in its routing table, with each $k$-bucket containing the IP address and node ID of up to $k$ other peers satisfying the ID prefix condition. 
We consider a set $S$ of (key, value) pairs that have been stored in the network; each (key, value) pair $(x, y) \in S$ is stored in $k$ peers whose IDs are closest to $x$ in XOR distance. 
We let $S_x$ denote the set of keys in $S$. 
Time is slotted into rounds, where in each round a randomly chosen node performs a lookup for a certain key. 
If a node $v \in V$ is chosen during a round, it issues a lookup query for key $x \in S_x$ where $x$ is chosen according to a demand distribution $p_v$, i.e., $p_v(x)$ is the probability key $x$ is queried. 
We focus primarily on recursive routing in this paper. 
When a node $v$ initiates a query for key $x$, it sends out the query to $\alpha$ closest (to $x$, in XOR distance) peers in its routing table. 
For any two nodes $u, w$, $l(u, w) \geq 0$ is the latency of sending or forwarding a query from $u$ to $w$. 
When a node $w$ receives a query for key $x$ and it has stored the value for $x$, the node returns the value back to the node $u$ from whom it received the query. 
Otherwise, the query is immediately forwarded to another node that is closest to $x$ in $w$'s routing table.  
When a node $w$ sends or forwards a value $y$ to a node $u$, it first takes time $\delta_w \geq 0$ to upload the value over the Internet followed by time $l(w, u)$ for the packets to propagate to $u$.
We do not model the time take to download the value, as download bandwidth is typically higher than upload bandwidth. 
Thus, for a routing path $v, u, w$ with $v$ being the query initiator and $w$ storing the desired value, the overall time taken for $v$ to receive the value is $l(v, u) + l(u, w) + \delta_w + l(w, u) + \delta_u + l(u,v)$.  
The above outlines our lookup model for the DHT application. 
For KBR, we follow the same model except only a single query (i.e., $\alpha = 1$) is sent by the initiating node. 
We assume each node has an access to the IP addresses and node IDs of a small number of random nodes in the network. 

\smallskip 
\noindent 
{\bf Problem statement.} 
For each of the KBR and DHT applications, our objective is to design a decentralized algorithm for computing each node's routing table such that the average time (averaged over the distribution of queries sent from the node) taken to perform a lookup is minimized at the node. 
We consider non-cooperative algorithms where a node computes its routing table without relying on help from other nodes.  

\section{\ourAlgorithm}
\label{s:algorithm}

\subsection{Overview}
$\ourAlgorithm$ is a fully decentralized and adaptive algorithm that learns a node's routing table to minimize lookup times, purely based on the node's past interactions with the network. 
\ourAlgorithm~is inspired by ideas from non-stationary and streaming multi-armed bandit problems applied to a combinatorial bandit setting~\cite{assadi2020exploration,cavenaghi2021non,nobari2019dba}. 
A \ourAlgorithm~node balances efficient routing table configurations it has seen in the past (exploitation) against new, unseen configurations (exploration) with potentially even better latency efficiency. 
For each query that is initiated or routed through a $\ourAlgorithm$ node, the node stores data pertaining to which peer(s) the query is routed to and how long it takes for a response to arrive. 
This data is used to periodically make a decision on whether to retain peers currently in the routing table, or switch to a potentially better set of peers. 
Treating the routing table as the decision variable of a combinatorial MAB problem leads to a large space and consequently inefficient learning. 
We therefore decompose the problem into $n$ independent subproblems, where the $i$-th subproblem learns only the entries of the $i$-th $k$-bucket. 
This decomposition is without loss of generality as each query is routed through peers in at most one $k$-bucket. 
In the following we therefore explain how a \ourAlgorithm~node can learn the entries of its $i$-th $k$-bucket.  

In \ourAlgorithm, a decision on a $k$-bucket (i.e., whether to change one or more entries of the bucket) is made each time after $b$ queries are routed via peers in the bucket (e.g., $b = 100$ in our experiments). 
We call the time between successive decisions on a $k$-bucket as an epoch. 
Before each decision, a performance score is computed for each peer in the bucket based on the data collected over the epoch for the bucket. 
Intuitively, the performance score for a peer captures how frequently queries are routed through the peer {\em and} how fast responses are received for those queries. 
By comparing the performance scores of peers in the bucket during the current epoch against the scores of peers in the previous epoch, \ourAlgorithm~discovers the more efficient bucket configuration which is then used as the $k$-bucket for the subsequent epoch.\footnote{To increase stability under churn, we may choose to replace only unresponsive peers---as in the original Kademlia protocol---in each epoch.}    
To discover new (unseen) bucket configurations, \ourAlgorithm~also explores random bucket configurations according to a user-defined schedule. 
In our implementation, one entry on the $k$-bucket is chosen randomly every other epoch. 
The overall template of \ourAlgorithm~is presented in Algorithm~\ref{algo:highlevel}.

\begin{algorithm}[t]
\DontPrintSemicolon
\SetKwInOut{Input}{input}\SetKwInOut{Output}{output}
\Input{
data $\mathcal{D}$ on queries sent during current epoch; 
peers $\Gamma_\text{curr}$ and $\Gamma_\text{prev}$ in $k$-bucket of current and previous epochs respectively; 
total score $\text{PrevScoreBucket}$ of previous $k$-bucket; 
flag $F$ indicating whether to explore in next epoch; 
list $\mathcal{L}$ of peers eligible to be included within $k$-bucket; 
security parameter $\rho$; 
}
\Output{updated set of peers $\Gamma_\text{next}$ for next epoch;}
\tcc{Score each peer in $\Gamma_\text{curr}$ using a scoring algorithm based on measurements collected during epoch}
score($u$) $\leftarrow$ \textsc{ScorePeer}$(u, \mathcal{D})$, for each peer $u \in \Gamma_\text{curr}$ \; 
\eIf {flag $F$ is true} {
\tcc{Replace worst peer with a random peer during next epoch} 
$u^* \leftarrow \text{argmin}_{u \in \Gamma}$ score$(u)$ \;
$\Gamma_\text{next} \leftarrow \Gamma_\text{curr} \backslash \{ u^* \} \cup \textsc{SelectRandomPeer}(\mathcal{L}, \rho)$ \; 
}
{
\tcc{Choose best peer set between current and previous epoch as decision for next epoch} 
\eIf {\textsc{ScoreBucket}$(\Gamma_\text{curr}, \mathcal{D})$ $>$ $\mathrm{PrevScoreBucket}$ } {
$\Gamma_\text{next} \leftarrow \Gamma_\text{curr}$
} {
$\Gamma_\text{next} \leftarrow \Gamma_\text{prev}$ \; 
}
} 
\caption{Algorithm outline for updating entries of $i$-th $k$-bucket of node $v$ in each epoch.}
\label{algo:highlevel}
\end{algorithm}

\subsection{Scoring Function}

During an epoch with $k$-bucket $\Gamma_\text{curr}$, let $q_1, q_2, \ldots, q_r $ be the set of queries that have been sent or relayed through one or more peers in the $k$-bucket. 
For each query $q_i, 1 \leq i \leq r$, let $d_i(u) \geq 0$ be the time taken to receive a response upon sending or forwarding the query through peer $u$ for $u \in \Gamma_\text{curr}$.  
If $q_i$ is not sent or forwarded through a peer $u \in \Gamma_\text{curr}$ we let $d_i(u) = \Delta$ where $\Delta \geq 0$ is a user-defined penalty parameter.\footnote{Notice that $d_i(u)$ is well-defined for both recursive and iterative routing.}  
A large value for $\Delta$ causes \ourAlgorithm~to favor peers that are frequently used in the $k$-bucket, while a small value favors peers from which responses are received quickly. 
In our experiments, we choose $\Delta$ to be a value that is slightly larger than the moving average of latencies of lookups going through the bucket. 
The function \textsc{ScoringFunction}$(u, \mathcal{D})$ to compute the score for a peer $u$ is then defined as 
$ 
\text{score}(u) = \textsc{ScoringFunction}(u, \mathcal{D}) = - \sum_{i = 1}^r d_i(u) ~ \forall u \in \Gamma_\text{curr}. 
$ 
The overall score for the $k$-bucket is then given as 
$ 
\textsc{ScoreBucket}(\Gamma_\text{curr}, \mathcal{D}) =  - \sum_{u \in \Gamma_\text{curr}} \text{score}(u) / |\Gamma_\text{curr}|. 
$
For a $k$-bucket that is empty, we define its score to be $-\Delta$. 
 
 \subsection{Random Exploration}
 
To discover new $k$-bucket configurations with potentially better performance than past configurations, a \ourAlgorithm~node includes randomly selected peers within its bucket through the \textsc{SelectRandomPeer}$()$ function as outlined in Algorithm~\ref{algo:highlevel}.  
The \ourAlgorithm~node maintains a list $\mathcal{L}$ of peers eligible to be included within its $k$-bucket, which satisfy the required node ID prefix conditions.
In addition to the peer IP addresses, we assume the node also knows the RTT to each peer in the list. 
For a random exploratory epoch, the node replaces the peer having the worst score from the previous epoch with a randomly selected peer from the list. 
The number of peers in the bucket that are replaced with a random peers can be configured to be more than one more generally.  

A key contribution in \ourAlgorithm~is how peers are sampled from the list of known peers to be included in the $k$-bucket.
Depending on the number of nodes in the network, and the index of the $k$-bucket, the number of eligible peers can vary with some peers close to the node while some farther away (in RTT sense).     
A na\"ive approach of sampling a node uniformly at random from the list, can eventually lead to a bucket configuration in which all peers are located close to the node. 
This is due to the algorithm `discovering' the proximity neighbor selection (PNS) protocol which has been demonstrated to have efficient latency performance compared to other heuristics~\cite{gummadi2003impact,baumgart2007s}. 
 However, as with PNS, the routing table learned with a uniformly random sampling strategy is prone to a Sybil attack as it relatively inexpensive to launch a vast number of Sybil nodes concentrated at a single location close to a victim node(s)~\cite{putman2018business,steiner2007exploiting}. 
While the PNS peer selection strategy does not have an efficient performance in all scenarios (e.g., if the node upload latencies are large; see \S\ref{s:evaluation}), in cases where it does, \ourAlgorithm~would be susceptible to attack. 
What we desire, therefore, is to learn a routing table configuration in which not all peers are located close to the node.
Such a routing table configuration may not be performance efficient (e.g., PNS may have a better latency performance in certain scenarios), but is more secure compared to PNS.  

%

We capture this intuition by introducing a security parameter $\rho \geq 0$, that is user-defined, to restrict the choice of peers that are sampled during exploration. 
For a chosen $\rho$ value, a \ourAlgorithm~node computes a subset $\mathcal{L}_{>\rho} \subseteq \mathcal{L}$ of peers to whom the RTT is greater than $\rho$ from the node. 
The $\textsc{SelectRandomPeer}(\mathcal{L}, \rho)$ then samples a peer uniformly at random from $\mathcal{L}_{>\rho}$. 
A high value for $\rho$ selects peers that are at a distance from the node, providing security against Sybil attacks at a cost of potentially reduced latency performance (and vice-versa).

\section{Evaluation}
\label{s:evaluation}

\subsection{Experiment Setup}
\label{s:expsetup}
We evaluate \ourAlgorithm~using a custom discrete-event simulator built on Python following the model presented in \S\ref{s:system model}.\footnote{We do not use the erstwhile popular OverSim \cite{baumgart2007oversim} and PeerSim \cite{montresor2009peersim} simulators, as they are outdated and no longer maintained by their authors. \ourAlgorithm~simulator is available at https://github.com/yunqizhang99/KadabraSim/.}

\smallskip 
\noindent 
{\bf Baselines.} 
Since the main focus of \ourAlgorithm~is on how to configure the routing table, we compare our algorithm against the following baselines with differing (i) routing table (bucket) population mechanisms, and (ii) peer selection methods during query forwarding: 

\begin{figure}[!tbp]
  \centering
   \subfloat[]{\includegraphics[width=0.5\textwidth]{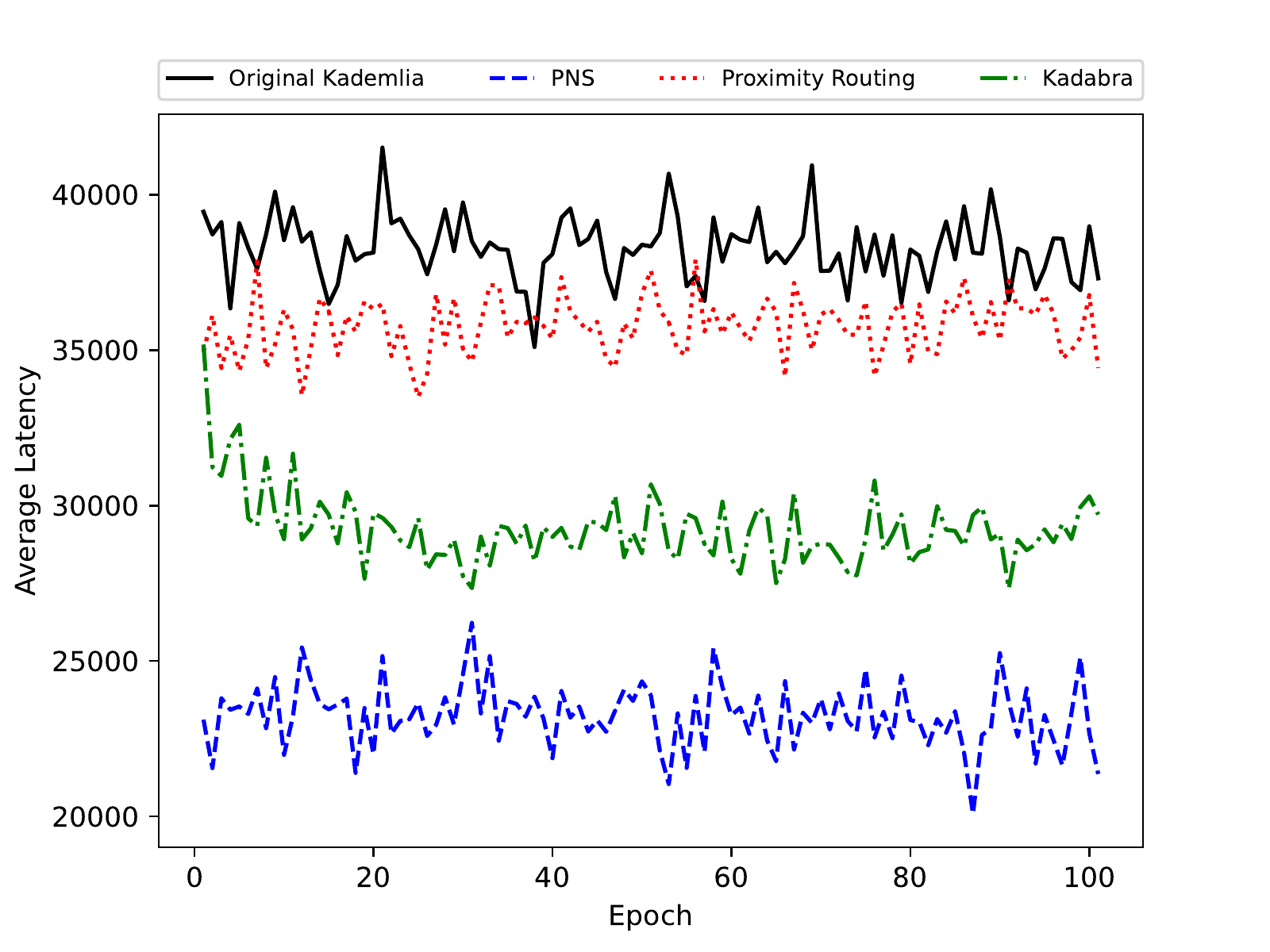}\label{fig:ednsc1latef1}}
    \subfloat[]
{\includegraphics[width=.5\textwidth]{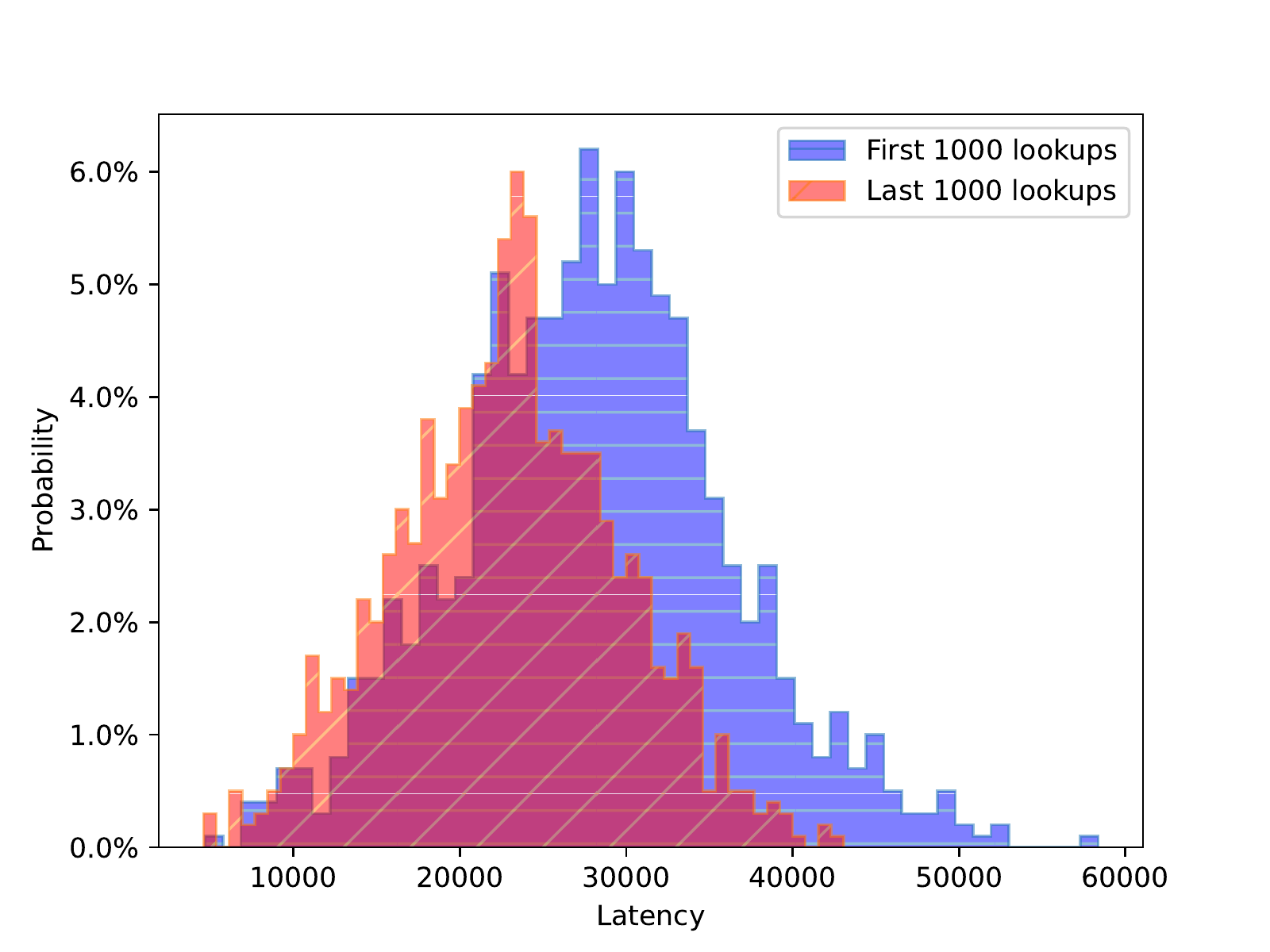}\label{fig:ednspdf1}}
  \caption{Nodes in a square under uniform demand: (a) Average latency of queries in each epoch for queries routed through the 1st $k$-bucket of an arbitrary node. (b) Histogram of lookup latencies in \ourAlgorithm~during the first and last 1000 rounds in a 10 million query run.
}
\end{figure}

\smallskip
\noindent
{\em (1) Vanilla Kademlia}~\cite{maymounkov2002kademlia,baumgart2007s}. 
    The original Kademlia protocol in which a node populates its buckets by randomly adding  peers from node ID ranges corresponding to the buckets. 
    When forwarding a query, the node chooses the peer whose node ID is closest (in XOR distance) to the query's target node ID from the appropriate bucket.

\smallskip
\noindent
{\em (2) Proximity routing (PR)}~\cite{gummadi2003impact,baumgart2007s}. In PR buckets are populated exactly as in vanilla Kademlia. 
    However, when routing a query the query is sent to the peer in the appropriate $k$-bucket that is closest to the node in RTT. 

\smallskip
\noindent
{\em (3) Proximity neighbor selection (PNS)}~\cite{gummadi2003impact,baumgart2007s}. In PNS the node picks peers which are closest to itself (in terms of RTT) from among eligible peers to populate each $k$-bucket. 
When forwarding a query, the node chooses the peer whose node ID is closest to the target node ID from the appropriate bucket.

\smallskip 
\noindent 
{\bf Network settings.} 
We consider two network scenarios: nodes distributed over a two-dimensional Euclidean space, and nodes distributed over a real-world geography. 

\smallskip 
\noindent 
{\em (1) Nodes in a square.} 
In this setting, 2048 nodes are assigned random locations within a 10000 $\times$ 10000 square. 
The latency $l(u,v)$ between any two nodes $u, v$ is given by 
$ l(u, v) = || u - v||_2 + w(u,v) $, where $|| u - v ||_2$ is the Euclidean distance between $u$ and $v$ on the square and $w(u,v)$ is random perturbation from an uniform distribution between 100 and 5000. 
Each node has a node latency ($\delta$ in \S\ref{s:system model}) sampled uniformly between 100 and 2000. 

\smallskip 
\noindent 
{\em (2) Nodes in the real world.}
We again consider 2048 nodes located in various cities around the world, as reported by Ethereum node tracker~\cite{ethnodetrack}. 
The latency between nodes in any pair of cities is obtained from a global ping latency measurement dataset~\cite{wondnetpings}.\footnote{For cities not included in the ping dataset, we measure the latency to the geographically closest city available in the dataset.} 
Each node has a node latency sampled from an exponential distribution of mean 1000ms. 

\begin{figure}[!tbp]
  \centering
  \subfloat[]{\includegraphics[width=0.5\textwidth]{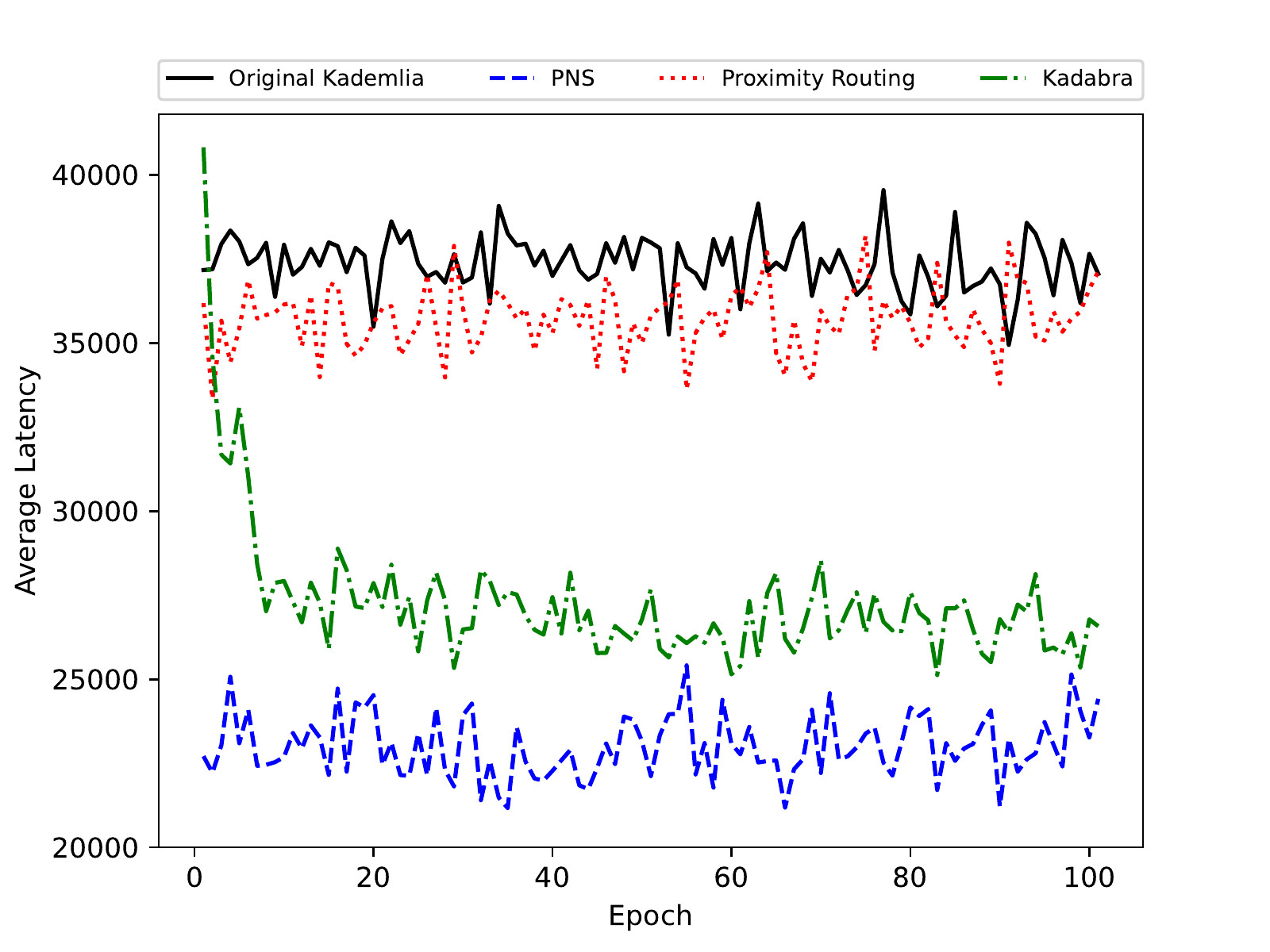}\label{fig:ednsc2late}}
  \hfill
  \subfloat[]{\includegraphics[width=0.5\textwidth]
{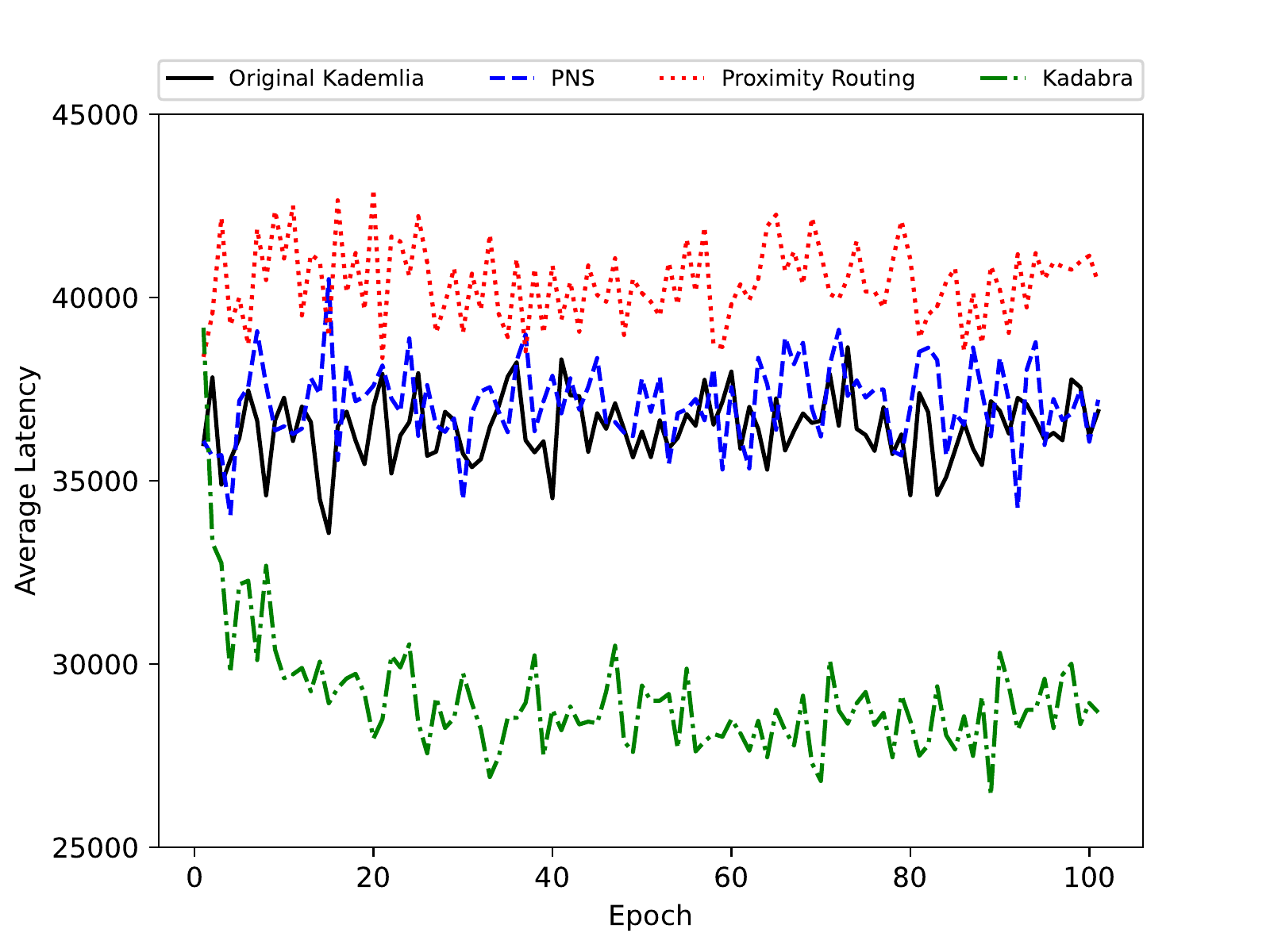}\label{fig:ednsc5late}}
  \caption{Nodes in a square: (a) Average latency during each epoch for queries routed through the 1st $k$-bucket of an arbitrary node with demand hotspots.  
  (b) Performance of a node within a region of high node latency nodes.}
\end{figure}

\smallskip 
\noindent 
{\bf Application and traffic patterns.} 
We first consider the KBR application under the following three traffic patterns: 

\smallskip 
\noindent 
{\em (1) KBR under uniform demand.}
In this setting, a node during a round (see \S\ref{s:system model}) issues a lookup to another node chosen uniformly at random from among the available nodes. 

\smallskip 
\noindent 
{\em (ii) KBR under demand hotspots.}
In this setting, there are 20\% of keys (nodes) that form the target destination for 80\% of lookups. 
The hotspot nodes are randomly chosen. 

\smallskip 
\noindent 
{\em (3) KBR under skewed network bandwidth.}
To model regions around the world with poor Internet speeds, we consider a subset of geographically close nodes whose node upload latencies (see \S\ref{s:system model}) are  twice as large as the average node latency in the network.  

In Appendices~\ref{apx:nodesinsquarekbr}--\ref{apx:iterativerouting}, we have presented additional results for the above settings, and have also considered the DHT application and iterative routing.

\subsection{Results}
\label{s:evalresults}

\noindent 
{\bf Nodes in a square.} 
Fig.~\ref{fig:ednsc1latef1} plots the average latency between forwarding a query through the 1st $k$-bucket and receiving a response during each epoch for an arbitrarily chosen node within the square. 
We observe that starting with a randomly chosen routing table configuration (at epoch 0), \ourAlgorithm~continuously improves its performance eventually achieving latencies that are 15\% better. 
Compared to the latencies in the original Kademlia protocol, \ourAlgorithm's latencies are lesser by more than 20\%. 
For this specific network setting, PNS shows the best performance (at the cost of poor security). 
We have used $\rho$ values of [400, 350, 300, 250, 200, 150, 100, 50, 0] for the different $k$-buckets (1st to last) in \ourAlgorithm, which results in slightly higher latencies compared to PNS.  

To show that all nodes in the network benefit from \ourAlgorithm, we conduct an experiment lasting for 10 million rounds (1 query per round from a random source to a random destination), with the sequence of first 1000 queries being identical to the sequence of the last 1000 queries.
Fig.~\ref{fig:ednspdf1} plots a histogram of the query latencies for the first and last 1000 queries in \ourAlgorithm.
We observe the 90-th percentile latency of \ourAlgorithm~during the last 1000 queries is lesser than that in the beginning by more than 24\%. 

Fig.~\ref{fig:ednsc2late} shows the average query latency over epochs for queries routed through the 1st $k$-bucket of an arbitrary node under hotspot demand. 
With certain keys being more popular than others, \ourAlgorithm~adapts the node routing tables biasing them for fast lookups of the popular keys---a capability that is distinctly lacking in the baselines. 
As a result, we observe \ourAlgorithm~outperforming the original Kademlia and PR by more than 25\%. 

To show that \ourAlgorithm~adapts to variations in the Internet capacities of nodes, we consider an experiment where nodes within an area (2000 $\times$ 2000 region in the center of the square) alone have a higher node latency (5000 time units) than the default node latency values. 
This setting models, for instance, low-income countries with below-average Internet speeds.
For a node within the high node latency region, PNS ends up favoring nearby peers also within that region which ultimately severely degrades the overall performance of PNS (Fig.~\ref{fig:ednsc5late}). 
\ourAlgorithm, on the other hand, is cognizant of the high node latencies in the region, and discovers $k$-bucket entries that provide more than 25\% improvement in latency performance compared to PNS. 

\begin{figure}[!tbp]
  \centering
  \subfloat[]{\includegraphics[width=0.5\textwidth]{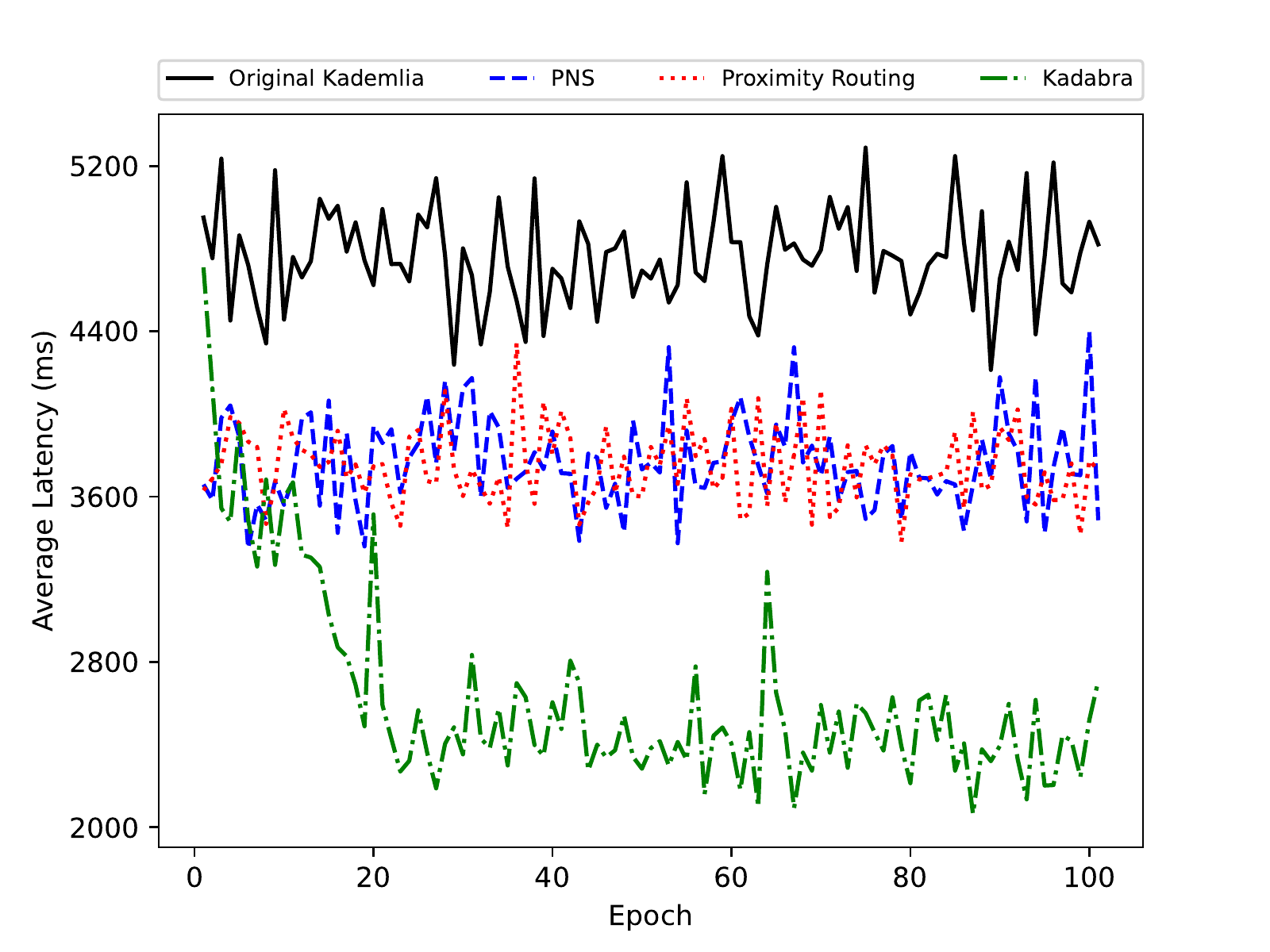}\label{fig:ndrwc1late}}
  \hfill
  \subfloat[]{\includegraphics[width=0.5\textwidth]{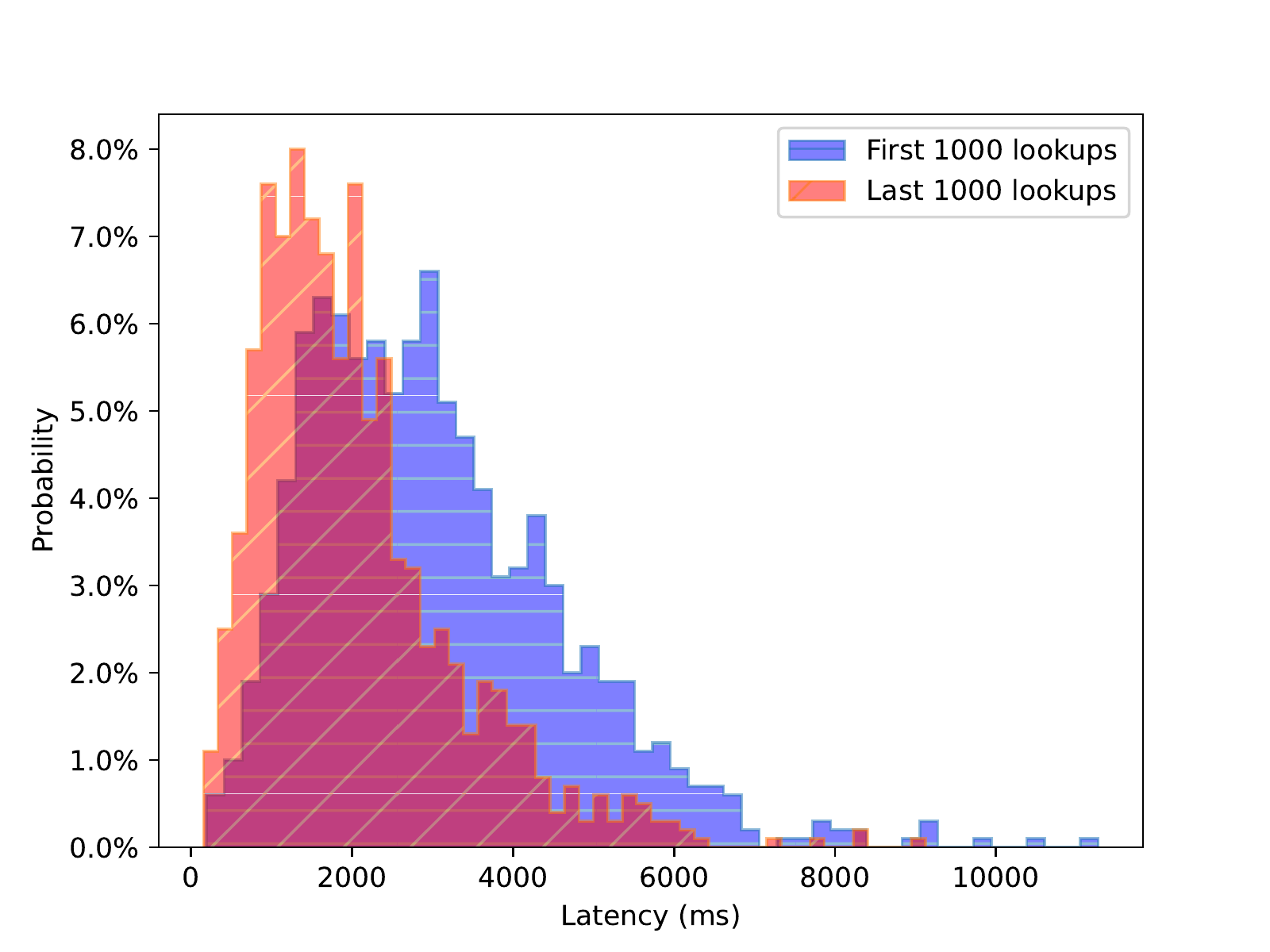}\label{fig:ndrwpdf}}
  \caption{Nodes in the real world: (a) Performance of queries routed through the 1st $k$-bucket of a node in Frankfurt. (b) Histograms of query latencies before and after learning in \ourAlgorithm~with 10 million lookups. }
\end{figure}

\begin{figure}[!tbp]
  \centering
  \subfloat[]{\includegraphics[width=0.498\textwidth]{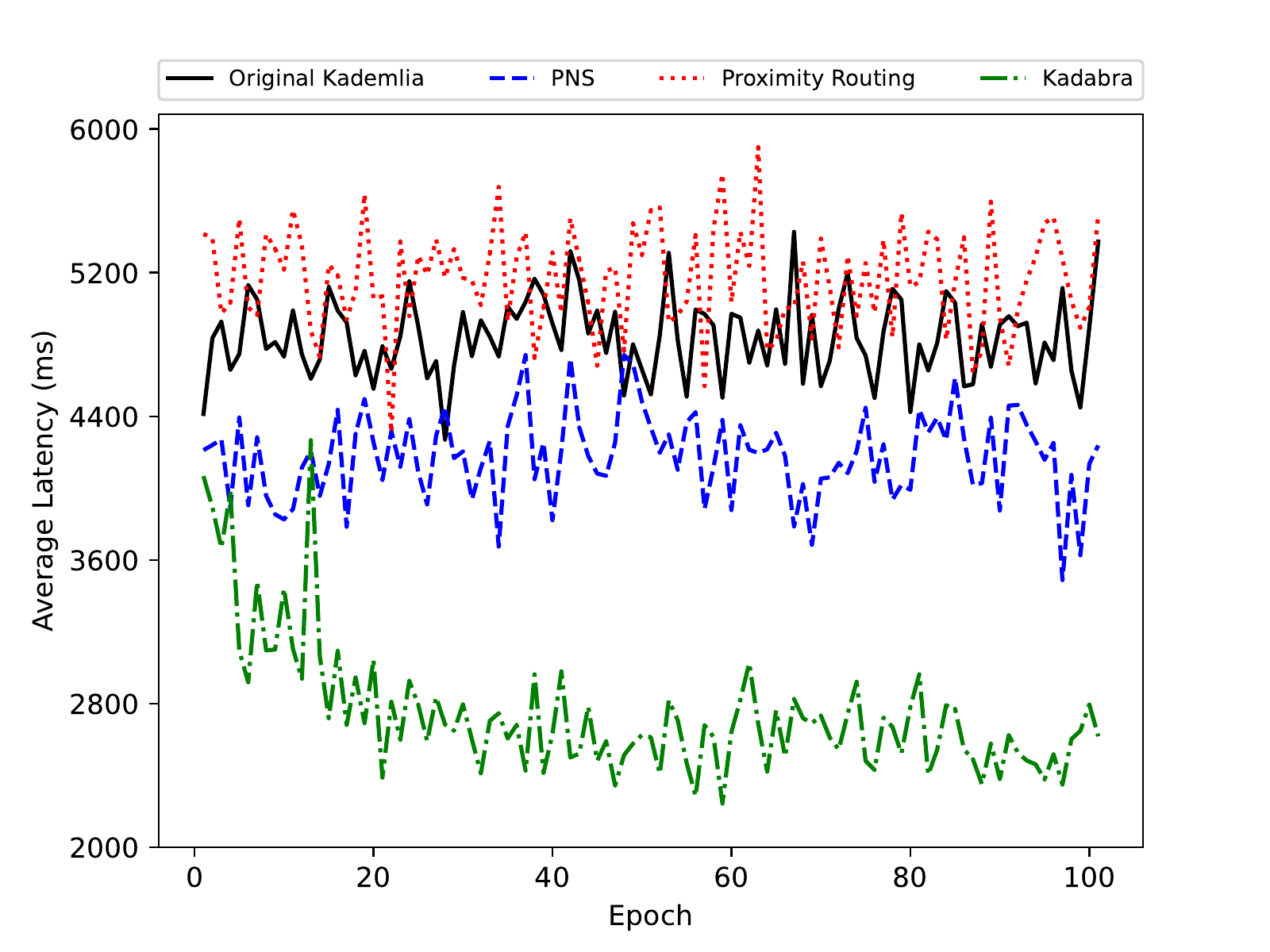}\label{fig:ndrwc2late}}
  \hfill
  \subfloat[]{\includegraphics[width=0.502\textwidth]{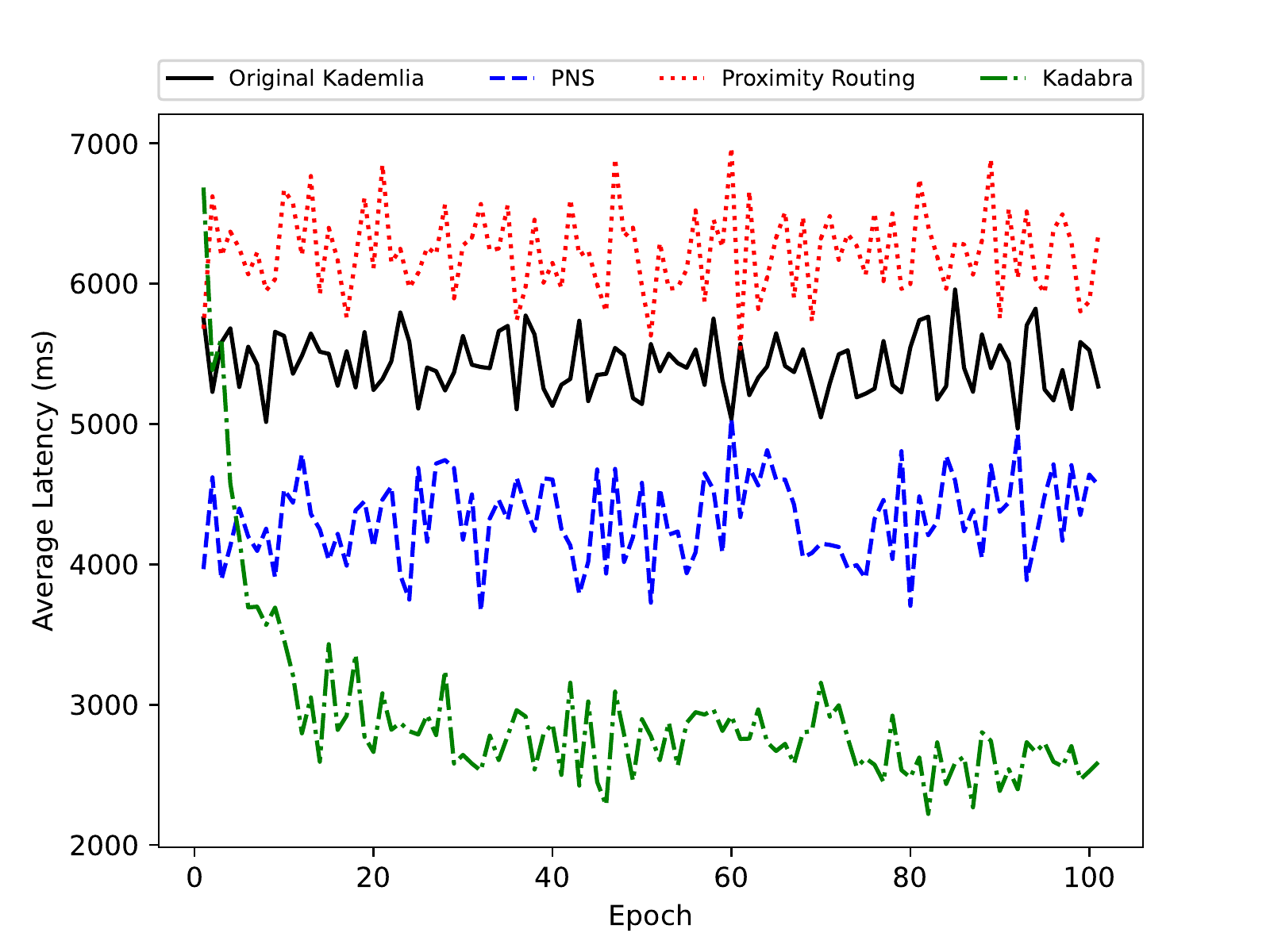}\label{fig:ndrwc5late}}
  \caption{Nodes in the real world: (a) Performance when there are demand hotspots. (b) Performance when 4\% of nodes near New York City have above average node latencies.}
\end{figure}

\noindent
{\bf Nodes in the real world.}
Unlike the square setting where nodes are uniformly spread out, in the real world setting nodes are concentrated around certain regions in the world (e.g., Europe or North America).
Moreover the node latencies are also chosen to reflect retrieval of large files~\cite{zichichi2020efficiency,trautwein2022design}. 
Fig.~\ref{fig:ndrwc1late} shows the latencies for queries routed through the 1st $k$-bucket of an arbitrary node (in this case, the node is located in Frankfurt). 
\ourAlgorithm~has 50\% lower latencies compared to the original Kademlia protocol and 35\% lower latencies compared to PNS and PR. 
This is because the baseline algorithms are not aware of the different node latencies of the peers, whereas \ourAlgorithm~is able to focus its search on peers having low node latencies.  
As in the square case, Fig.~\ref{fig:ndrwpdf} shows the benefit of \ourAlgorithm~extends to the entire network.  

Fig.~\ref{fig:ndrwc2late} shows performance when there are demand hotspots. 
Compared to uniform demand, both PNS and PR worsen in performance increasing the gap to \ourAlgorithm. 
A similar trend is observed in Fig.~\ref{fig:ndrwc5late} when we consider a region of nodes (near New York City in our experiments) and set their node latency to double the default average value. 
While even \ourAlgorithm~shows a slight degradation, it is still more than 40\% more efficient compared to PNS.

\begin{figure}[!tbp]
  \centering
  \subfloat[]{\includegraphics[width=0.496\textwidth]{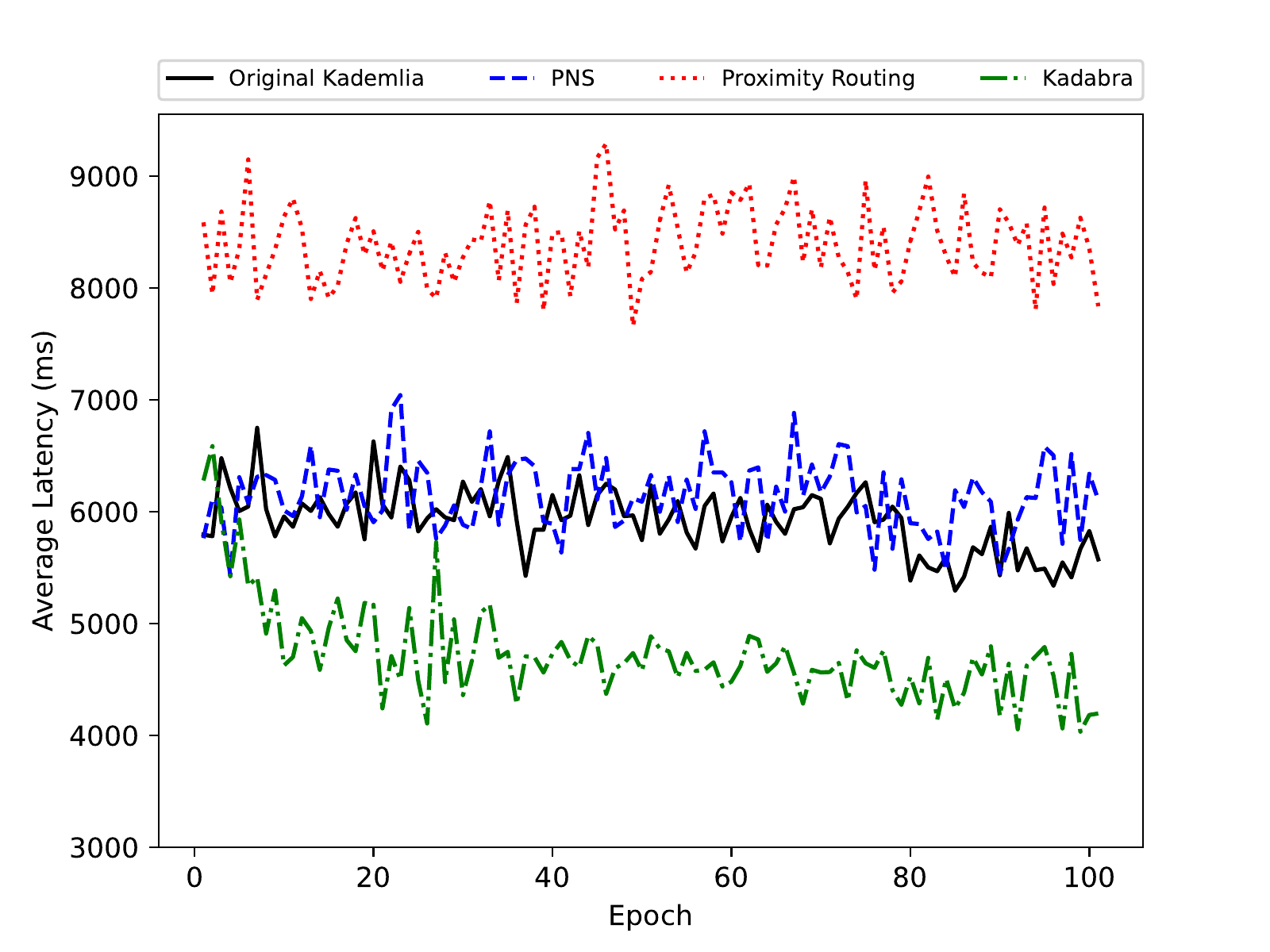}\label{fig:sec1late}}
  \hfill
  \subfloat[]{\includegraphics[width=0.504\textwidth]{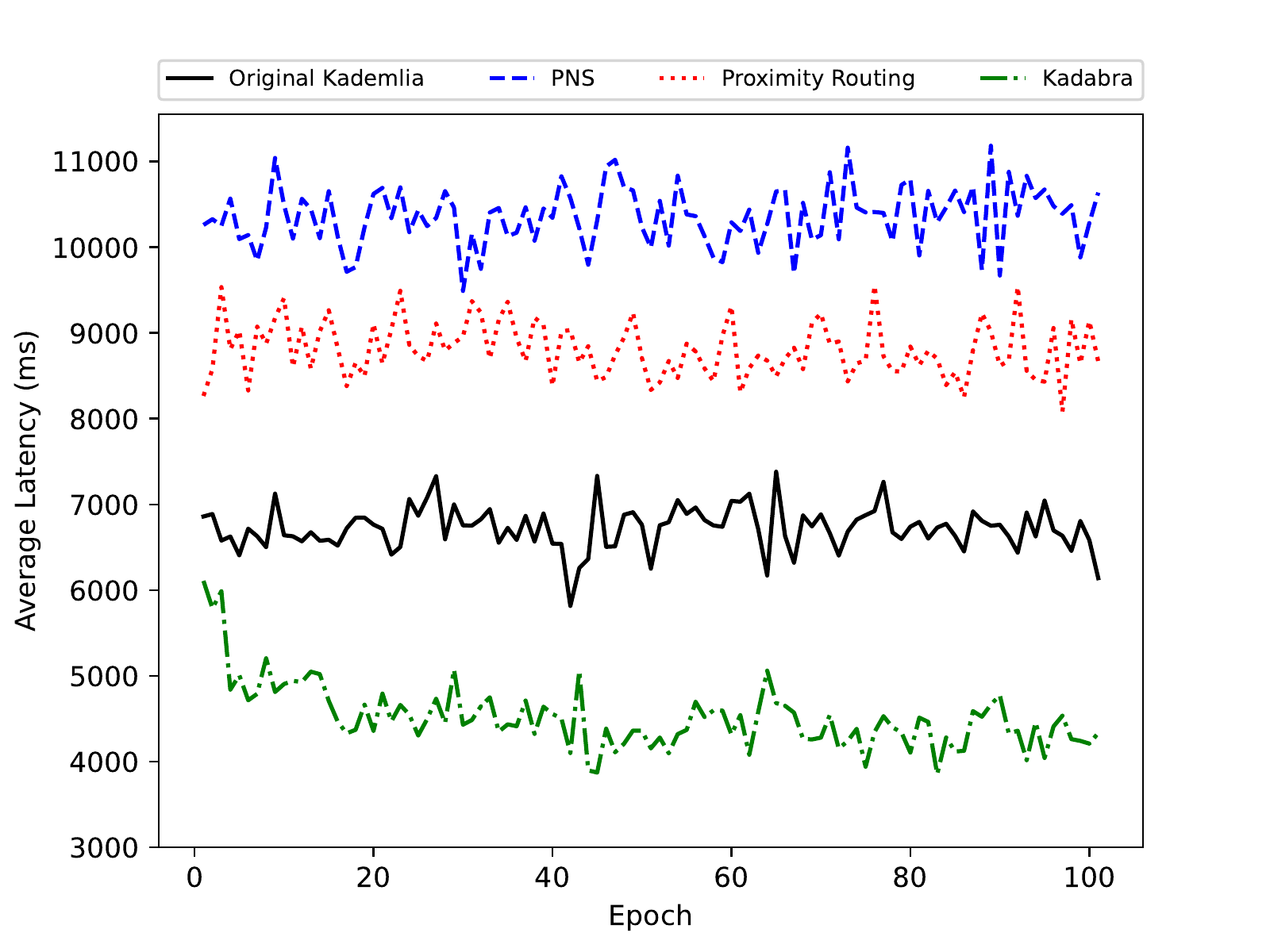}\label{fig:sec2late}}
  \caption{Nodes in the real world: \ourAlgorithm~outperforms baselines even when 20\% of the nodes are adversarial in the network. (a) Adversarial nodes are randomly located. (b) Adversarial nodes are concentrated in one region close to the victim node. Performance is measured at the victim node.}
\end{figure}

In addition, we evaluate the security of $\ourAlgorithm$ by setting 20\% of the nodes as adversarial, which deliberately delay queries passing through them by $3\times$ their default node latencies. 
While all algorithms degrade in this scenario, Fig.~\ref{fig:sec1late} shows that when the adversarial nodes are located at random cities \ourAlgorithm~discovers routes which avoid the adversarial nodes resulting in overall quicker lookups. 
Even when the adversarial nodes are concentrated in one region close to a victim node, Fig.~\ref{fig:sec2late} shows how a victim running \ourAlgorithm~can effectively bypass the adversarial nodes while PNS takes a huge performance loss at more that $2\times$ the latency of \ourAlgorithm. 

\section{Related Work}
A great many number of prior works have studied how to speedup DHTs by being aware of the peer locations in the underlying physical Internet~\cite{jimenez2011sub,kaune2008embracing}.
However, all of these works propose hand-crafted heuristics which do not adapt to network heterogeneity. 
Using parallel lookups and increasing the number of content replicas are some of the early methods. 
R/Kademlia enhances Kademlia routing with recursive overlay routing instead of iterative routing from vanilla Kademlia \cite{heep2010r}. 
Some algorithms utilize caching to accelerate lookups by identifying hotspots \cite{guangmin2009improved} and lowering the load on congested nodes \cite{einziger2016shades}. 
Heck et al.~\cite{heck2017evaluating} evaluate the network resilience of Kademlia. 
Jain et al.~\cite{jain2003study} compare performance of various DHTs against measurement-based overlays.  
In Kanemitsu et al.~\cite{kanemitsu2021kadrtt}, the authors propose KadRTT which uses RTT-based target selection and ID arrangement to accelerate lookups. 
Ratnasamy et al.~\cite{ratnasamy2002topologically} use landmark nodes and binning to optimize latencies in overlay networks. 
Steiner et al.~\cite{steiner2008faster}  proposes an integrated content lookup protocol to reduce content retrieval times in Kad, a popular file-sharing application built using Kademlia. 
Stutzback et al.~\cite{stutzbach2006improving} advocate for parallel lookups and study optimal system parameters in Kad. 
Zhu et al.~\cite{zhu2014improved} presents a storage algorithm for Kademlia against load imbalance. 
To the best of our knowledge, \ourAlgorithm~is the first effort to accelerate DHTs through a data-driven approach. 

\section{Conclusion}

We have presented \ourAlgorithm, a decentralized data-driven approach to learning the routing tables in Kademlia for accelerated lookups. 
Unlike existing heuristics, \ourAlgorithm~is cognizant to heterogeneity in network conditions resulting in routing tables that are tuned to the network and demand patterns. 
Our proposed protocol is also secure against Sybil, Eclipse and adversarial routing attacks. 
While these attacks are important, a thorough analysis of \ourAlgorithm's robustness against other known attacks~\cite{koutrouli2012taxonomy} is a direction for future work. 
In our experiments, we observe \ourAlgorithm~typically converges in a few epochs. 
Testing \ourAlgorithm's convergence and performance in a real world network (e.g., IPFS and Swarm) and obtaining a theoretical understanding on the convergence are also important directions for future work. 


%
%
%
\bibliographystyle{splncs04}
\bibliography{mybibliography}
\appendix 

\section{Nodes in a square - KBR}
\label{apx:nodesinsquarekbr}

\begin{figure}[!tbp]
  \centering
  \subfloat[]{\includegraphics[width=0.4\textwidth]{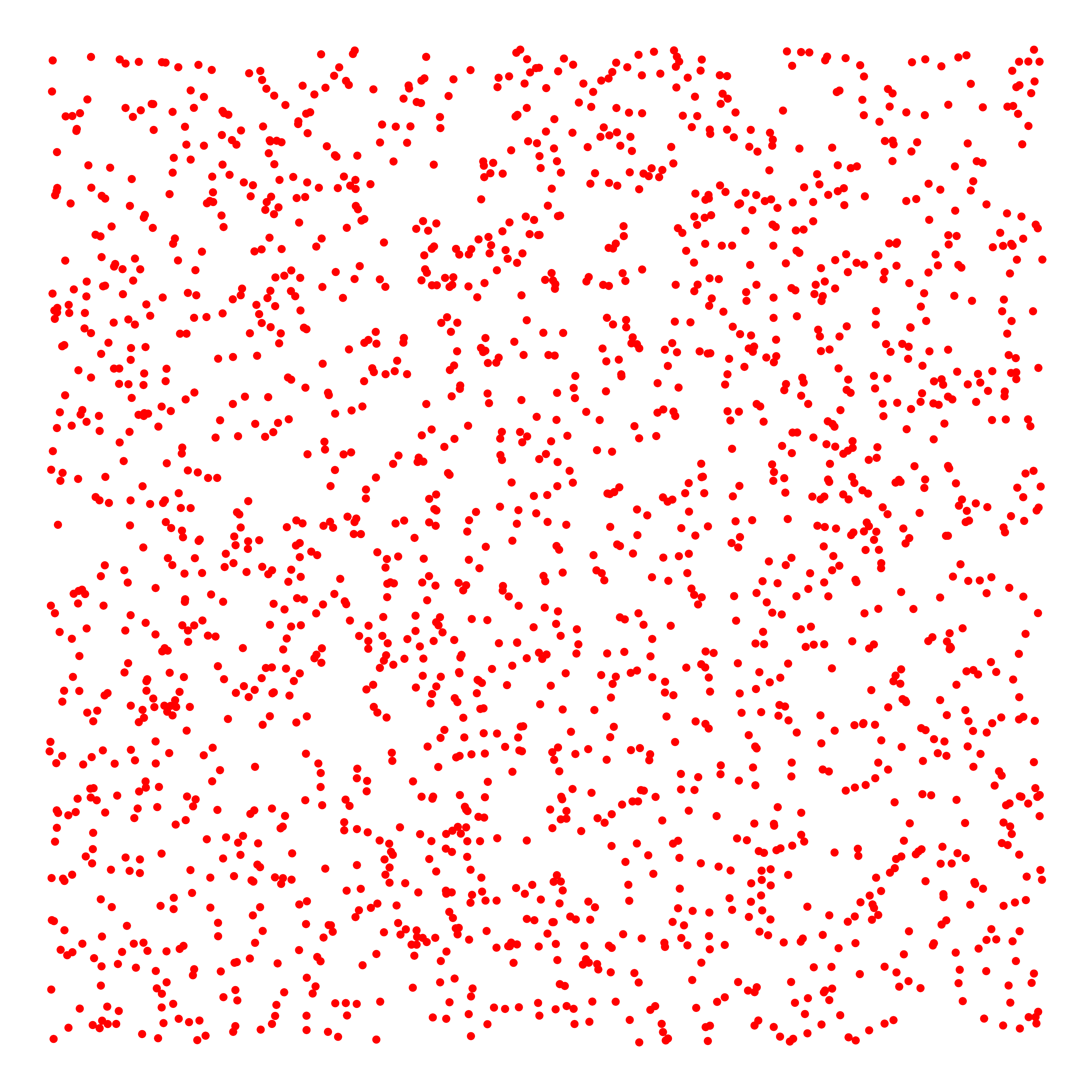}\label{fig:f1}}
  \hfill
  \subfloat[]{\includegraphics[width=0.46\textwidth]{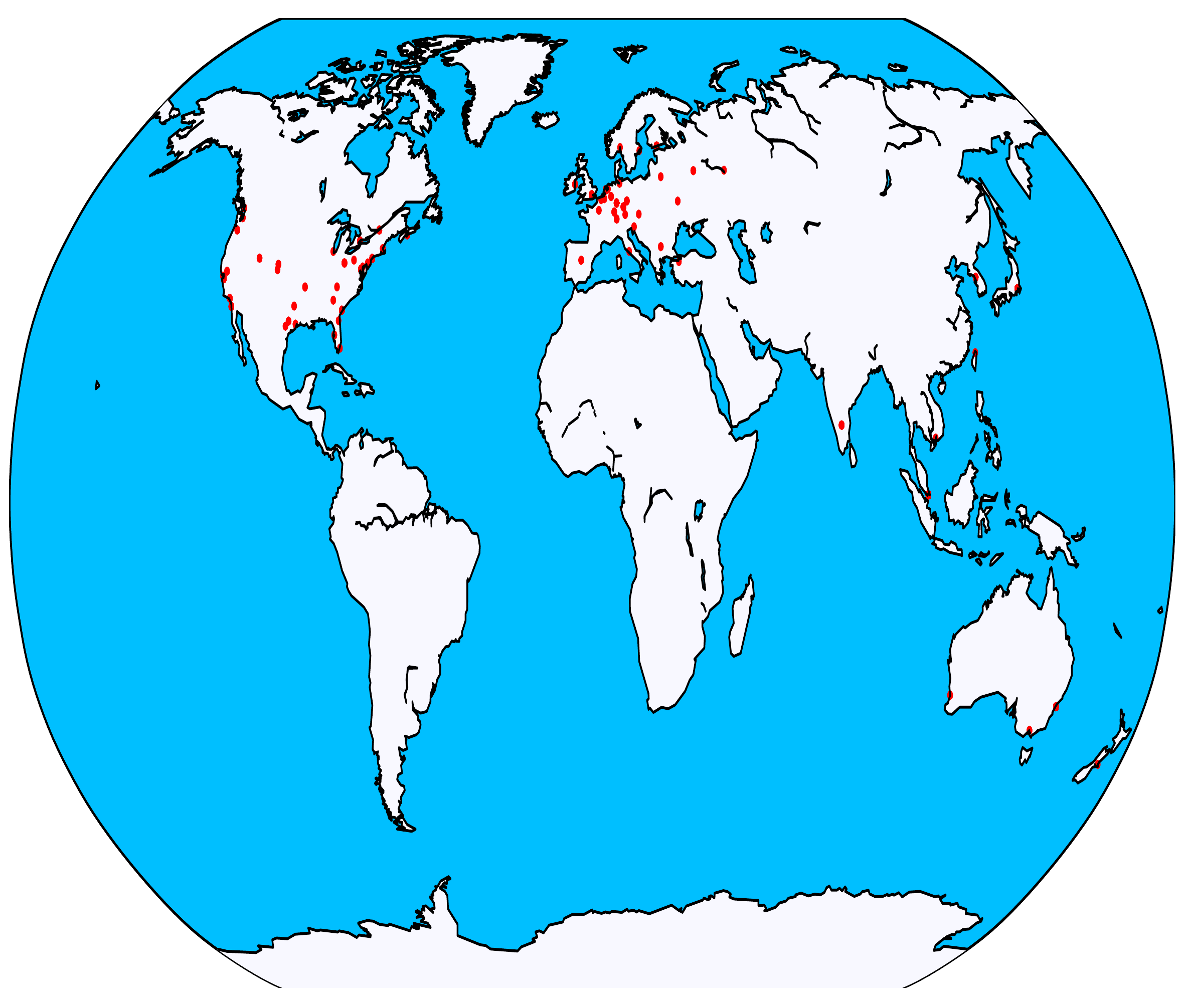}\label{fig:f2}}
  \caption{(a) 2048 nodes randomly distributed within a square. (b) 2048 nodes at cities around the world. A dot may represent multiple nodes due to high geographic concentration. }
\end{figure}

In this section, we present additional results for the KBR application when nodes are distributed randomly on a square.

\begin{figure}[!tbp]
  \centering
 \subfloat[Node A]{\includegraphics[width=.31\textwidth]{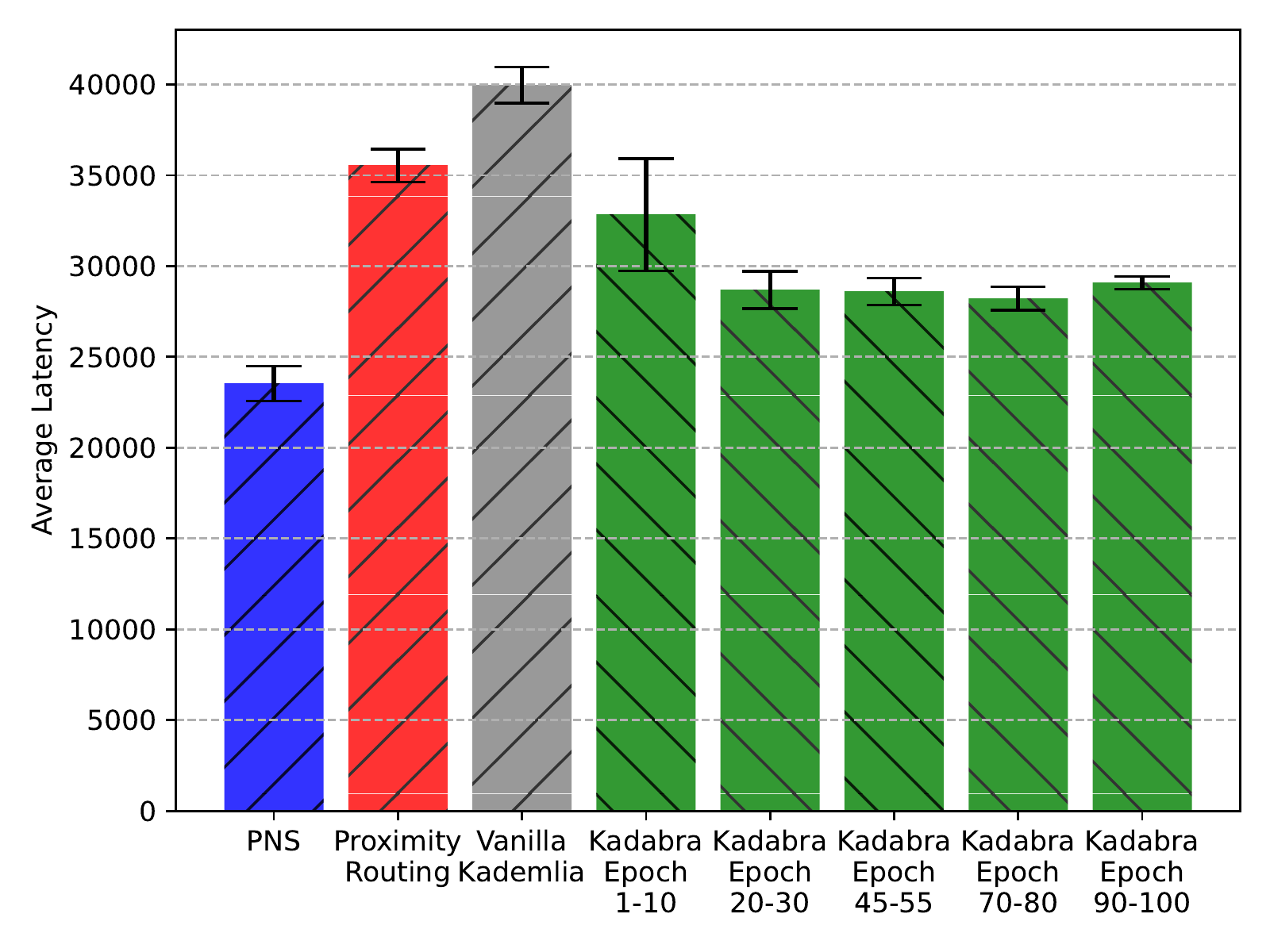}\quad}
\subfloat[Node B]{\includegraphics[width=.31\textwidth]{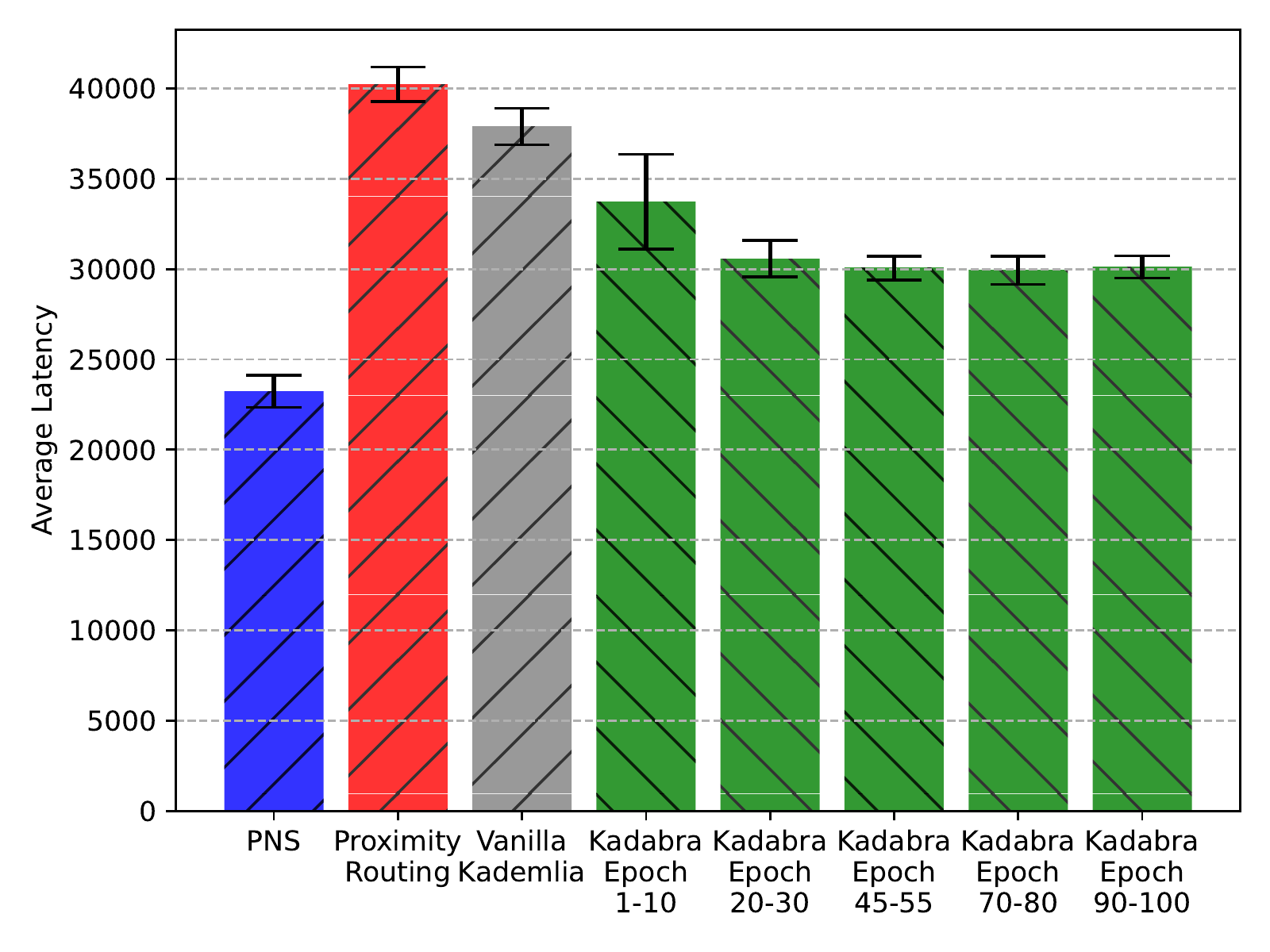}\quad}
\subfloat[Node C]{\includegraphics[width=.31\textwidth]{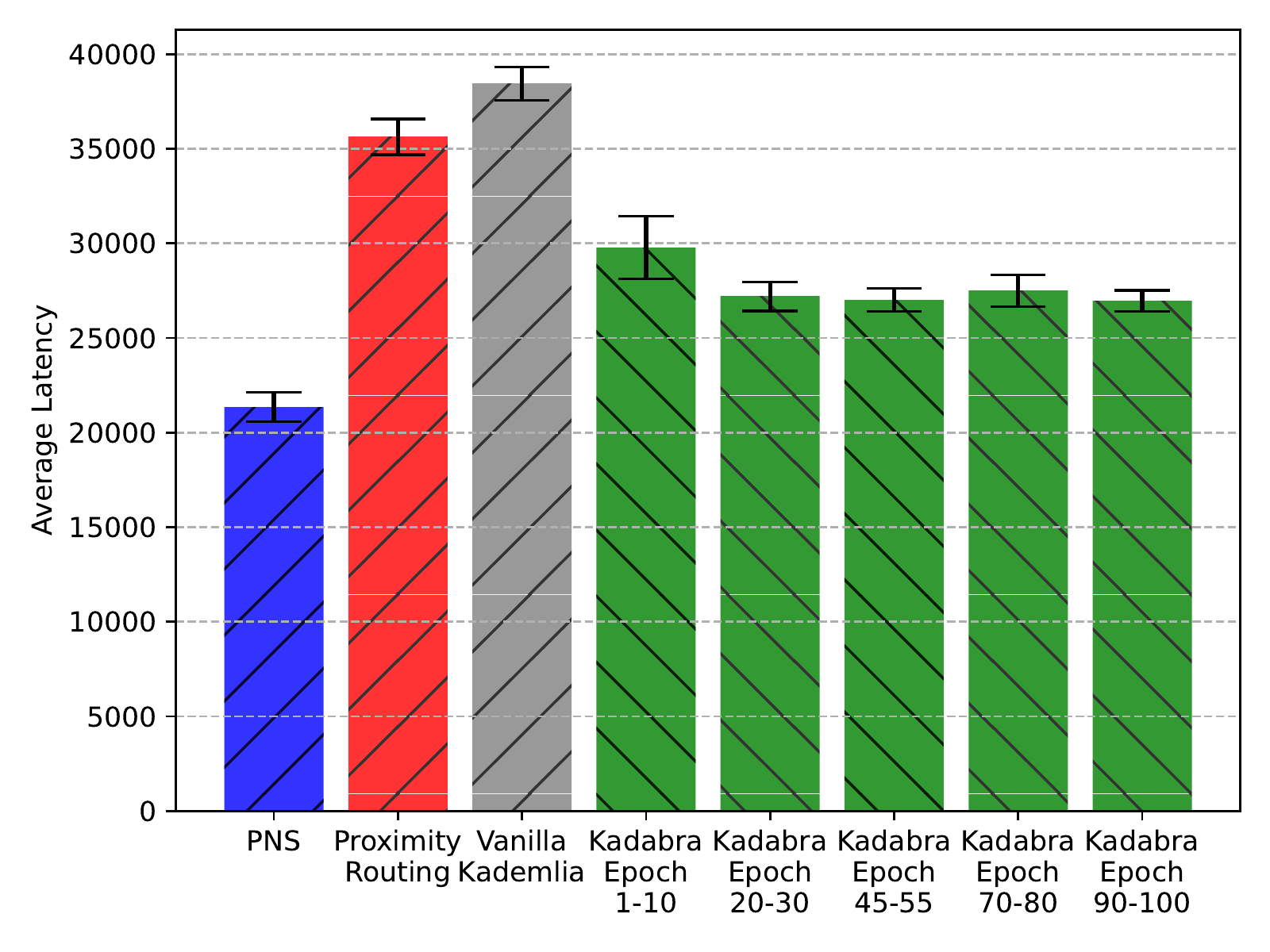}}

\medskip

\subfloat[Node D]{\includegraphics[width=.31\textwidth]{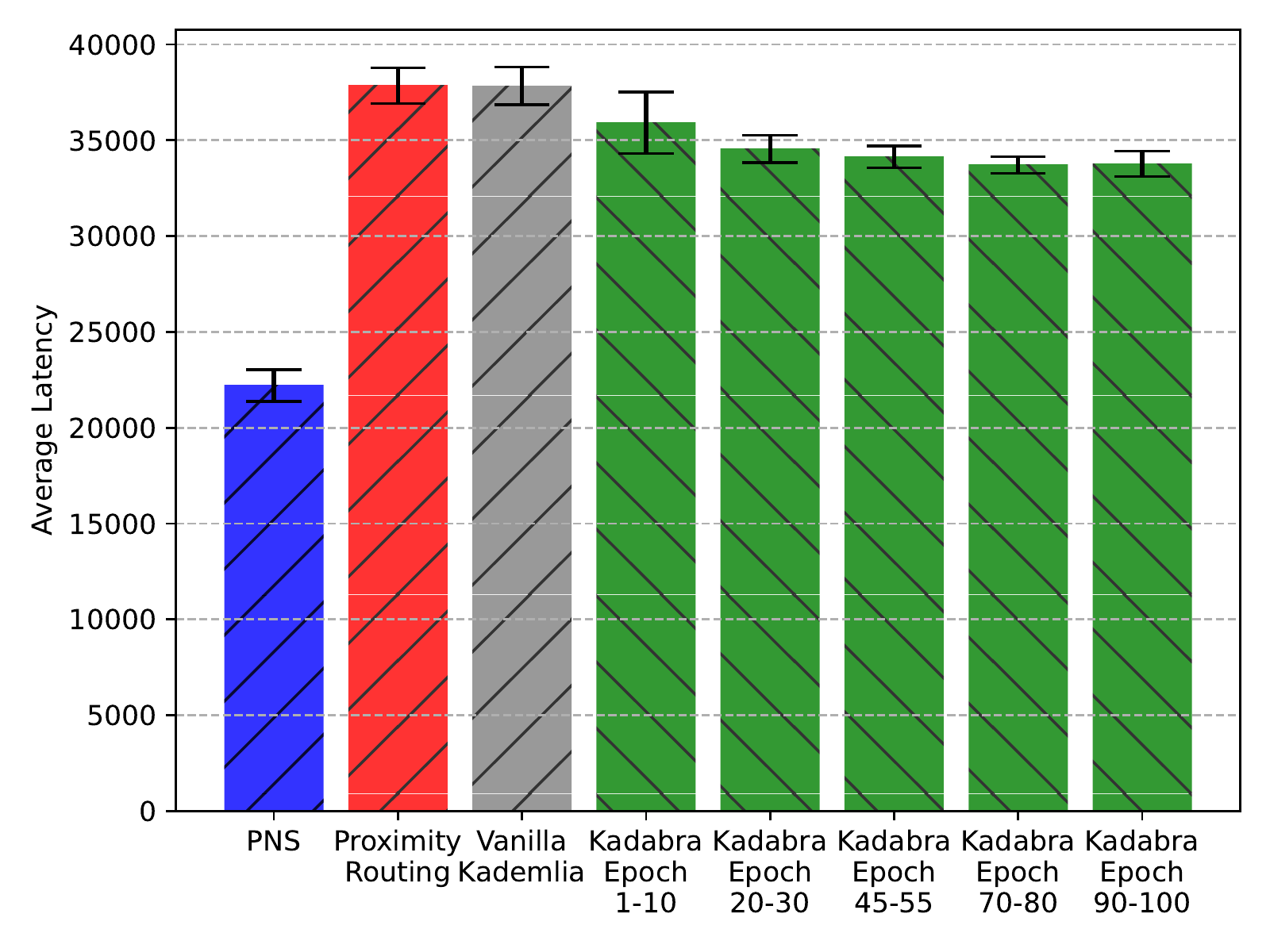}\quad}
\subfloat[Node E]{\includegraphics[width=.31\textwidth]{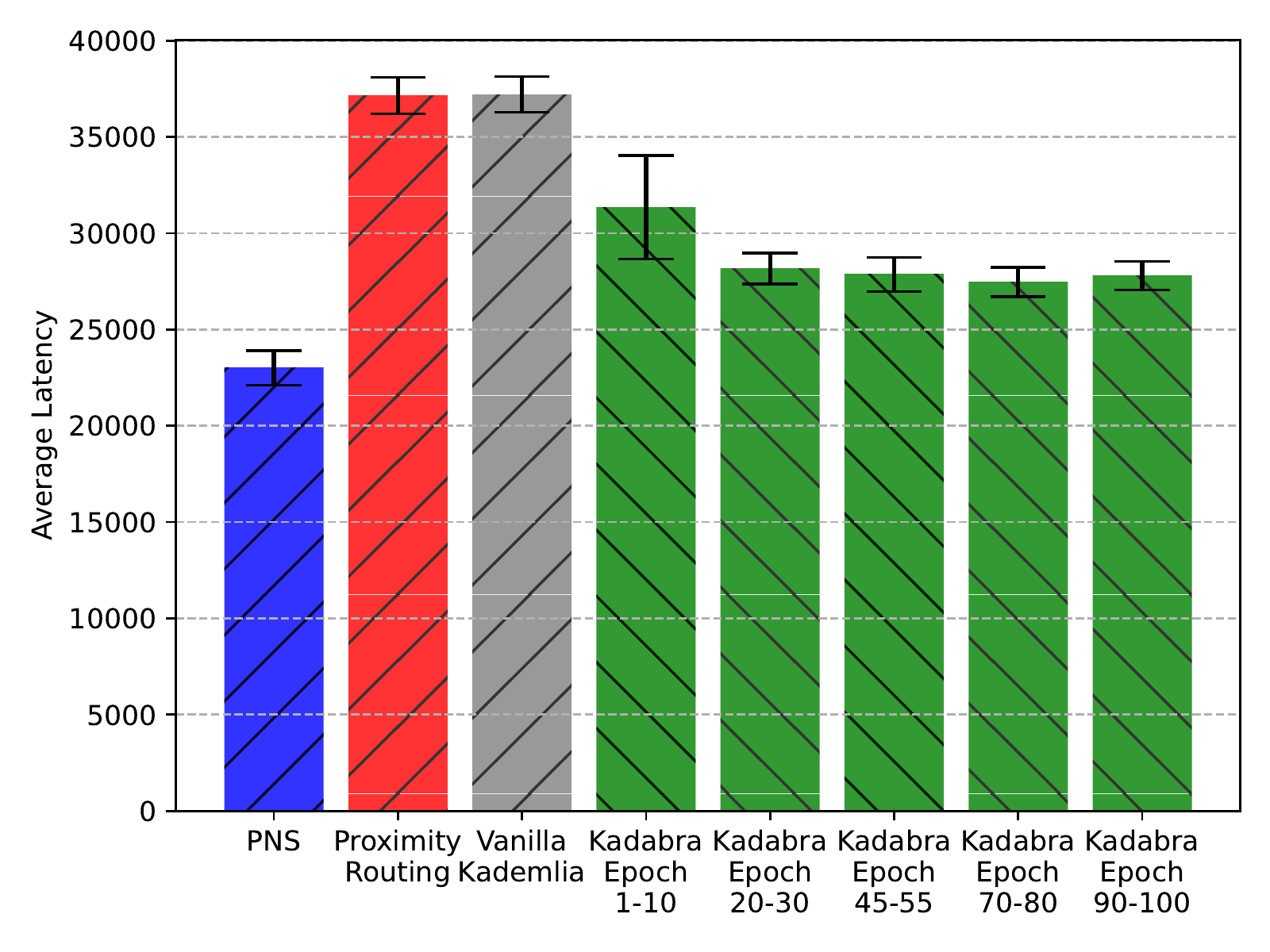}}
  \caption{Nodes in a square: We randomly sample five nodes in the square and compare the performance of $\ourAlgorithm$ and the baseline algorithms at each node under uniform demand.}
  \label{ednssc1obs5}
\end{figure}

\begin{figure}[!tbp]
  \centering
  \subfloat[Vanilla: 24225]{\includegraphics[width=0.25\textwidth]{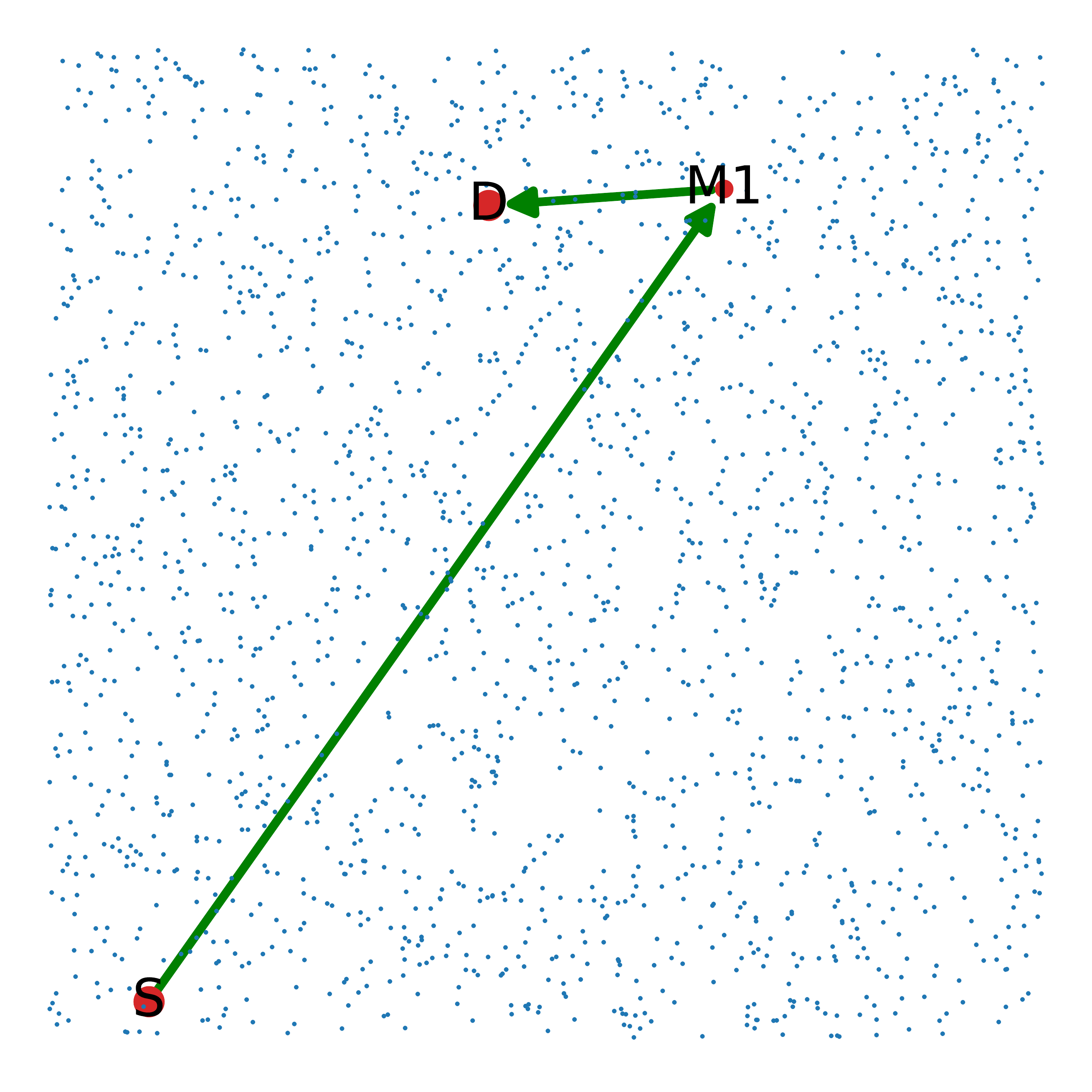}\label{fig:f1}}
  \hfill
  \subfloat[PR: 27124]{\includegraphics[width=0.25\textwidth]{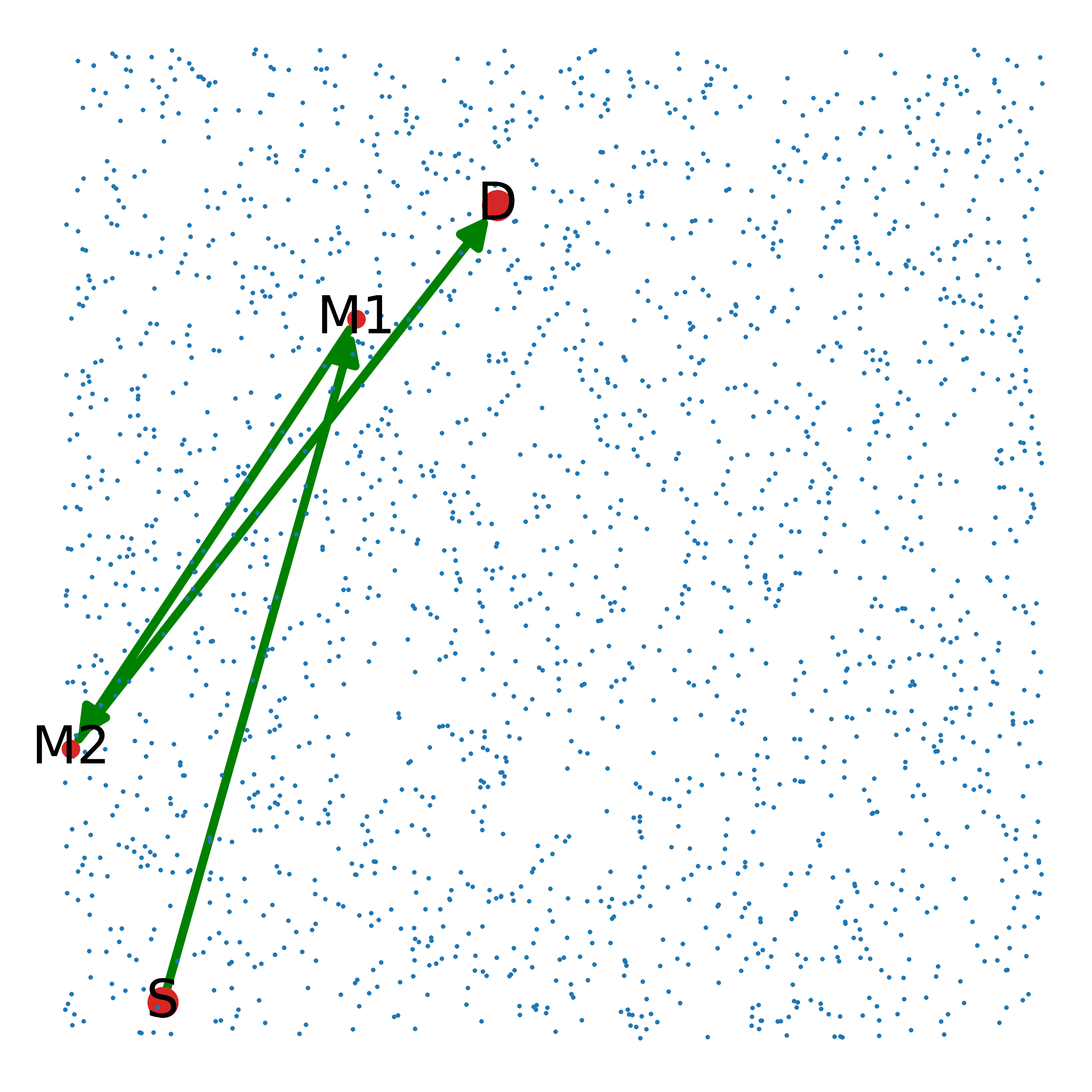}\label{fig:f2}}
  \subfloat[PNS: 21976]{\includegraphics[width=0.25\textwidth]{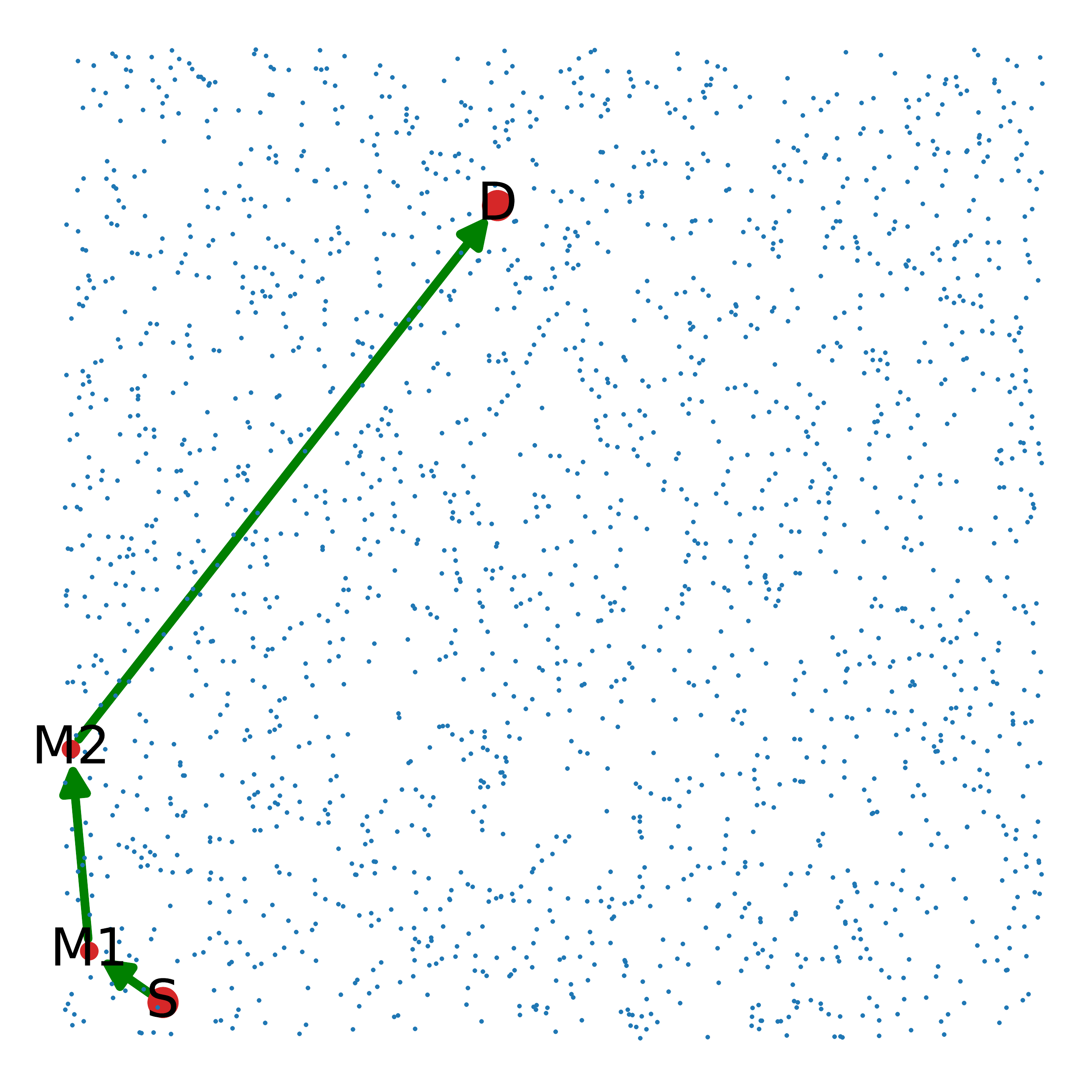}\label{fig:f2}}
  \subfloat[$\ourAlgorithm$: 15674]{\includegraphics[width=0.25\textwidth]{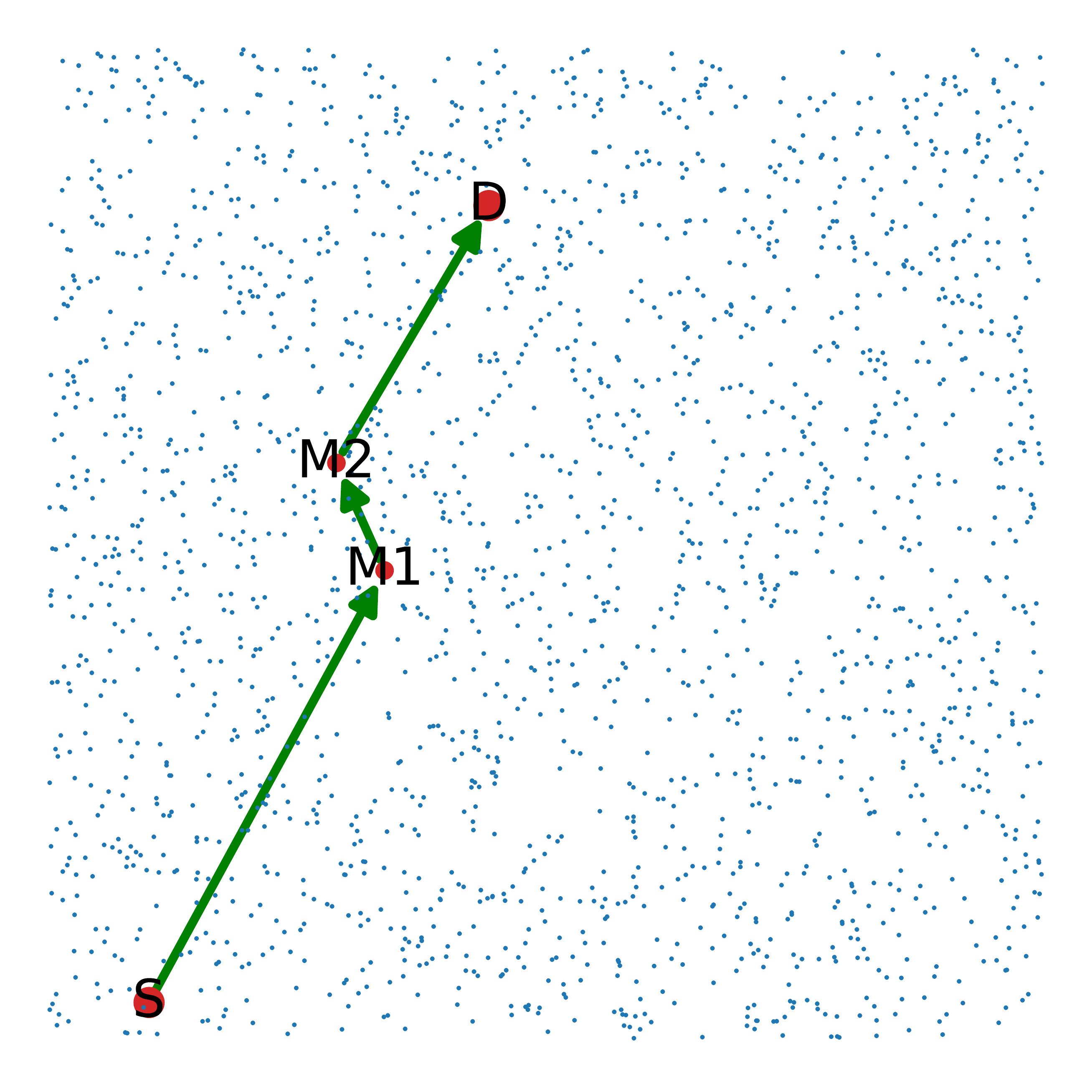}\label{fig:f2}}
  \caption{Nodes in a square: Paths and latencies (below the plots) of an example lookup. \ourAlgorithm~is trained for 50 epoch under uniform demand. \(Mi\) refers to a node that is on the path but is not the source or destination node.}
  \label{fig:ednssce1pathcase5}
\end{figure}

In \S\ref{s:evalresults}, for uniform traffic demand we have presented how the average latency varies with epochs for an arbitrarily chosen node. 
To show that the presented behavior is general, and not occurring only at a few nodes, in Fig.~\ref{ednssc1obs5} we show performance of \ourAlgorithm~and baselines at five randomly chosen nodes. 
In all cases, we observe a similar qualitative behavior. 
Fig.~\ref{fig:ednssce1pathcase5} presents an example of the paths taken for a lookup from the same source to the same destination on different heuristics. 
\ourAlgorithm~is able to achieve significantly lower path latency by choosing a relatively straight path with low node latencies. 
In the figure, node D's node latency is 1000. 
For vanilla Kademlia, node M1's node latency is 1400. 
For PR, node M1's node latency is 800 and M2's node latency is 1400. 
For PNS, node M1's node latency is 2000 and M2's node latency is 1400. For $\ourAlgorithm$, node M1's node latency is 100 and M2's node latency is 100.

\begin{figure}[!tbp]
  \centering
 \subfloat[Node A]{\includegraphics[width=.31\textwidth]{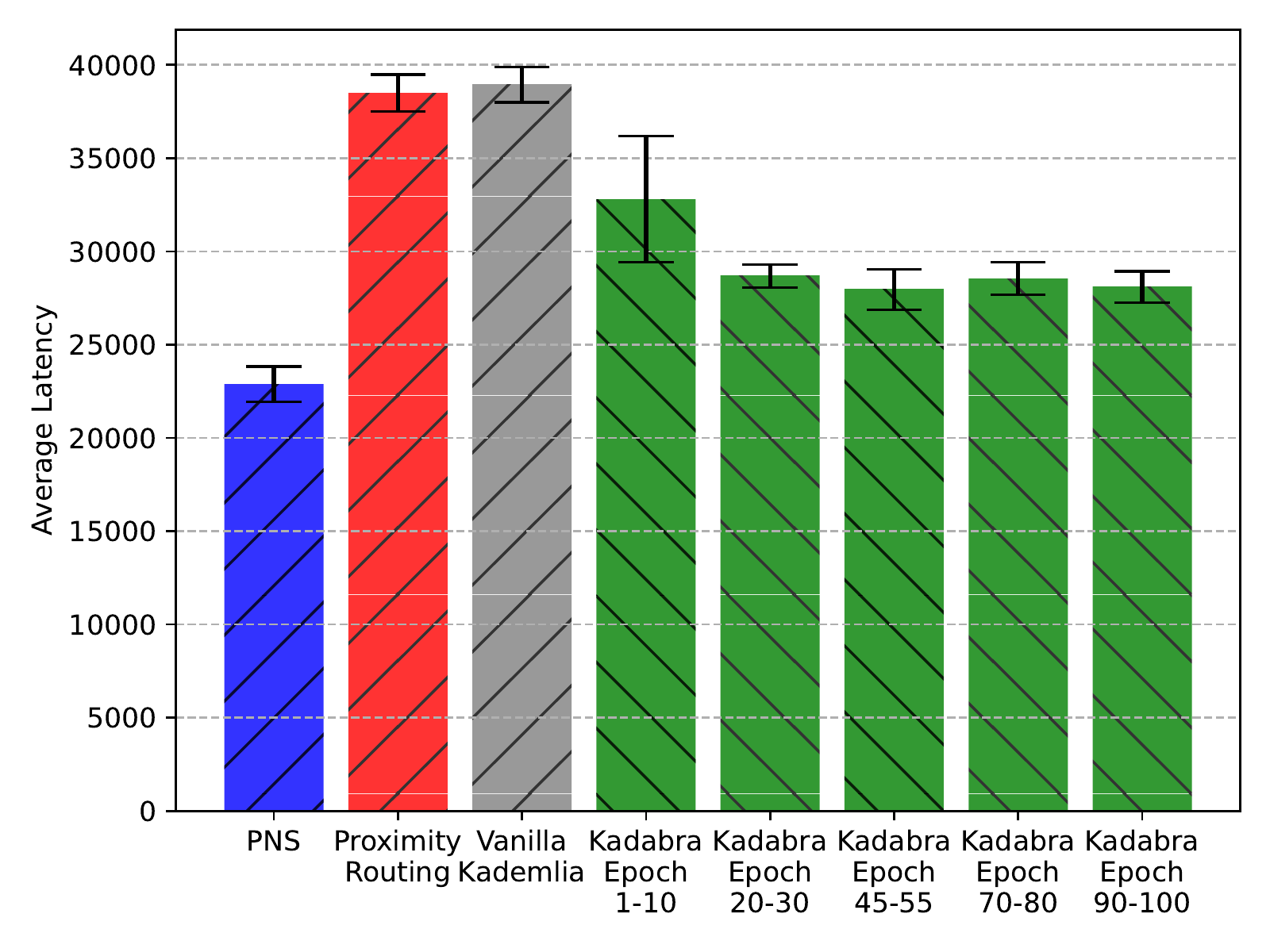}\quad}
\subfloat[Node B]{\includegraphics[width=.31\textwidth]{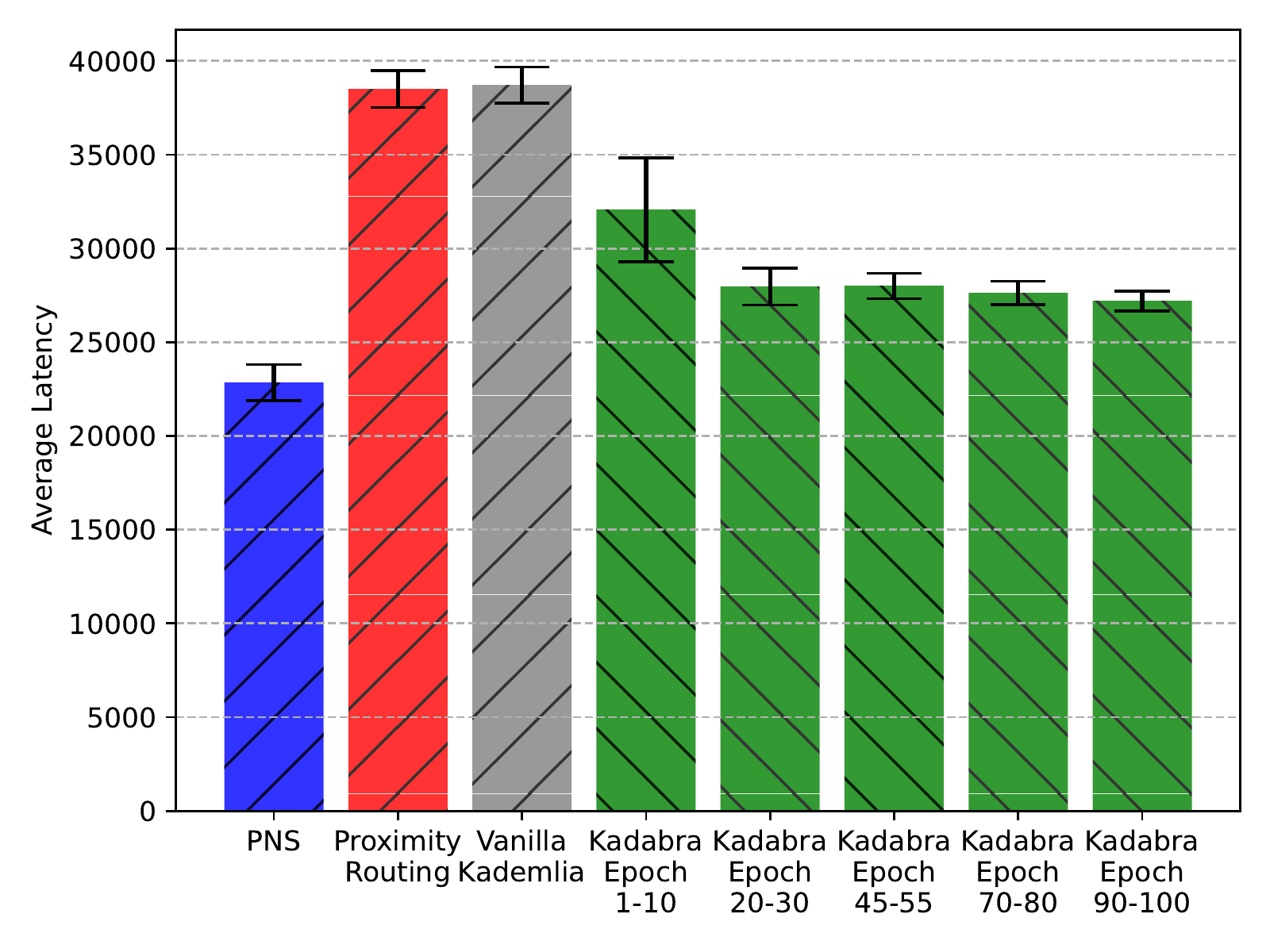}\quad}
\subfloat[Node]{\includegraphics[width=.31\textwidth]{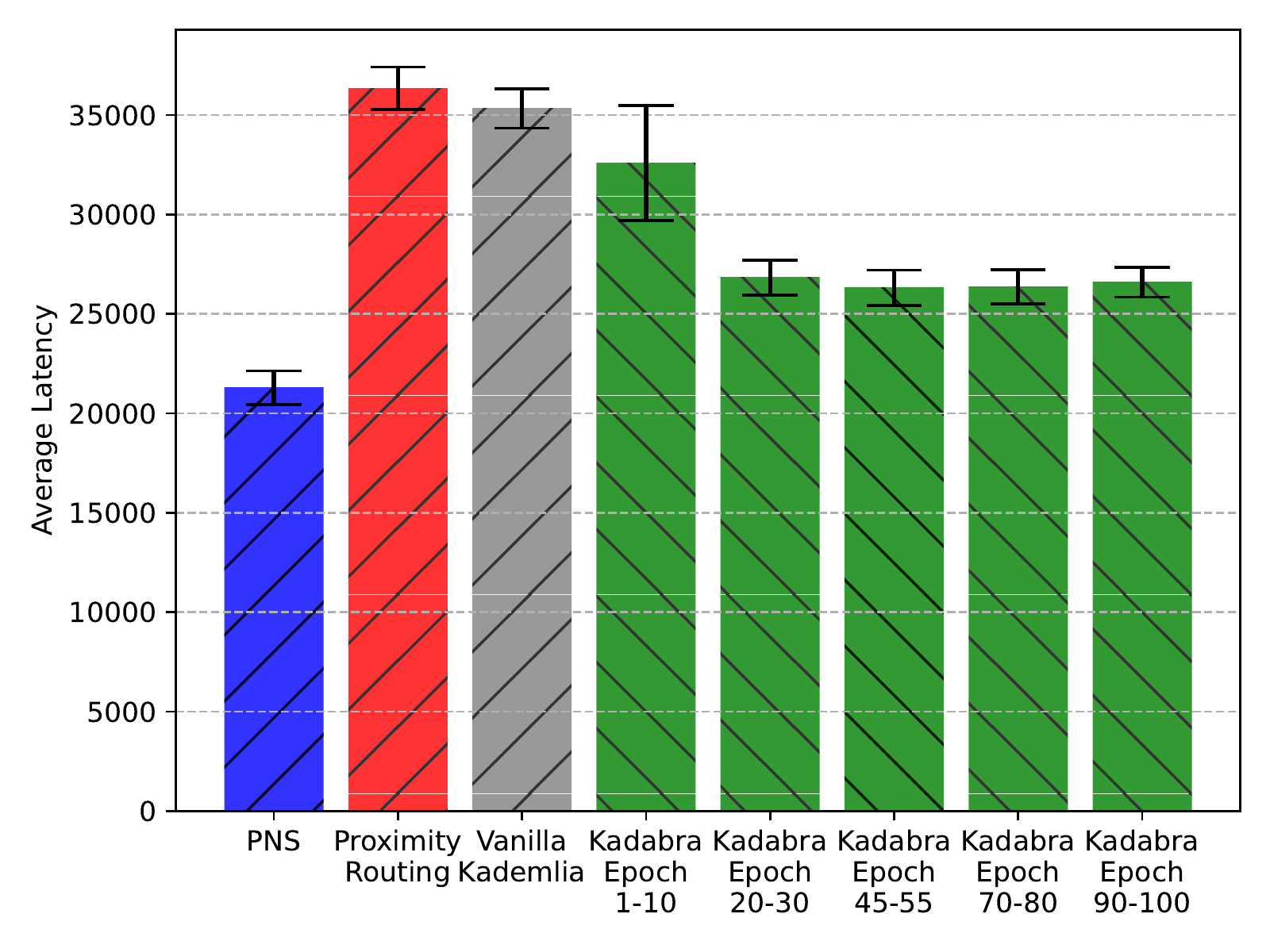}}

\medskip

\subfloat[Node D]{\includegraphics[width=.31\textwidth]{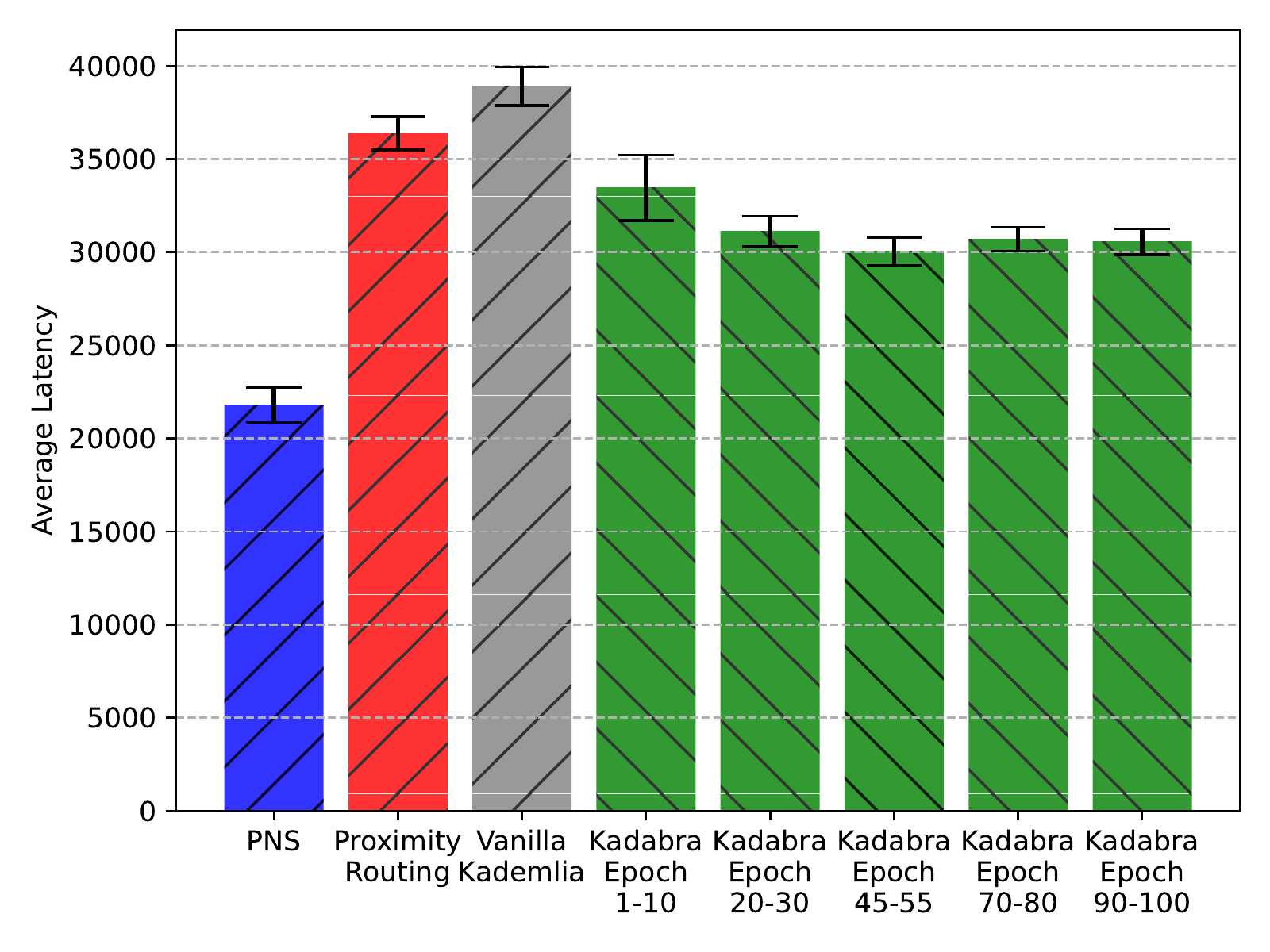}\quad}
\subfloat[Node E]{\includegraphics[width=.31\textwidth]{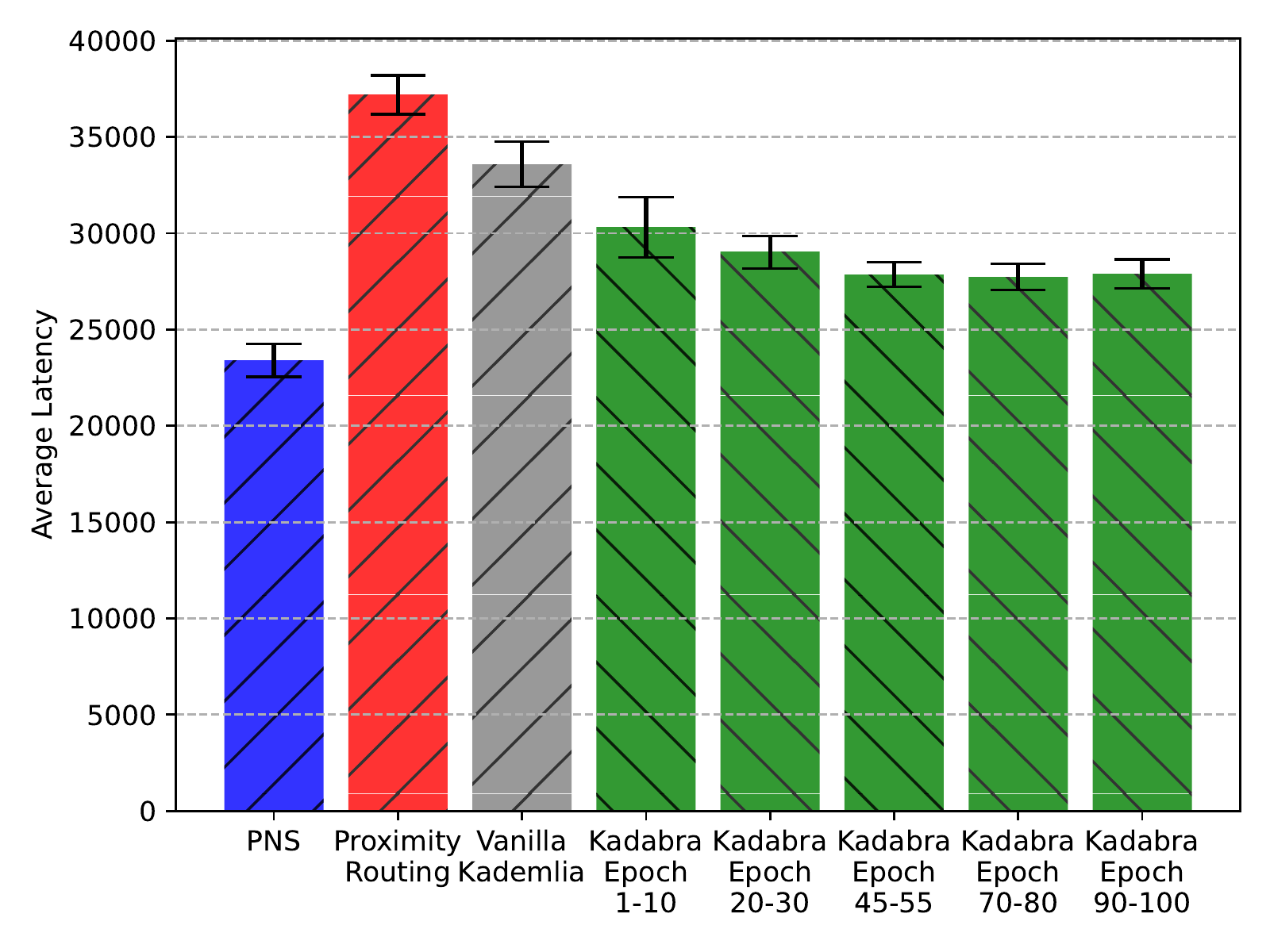}}
  \caption{Nodes in a square: We randomly sample five nodes in the square and compare the performance of $\ourAlgorithm$ and the baseline algorithms at each node under demand hotspots.}
  \label{fig:ednssc2nodes5}
\end{figure}

\begin{figure}[!tbp]
  \centering
  \subfloat[Vanilla: 25746]{\includegraphics[width=0.25\textwidth]{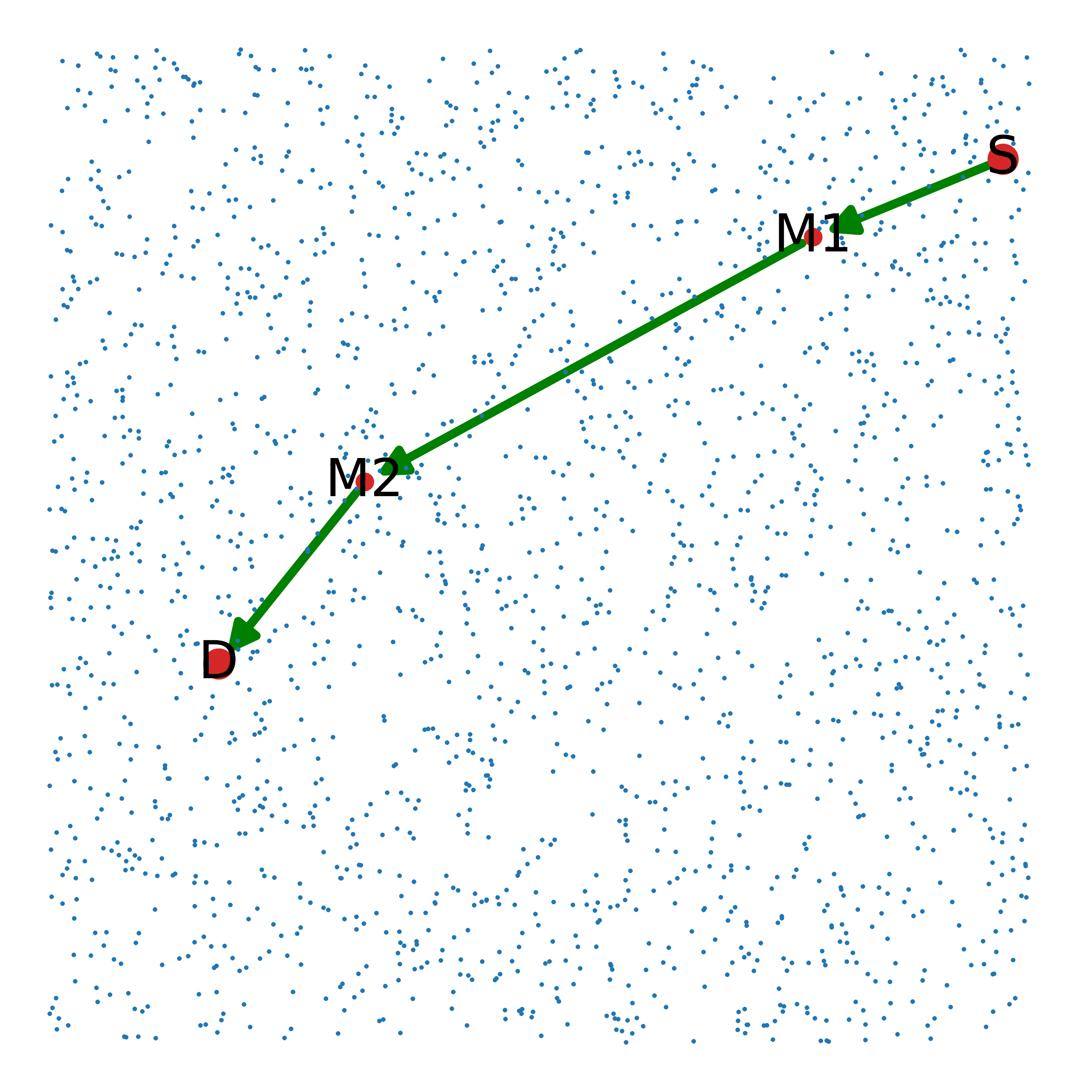}\label{fig:f1}}
  \hfill
  \subfloat[PR: 21773]{\includegraphics[width=0.25\textwidth]{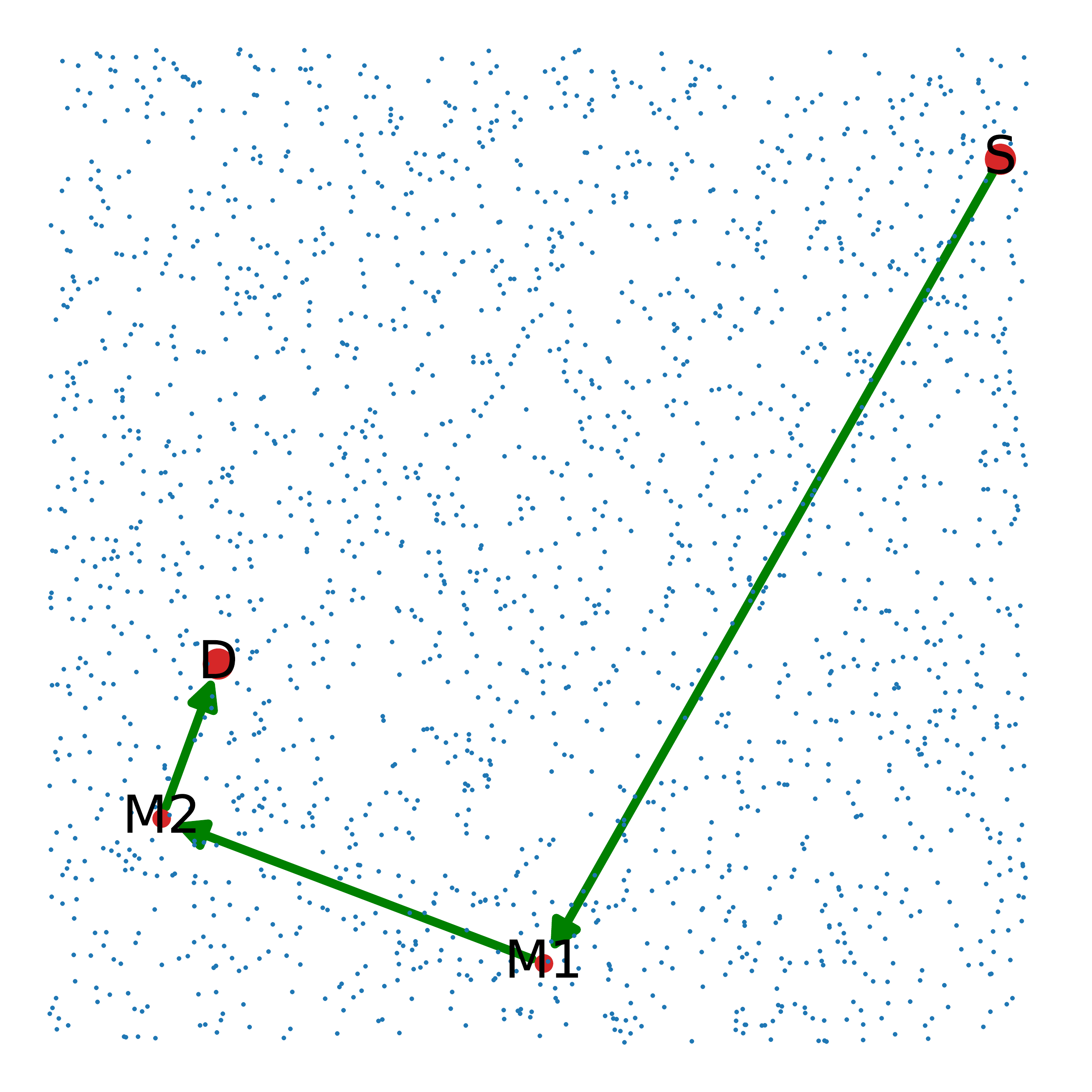}\label{fig:f2}}
  \subfloat[PNS: 17186]{\includegraphics[width=0.25\textwidth]{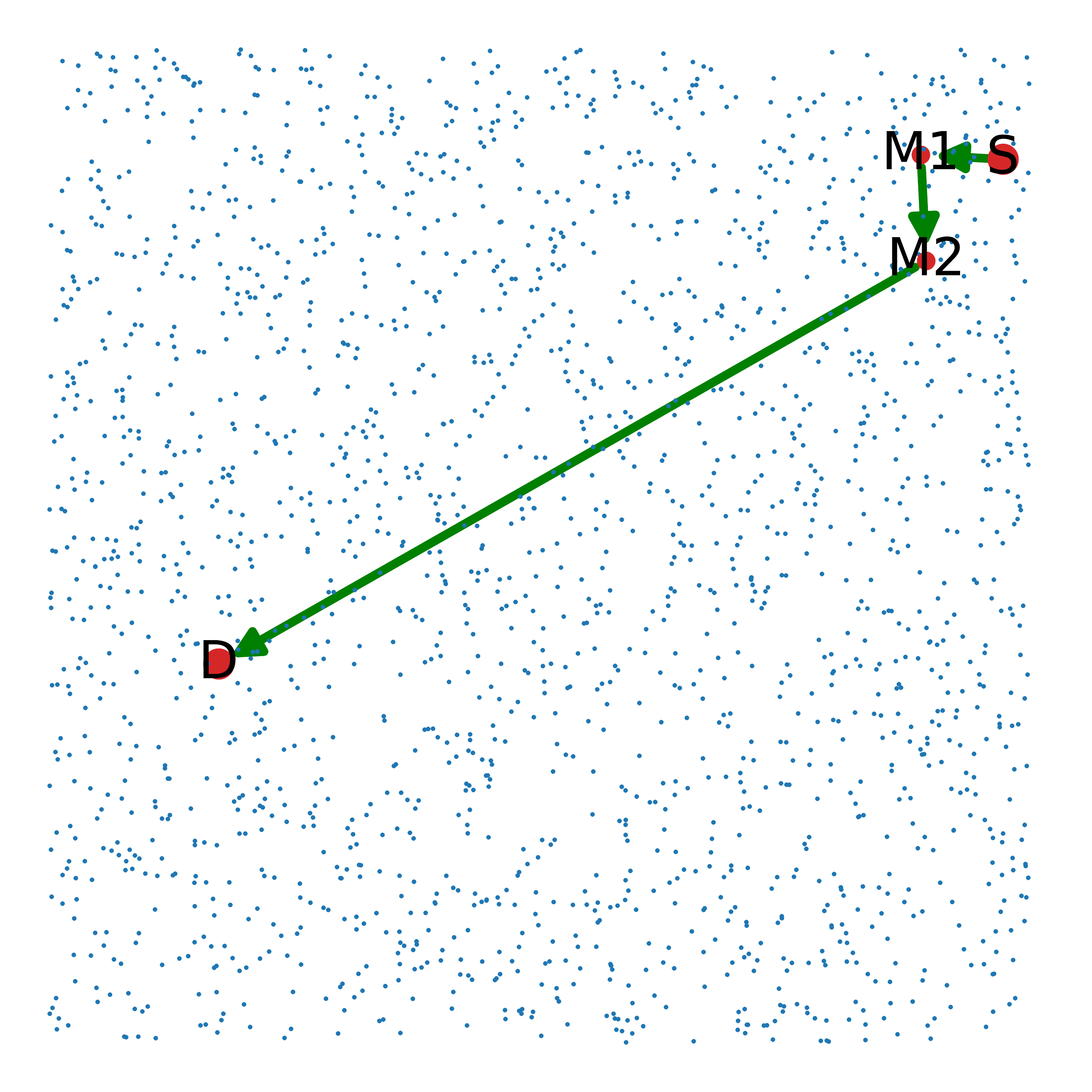}\label{fig:f2}}
  \subfloat[$\ourAlgorithm$: 18714]{\includegraphics[width=0.25\textwidth]{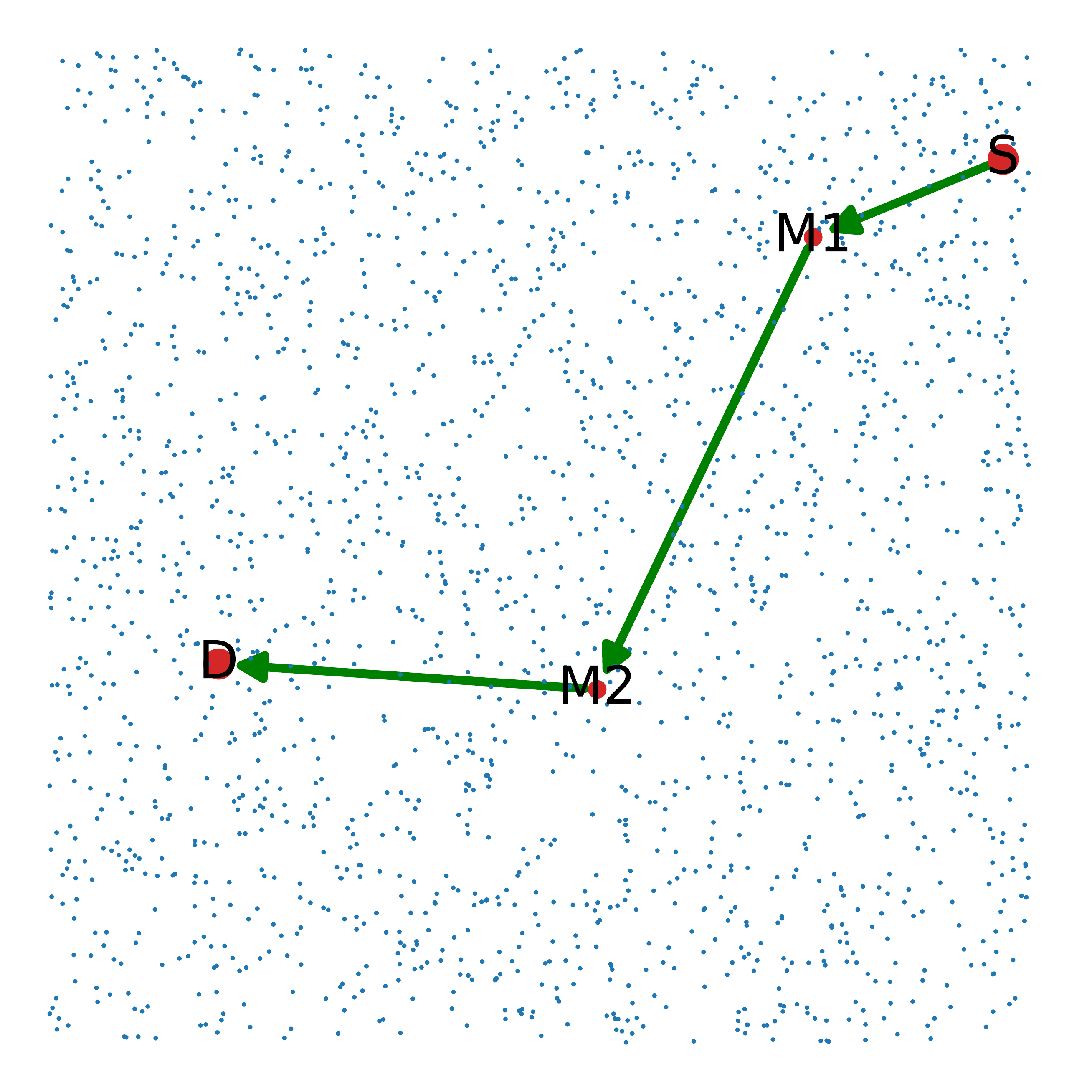}\label{fig:f2}}
  \caption{Nodes in a square: Paths and latencies (below the plots) of a sample lookup under hotspot demand.  The source node (\(S\)) and the destination node (\(D\)) are the same pair of nodes for all protocols. \(Mi\) refers to a node that is on the path but is not the source or destination node.  }
  \label{fig:ednssce2path5}
\end{figure}

Similarly, in Fig.~\ref{fig:ednssc2nodes5} we plot the performance under hotspot demand measured from five randomly chosen nodes. 
Here too, we observe the qualitative behavior is the same across the five nodes. 
Fig.~\ref{fig:ednssce2path5} shows the paths taken by the different heuristics on the same sample query. 
\ourAlgorithm~has been trained for 50 epochs. 
We observe \ourAlgorithm's path latency is $>25\%$ more efficient than the original Kademlia protocol. 
In the figure, node D's node latency is 800. 
For vanilla Kademlia, node M1's node latency is 400 and node M2's node latency is 1100. 
For PR, node M1's node latency is 100 and node M2's node latency is 1800. For PNS, node M1's node latency is 1700 and node M2's node latency is 1200. 
For $\ourAlgorithm$, node M1's node latency is 400 and node M2's node latency is 400.

\section{Nodes in a square - DHT}
\label{apx:squaredht}

Next, we consider the DHT application in which a (key, value) pair is stored on 3 nodes. 
When a node initiates a query for the key, it sends out queries on $\alpha = 2$ independent paths. 
The overall latency lookup latency is the time between sending out the queries and the earliest time when a response arrives on any of the paths. 
As in the KBR application, we consider three traffic settings: 

\smallskip 
\noindent 
{\em (1) DHT under uniform demand.}


\smallskip 
\noindent 
{\em (2) DHT under demand hotspots.}


\smallskip 
\noindent 
{\em (3) DHT under skewed network bandwidth.}

\begin{figure}[!tbp]
  \centering
  \subfloat[]{\includegraphics[width=0.5\textwidth]{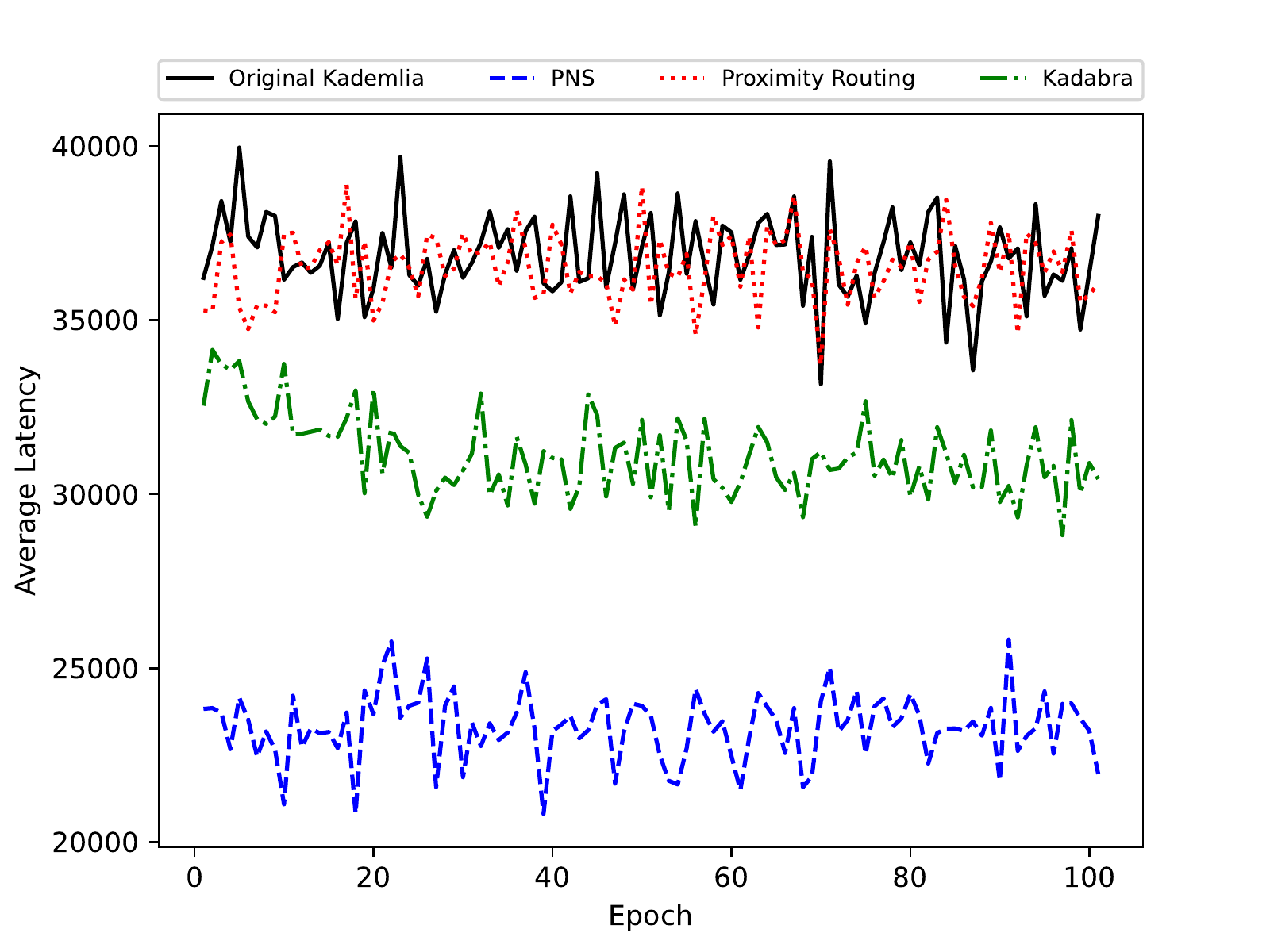}\label{fig:ednsc3late}}
  \hfill
  \subfloat[]{\includegraphics[width=0.5\textwidth]{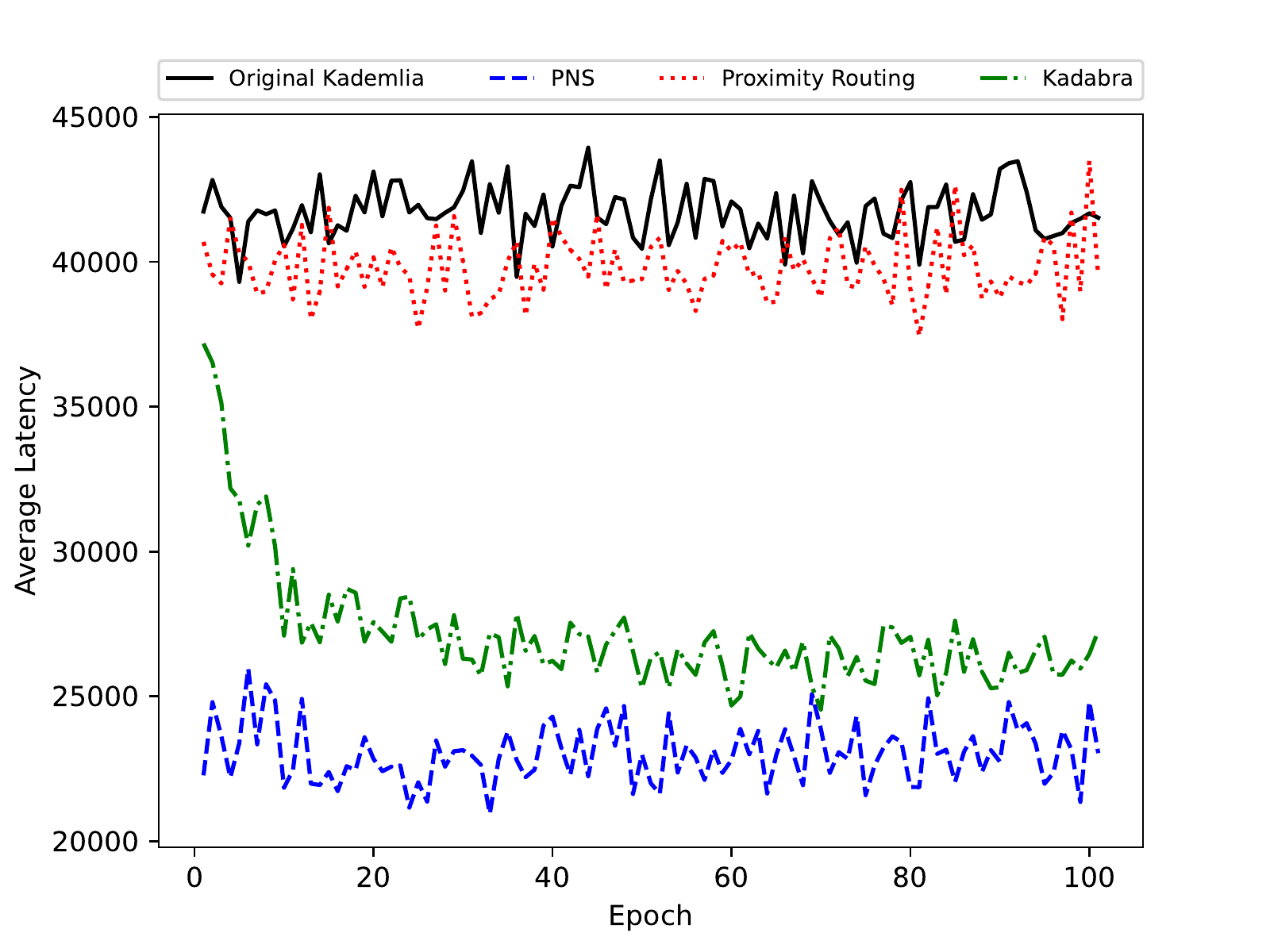}\label{fig:ednsc4late}}
  \caption{Nodes in a square: (a) Performance under uniform demand. (b) Performance under demand hotspots.}
\end{figure}

These settings are similar to the KBR case, and hence we do not elaborate them. 
Fig.~\ref{fig:ednsc3late} and~\ref{fig:ednsc4late} show the performance, at an arbitrarily chosen node, of queries sent through the 1st $k$ bucket under uniform demand and hotspot demand respectively. 
Similar to KBR, with hotspot nodes, the improvement we gain from $\ourAlgorithm$~is larger than the scenario without hotspot nodes.


\begin{figure}[!tbp]
  \centering
    \includegraphics[height=6cm]{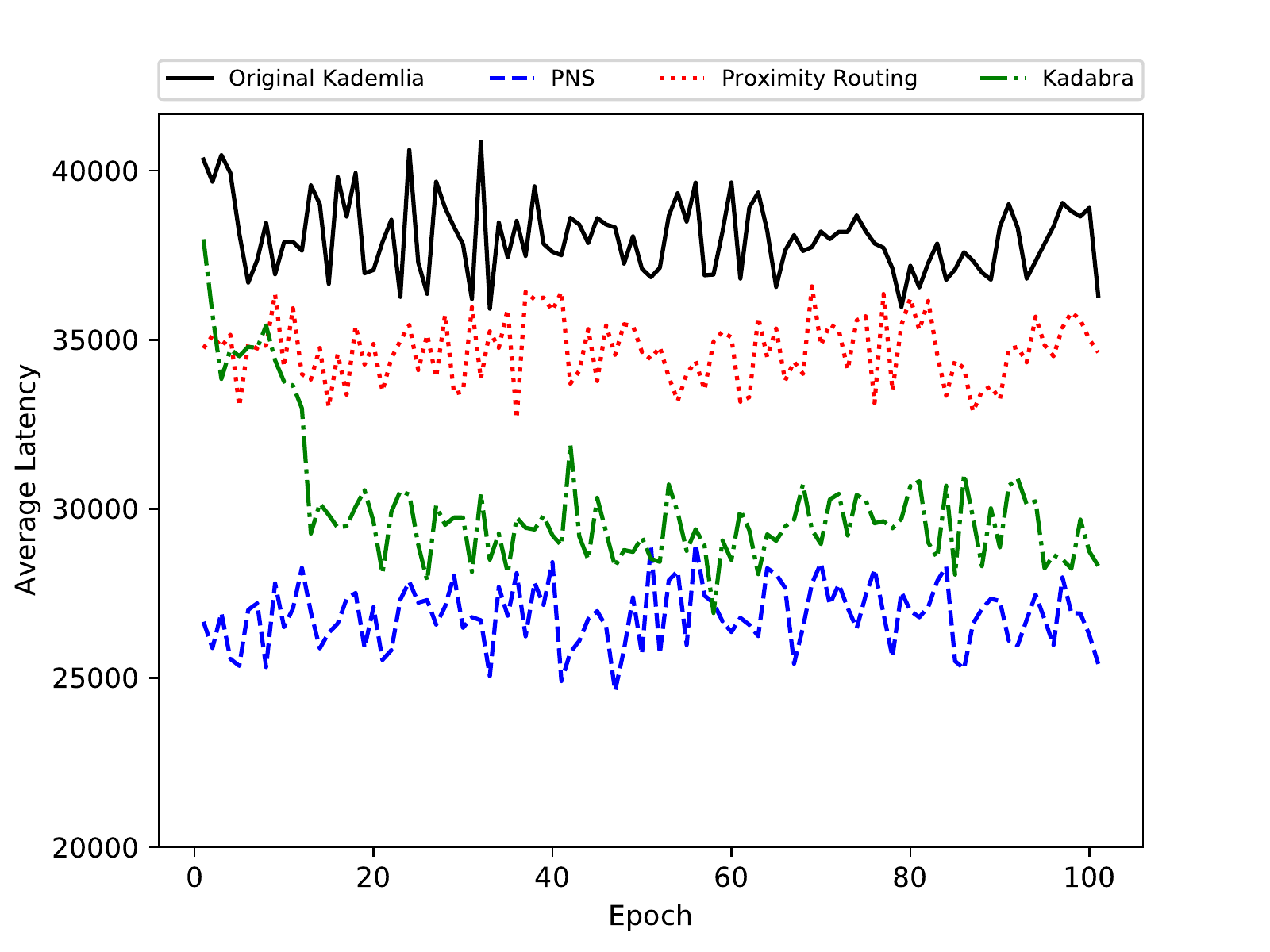}
    \caption{Nodes in a square with DHT application: Performance when a region of nodes have higher node latencies than average. Measurements are taken from within the high node latency area.}
    \label{fig:ednssce6}
\end{figure}

Fig.~\ref{fig:ednssce6} shows the case where nodes within a small 2000 $\times$ 2000 region at the center of the square have high node latencies than default values. 
Unlike KBR, in the DHT case PNS and \ourAlgorithm~are much closer. 

\section{Nodes in the real world - KBR}

In this section, we provide supplementary results for the cases considered under KBR application when nodes are distributed over a real world geography. 

\begin{figure}[!tbp]
  \centering
 \subfloat[Node A - New York]{\includegraphics[width=.31\textwidth]{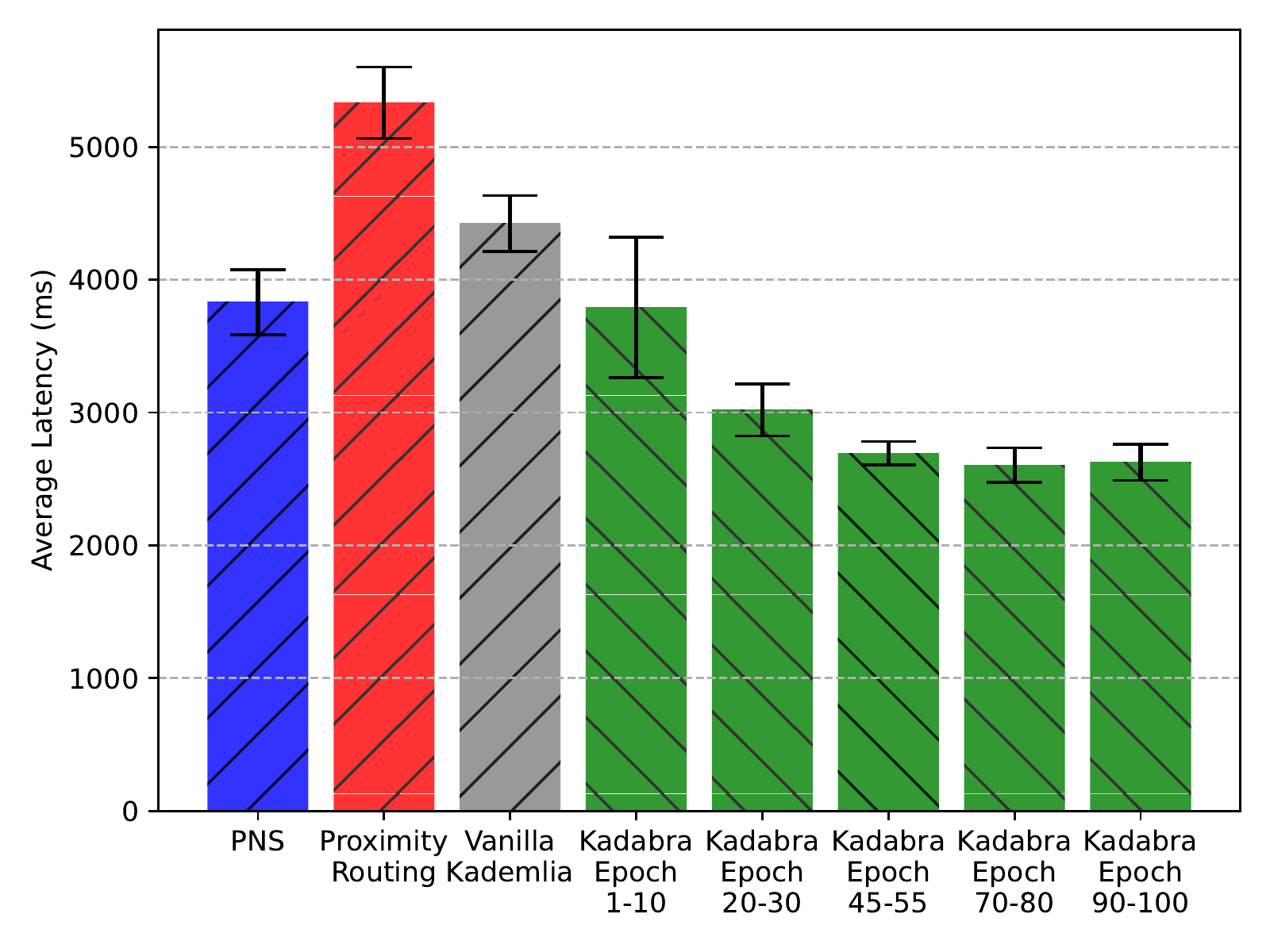}\quad}
\subfloat[Node B - San Jose]{\includegraphics[width=.31\textwidth]{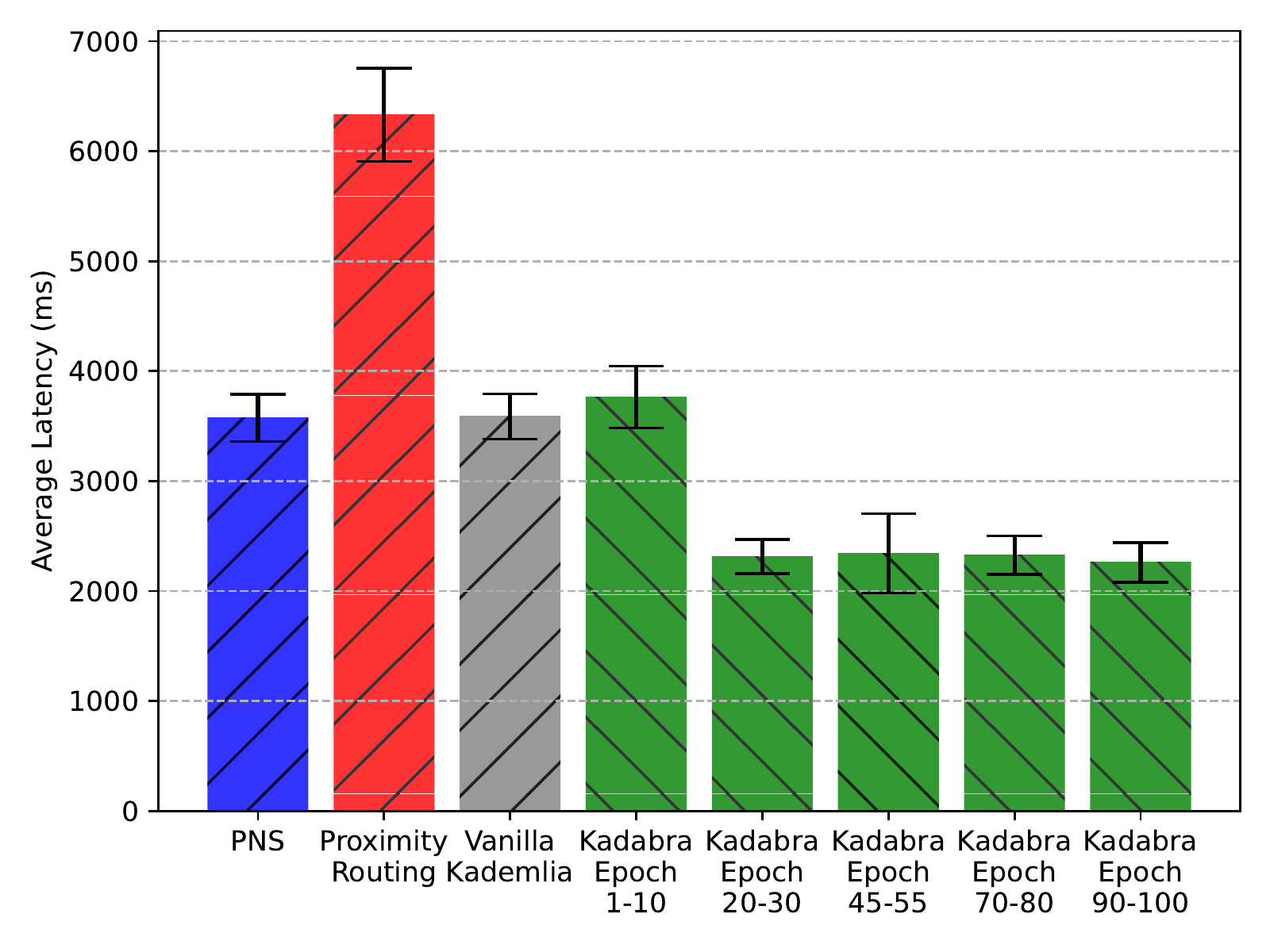}\quad}
\subfloat[Node C - Toronto]{\includegraphics[width=.31\textwidth]{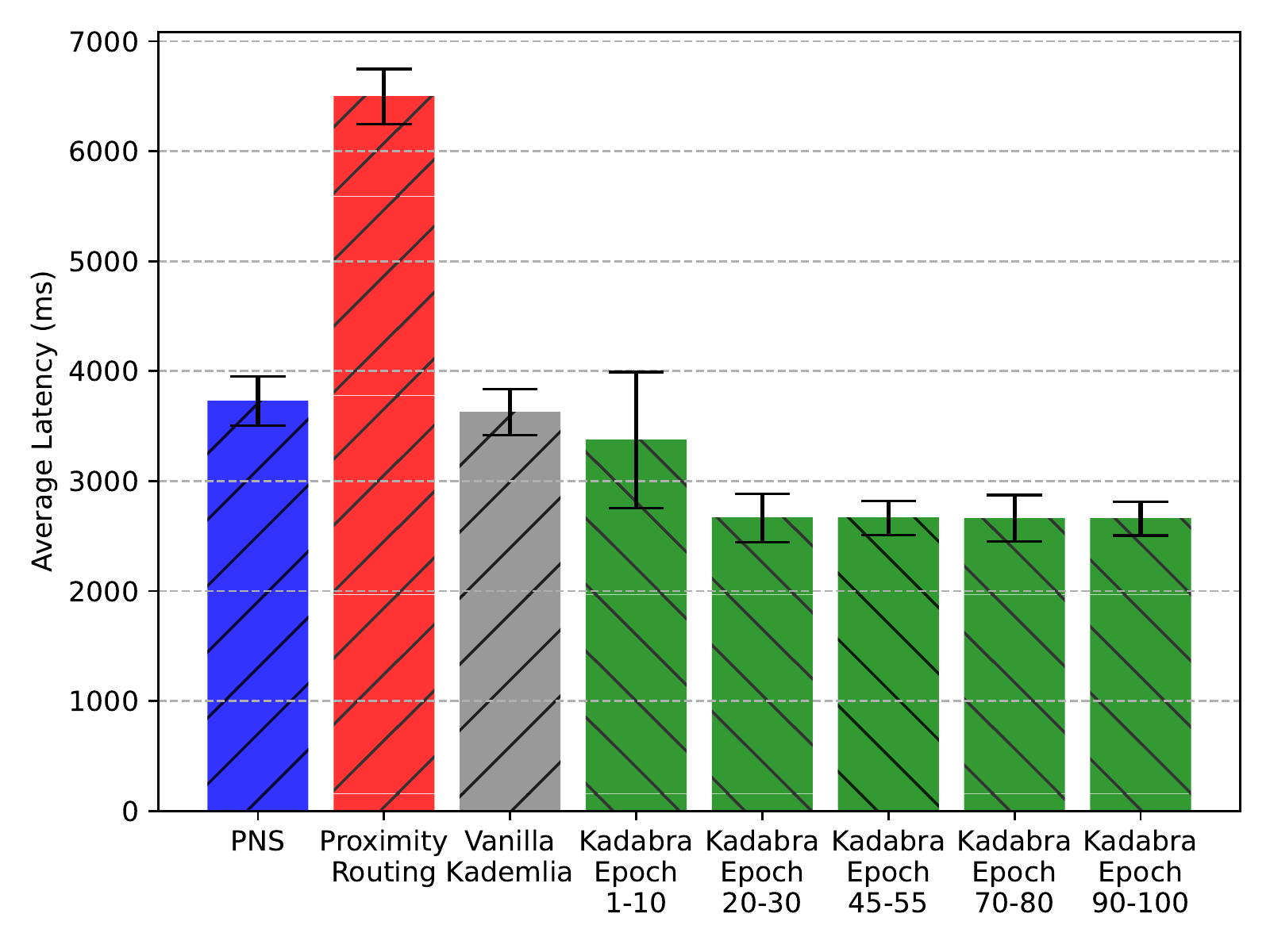}}

\medskip

\subfloat[Node D - Frankfurt]{\includegraphics[width=.31\textwidth]{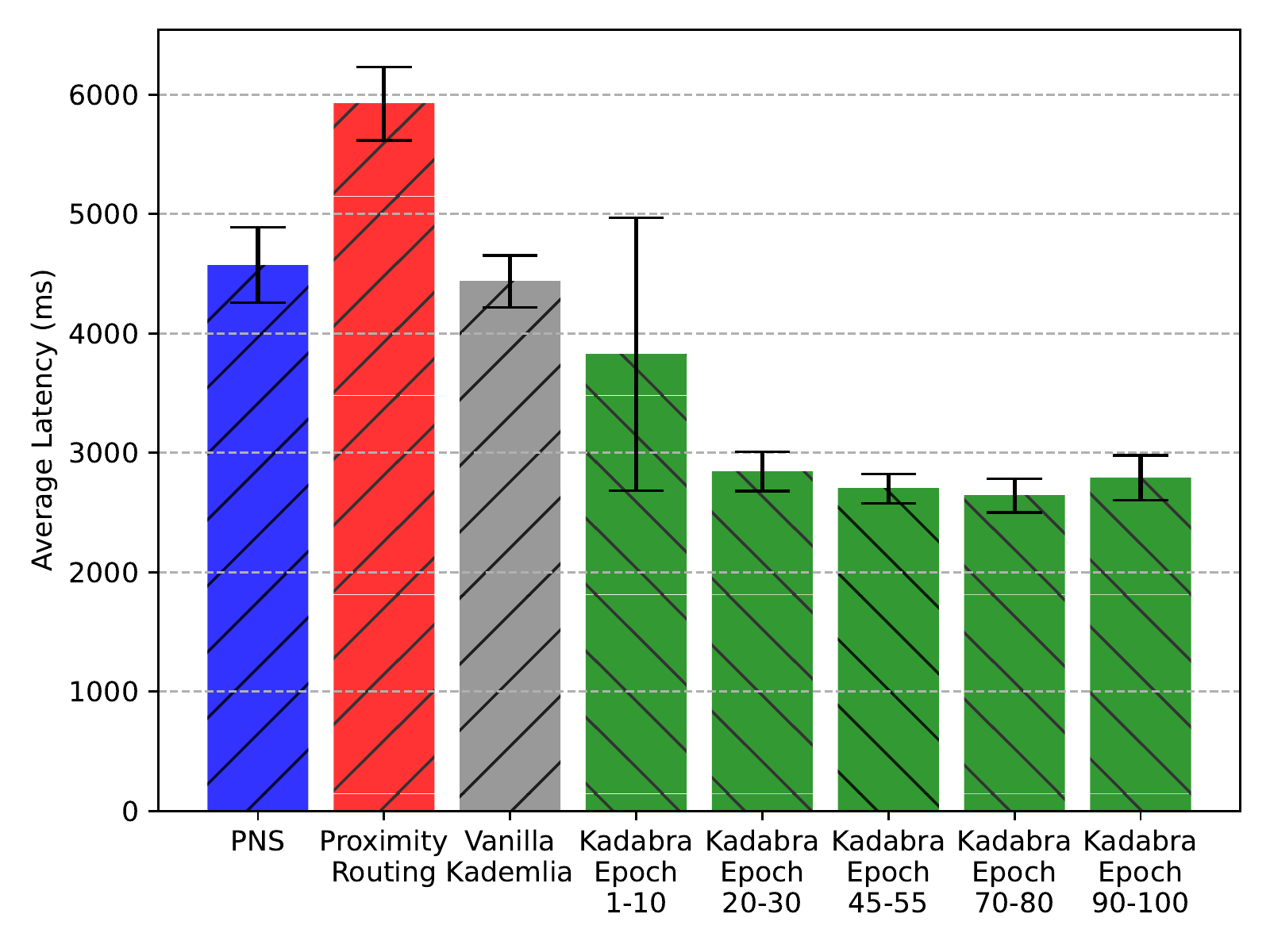}\quad}
\subfloat[Node E - London]{\includegraphics[width=.31\textwidth]{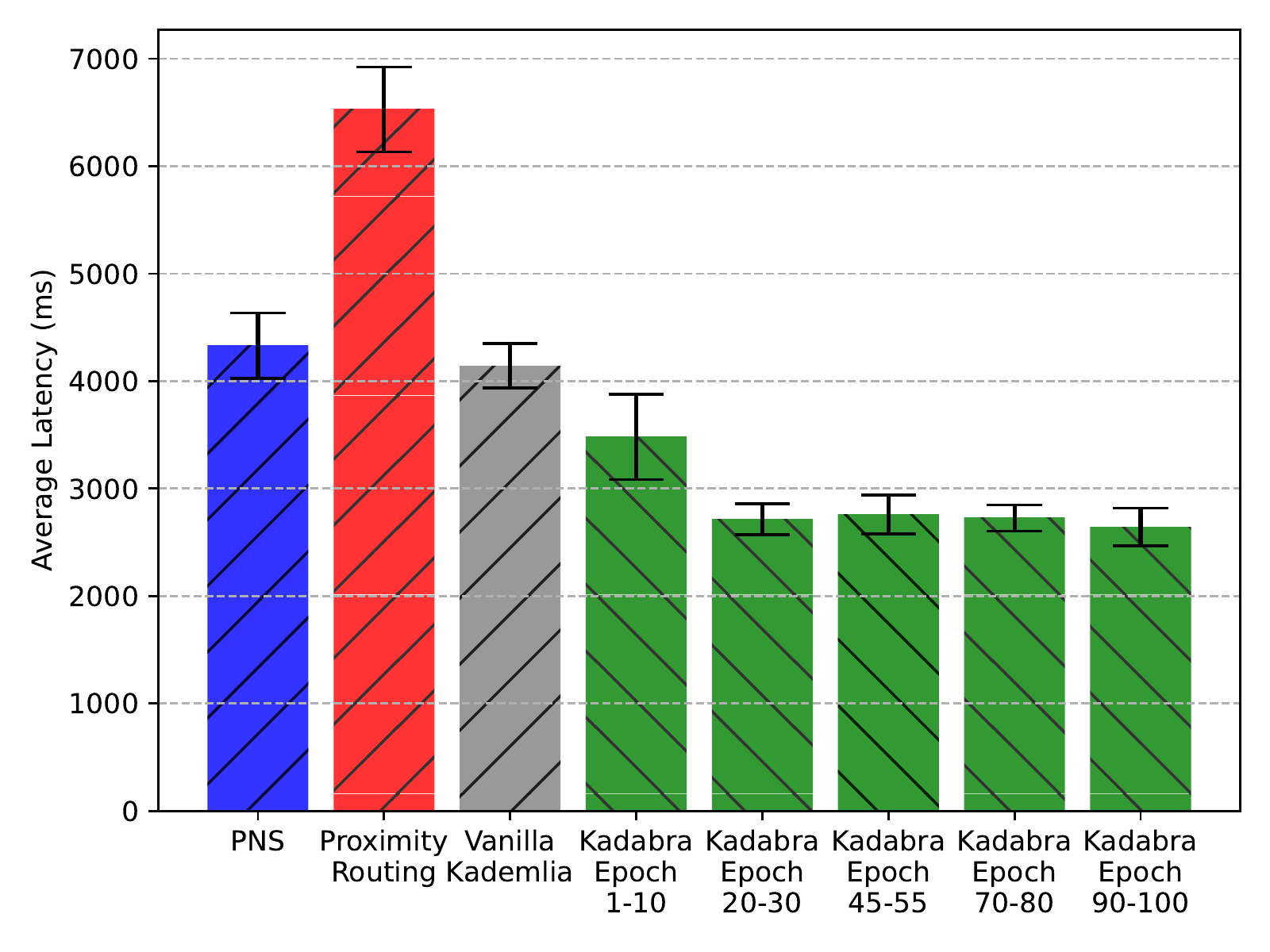}}
  \caption{Nodes in the real world: Performance under uniform demand at 
 five randomly sampled nodes around the world. }
\label{fig:ndrwsc15nodes}
\end{figure}

\begin{figure}[!tbp]
  \centering
 \subfloat[Node A - New York]{\includegraphics[width=.31\textwidth]{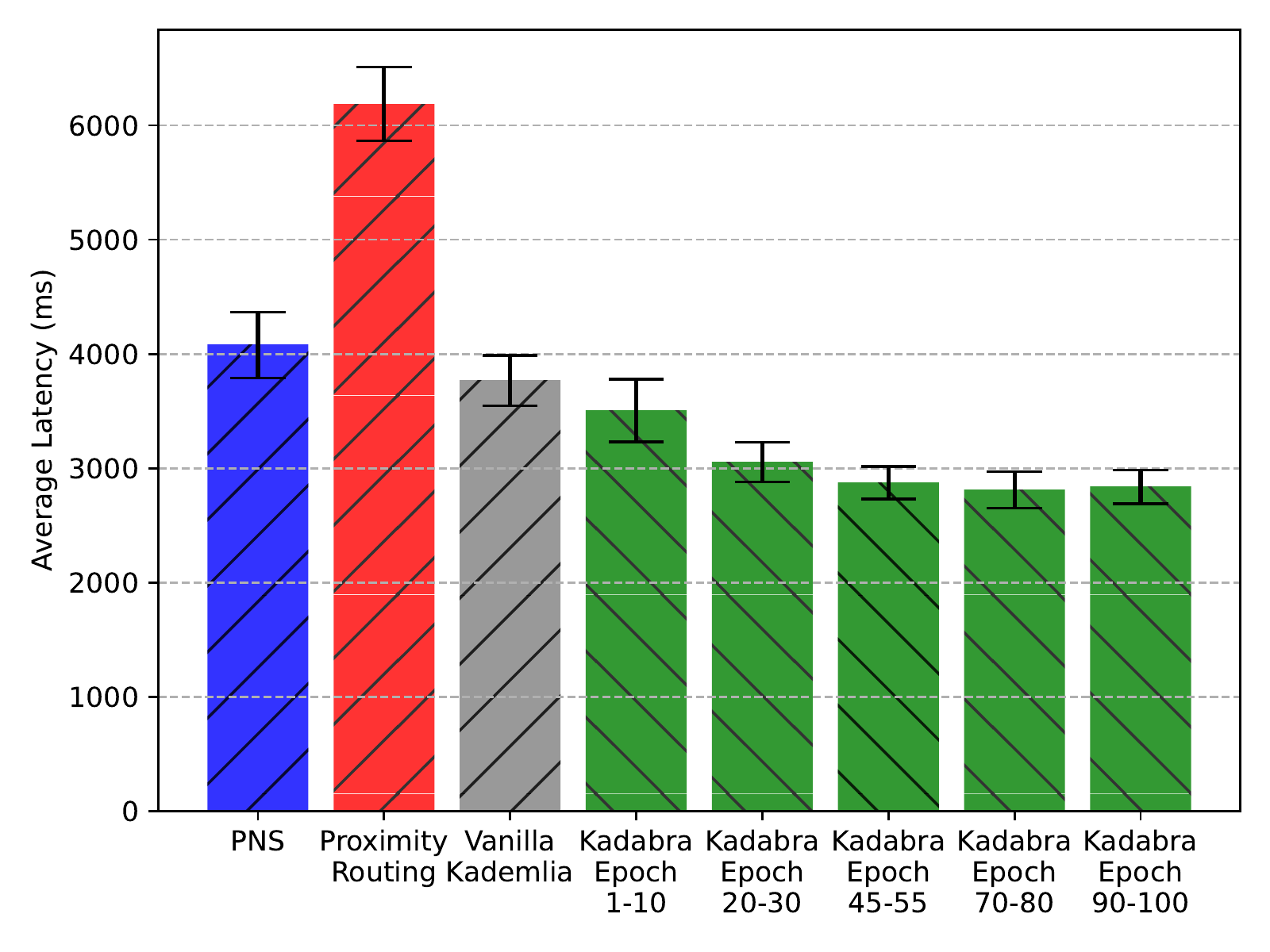}\quad}
\subfloat[Node B - San Jose]{\includegraphics[width=.31\textwidth]{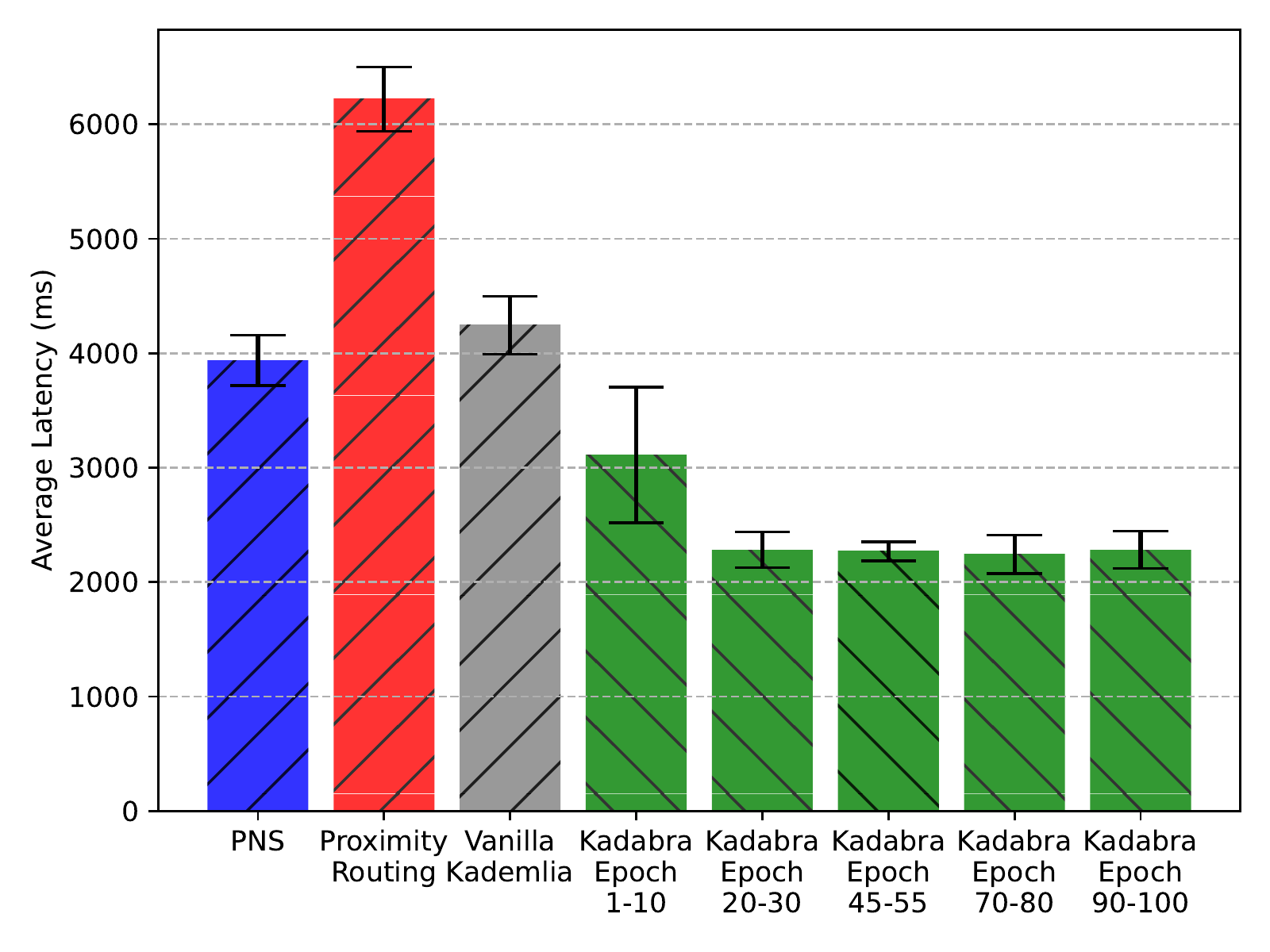}\quad}
\subfloat[Node C - Toronto]{\includegraphics[width=.31\textwidth]{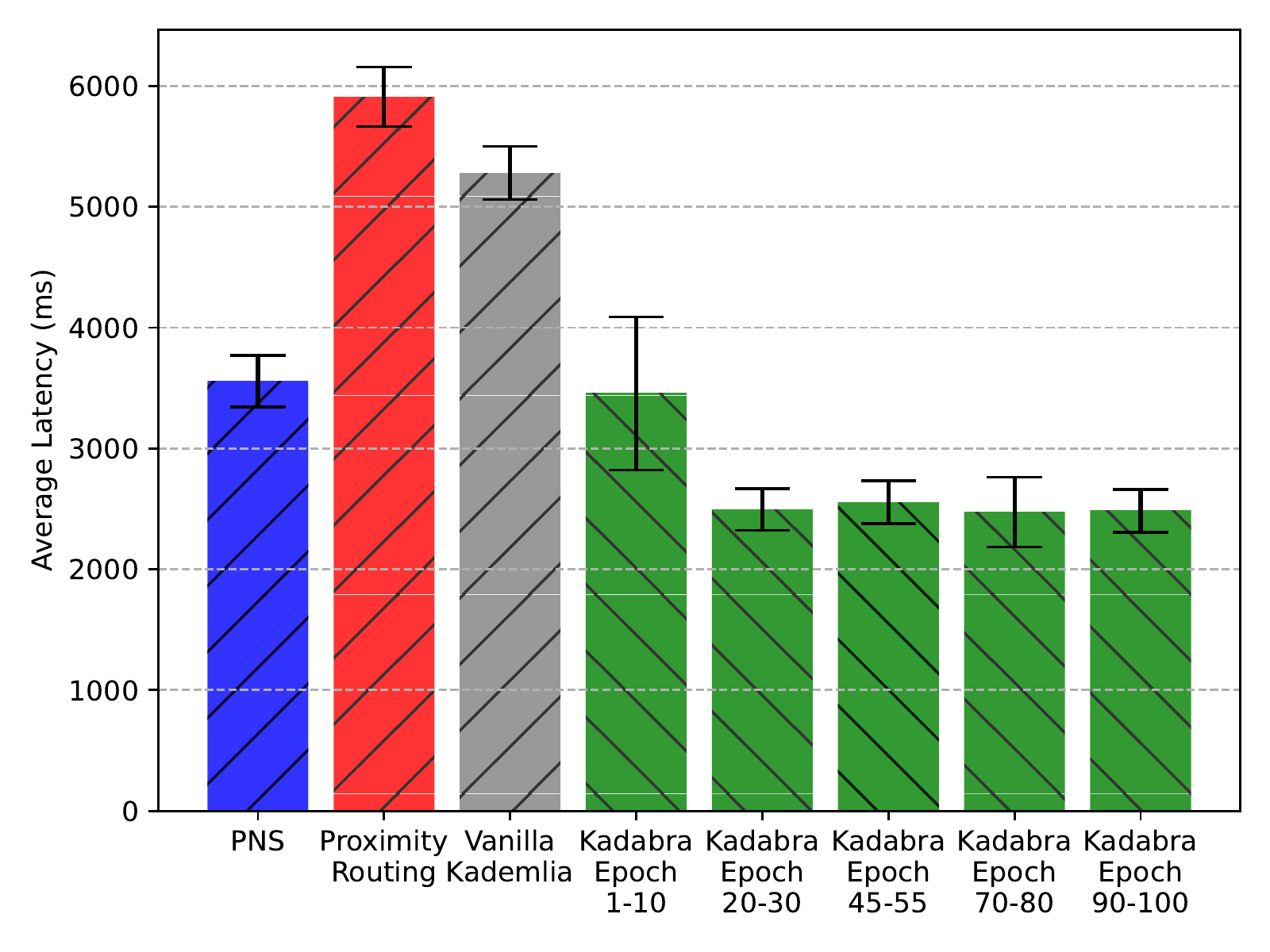}}

\smallskip

\subfloat[Node D - Frankfurt]{\includegraphics[width=.31\textwidth]{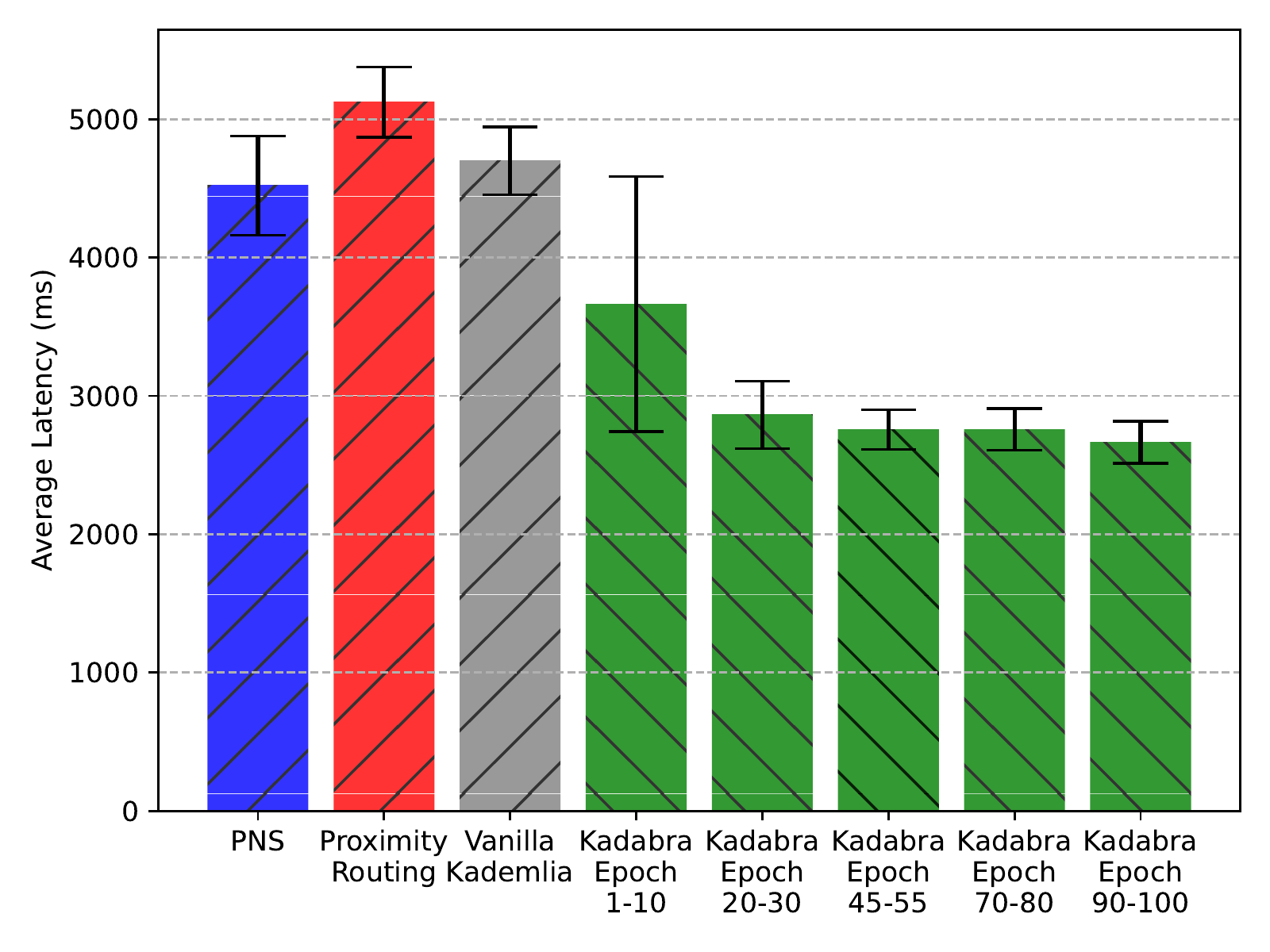}\quad}
\subfloat[Node E - London]{\includegraphics[width=.31\textwidth]{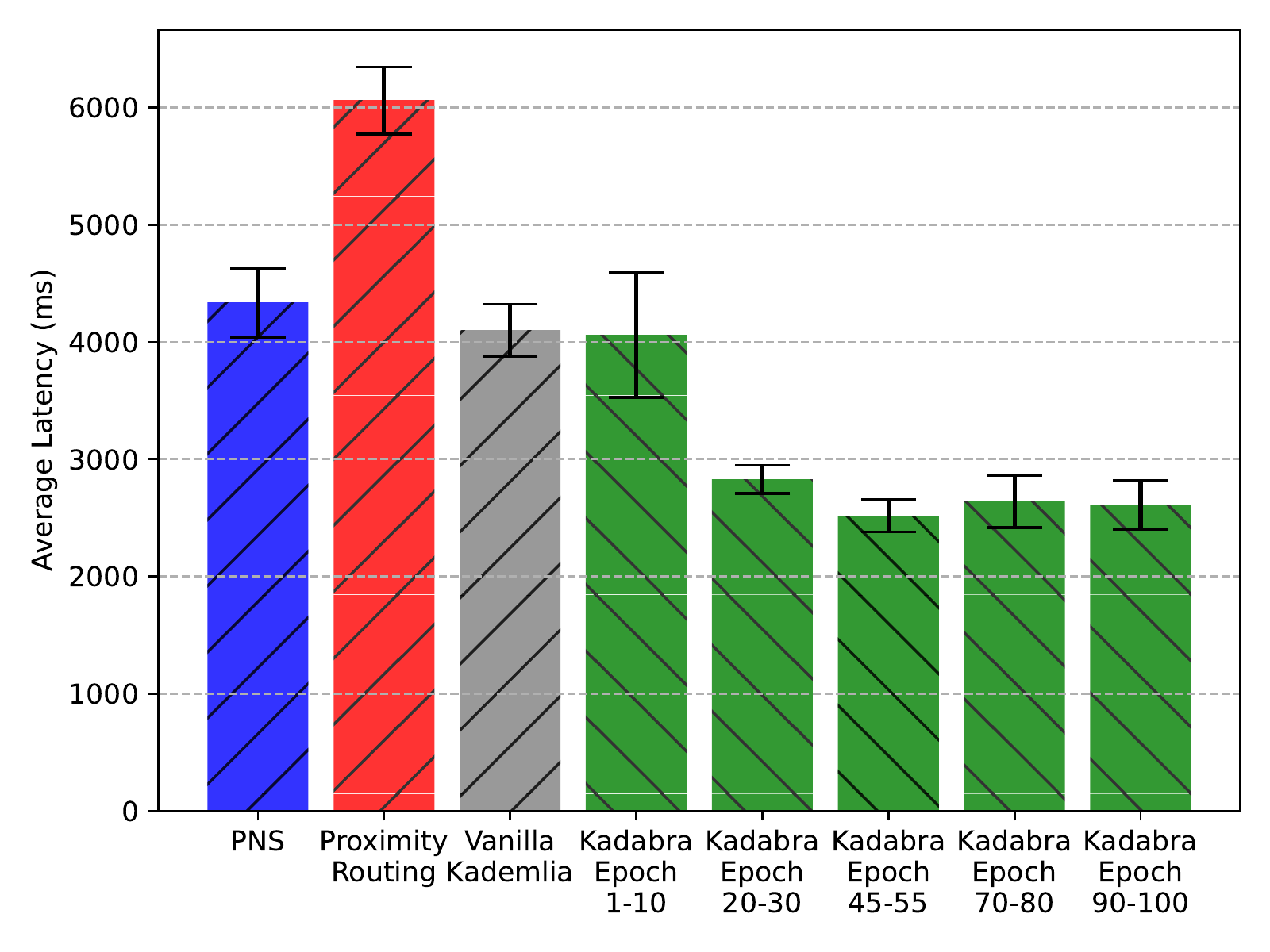}}
  \caption{Nodes in the real world: Performance under demand hotspots at five randomly sampled nodes around the world.}
  \label{fig:ndrwsce2node5}
\end{figure}

Fig.~\ref{fig:ndrwsc15nodes} and~\ref{fig:ndrwsce2node5} show performance at five randomly chosen nodes, for uniform demand and hotspot demand settings. 
The behavior observed in these plots are consistent with our discussion in \S\ref{s:evalresults}. 

\begin{figure}[!tbp]
  \centering
  \subfloat[]{\includegraphics[width=0.5\textwidth]{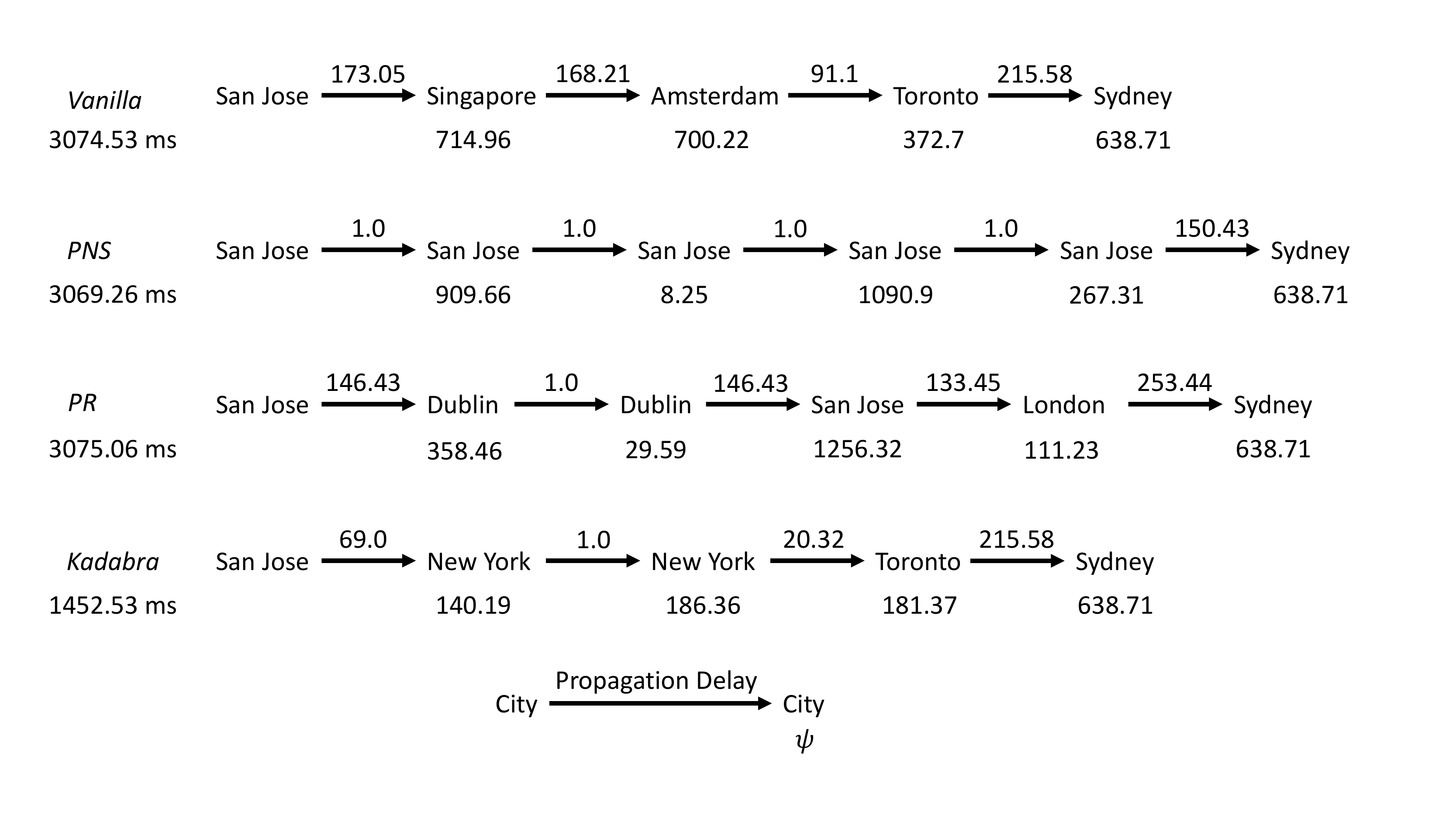}\label{fig:ndrwpathcas1}}
  \hfill
  \subfloat[]{\includegraphics[width=0.5\textwidth]{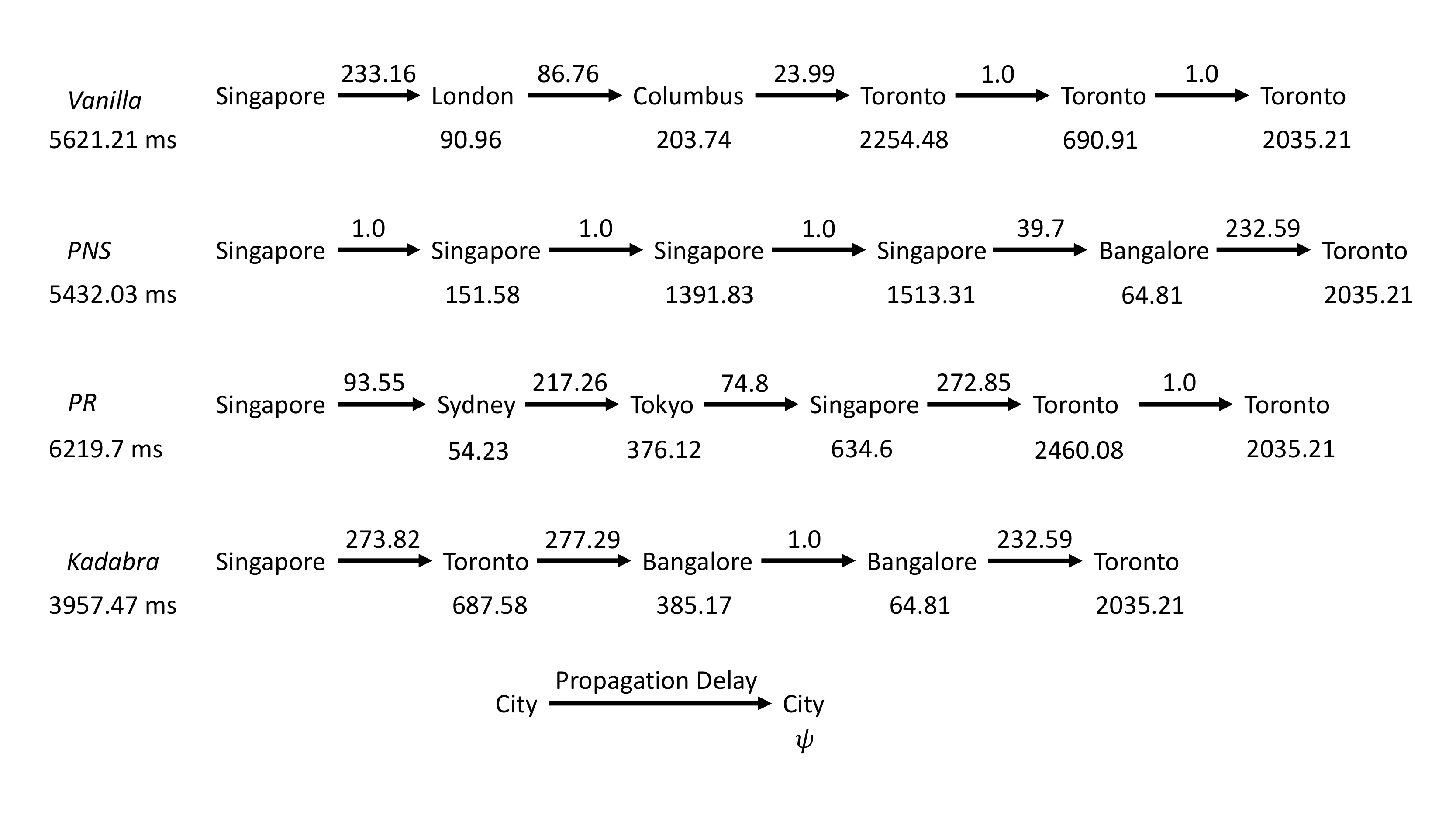}\label{fig:ndrwpathcase2}}
  \caption{Nodes in the real world: Paths and latencies of a sample query under (a) uniform demand, and (b) hotspot demand. }
\end{figure}

Fig.~\ref{fig:ndrwpathcas1} and~\ref{fig:ndrwpathcase2} show the paths and their corresponding latencies of a sample query, under uniform traffic and hotspot traffic respectively. 
Some cities are repeated on the paths are there may be multiple nodes within the same city.
The latency between two nodes in the same city is set to 1 ms in our experiments. 
Compared to baselines, \ourAlgorithm~chooses efficient paths with less number of hops and low node and link latencies. 

\section{Nodes in the real world - DHT}
\label{apx:rwdht}

\begin{figure}[!tbp]
  \centering
  \subfloat[]{\includegraphics[width=0.5\textwidth]{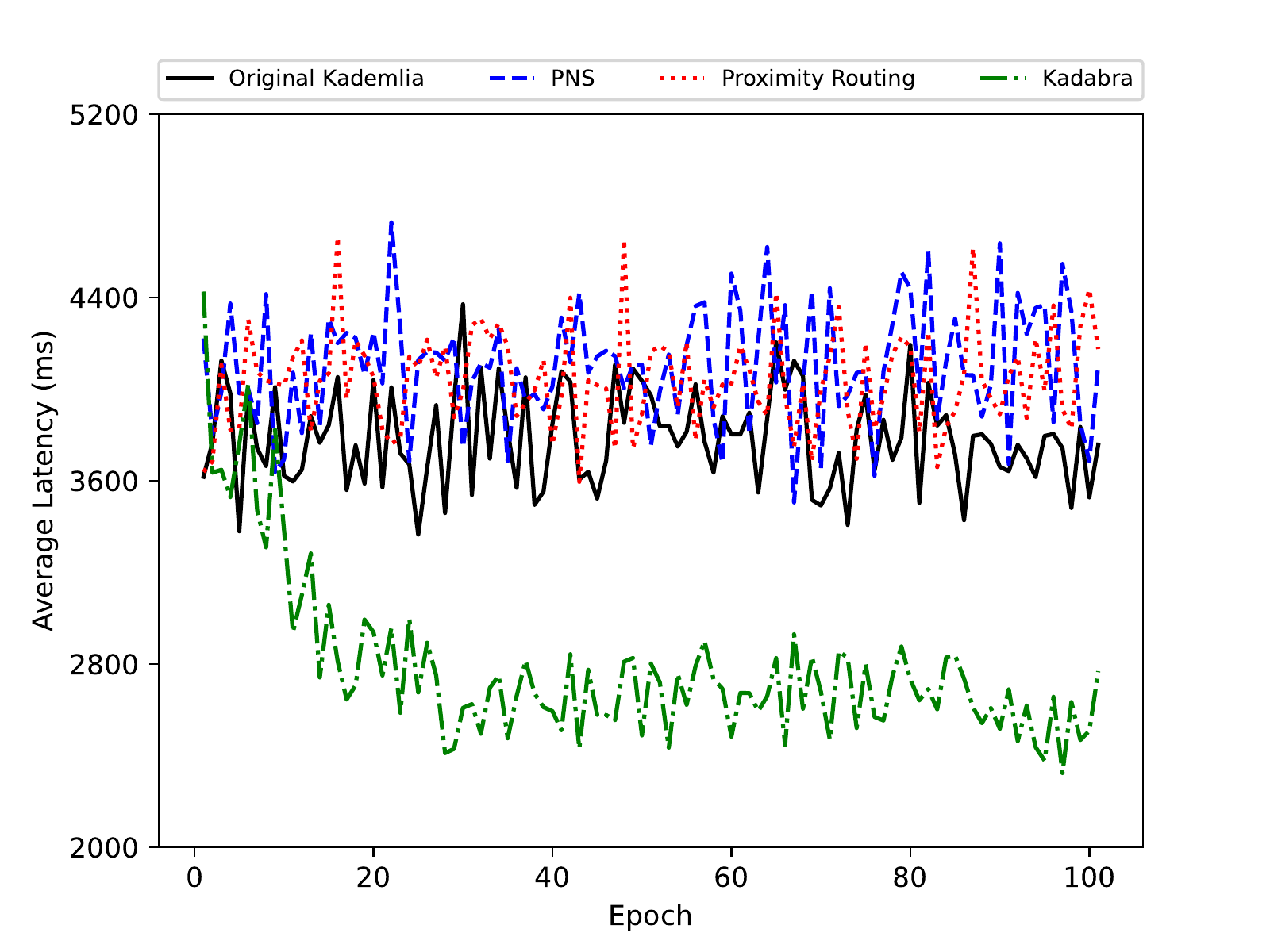}\label{fig:ndrwc3late}}
  \hfill
  \subfloat[]{\includegraphics[width=0.5\textwidth]{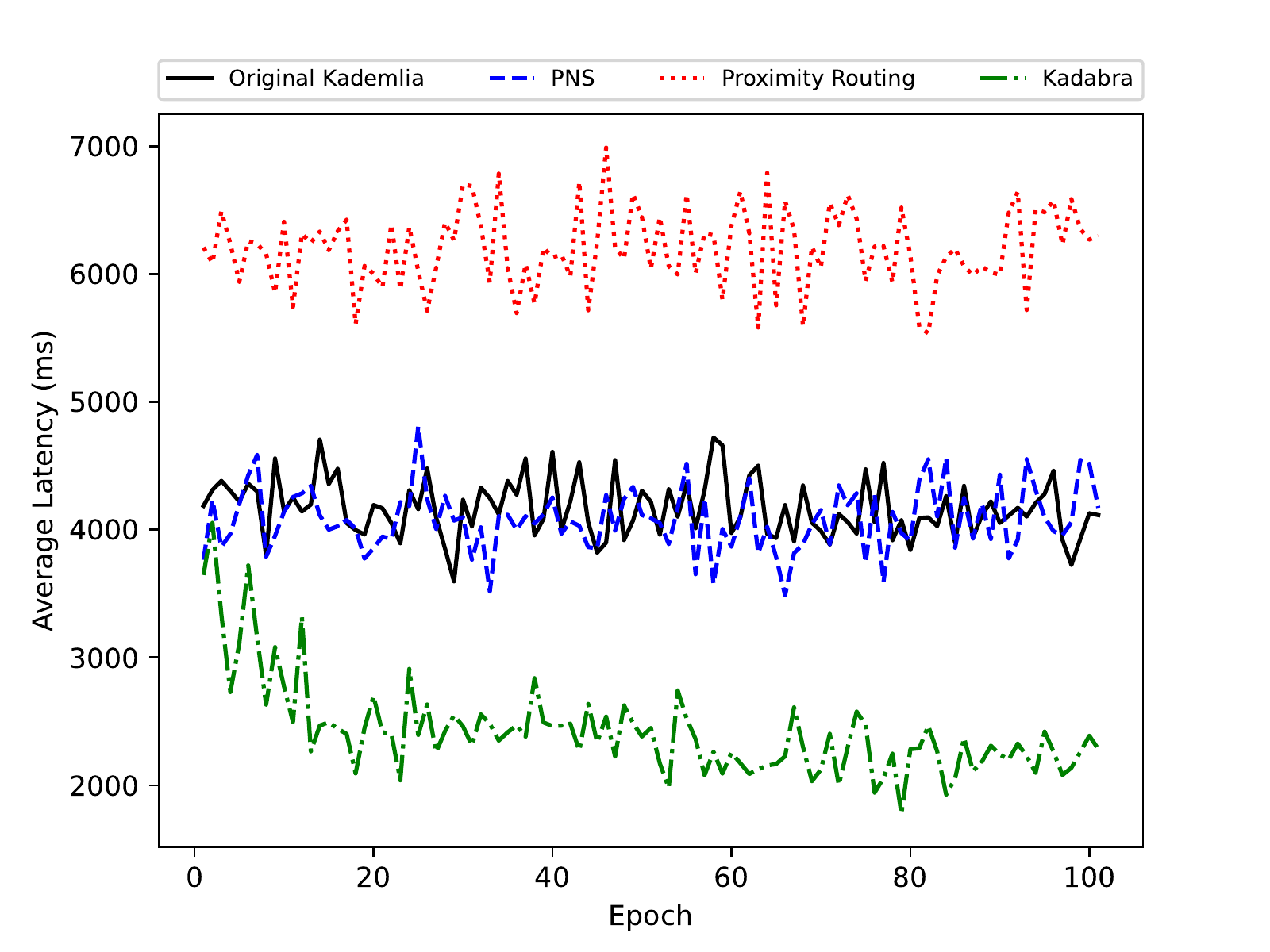}\label{fig:ndrwc4late}}
  \caption{Nodes in the real world with DHT application: (a) Performance under uniform demand. (b) Performance under hotspot demand.}
\end{figure}

This section considers the DHT application for the setting of nodes in the real world. 
Fig.~\ref{fig:ndrwc3late} and~\ref{fig:ndrwc4late} show performance under uniform traffic and hotspot traffic demands respectively. 
In each case, \ourAlgorithm~is exhibits lower latencies than the baselines. 
\ourAlgorithm~is more than 30\% faster compared to PNS in the two cases.

\begin{figure}[!tbp]
  \centering
    \includegraphics[height=6cm]{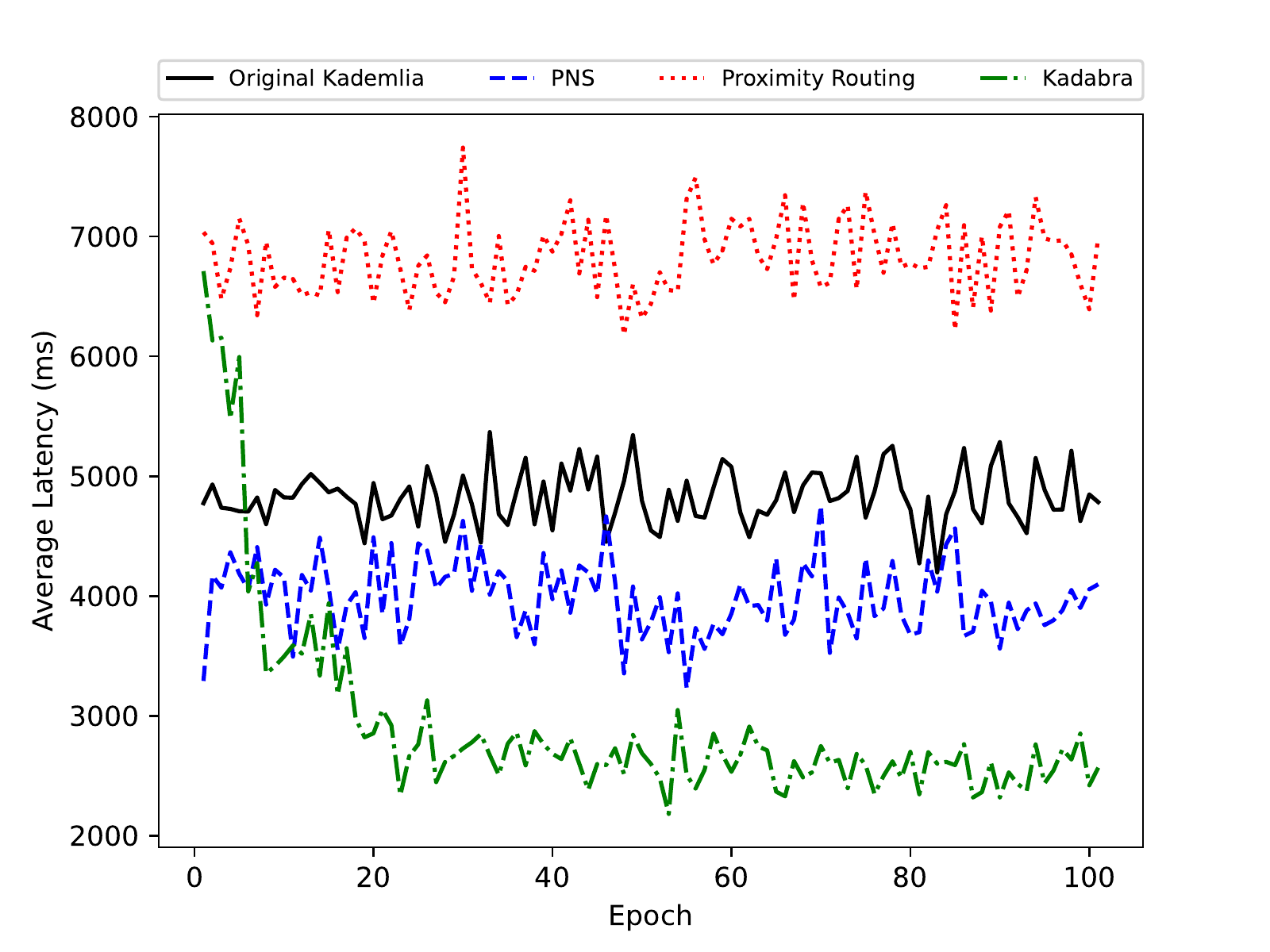}
    \caption{Nodes in the real world with DHT application: 4\% of nodes have a high node latency compared to network-wide average. Measurements are taken from one of the nodes within the high node latency area.}
    \label{fig:ndrwc6late}
\end{figure}

Fig.~\ref{fig:ndrwc6late} considers a setting where a region of nodes around New York have node latency that is twice the default average value. 
Here too, we observe \ourAlgorithm~outperforms the other baselines. 

\begin{figure}[!tbp]
    \centering
    \includegraphics[height=5cm]{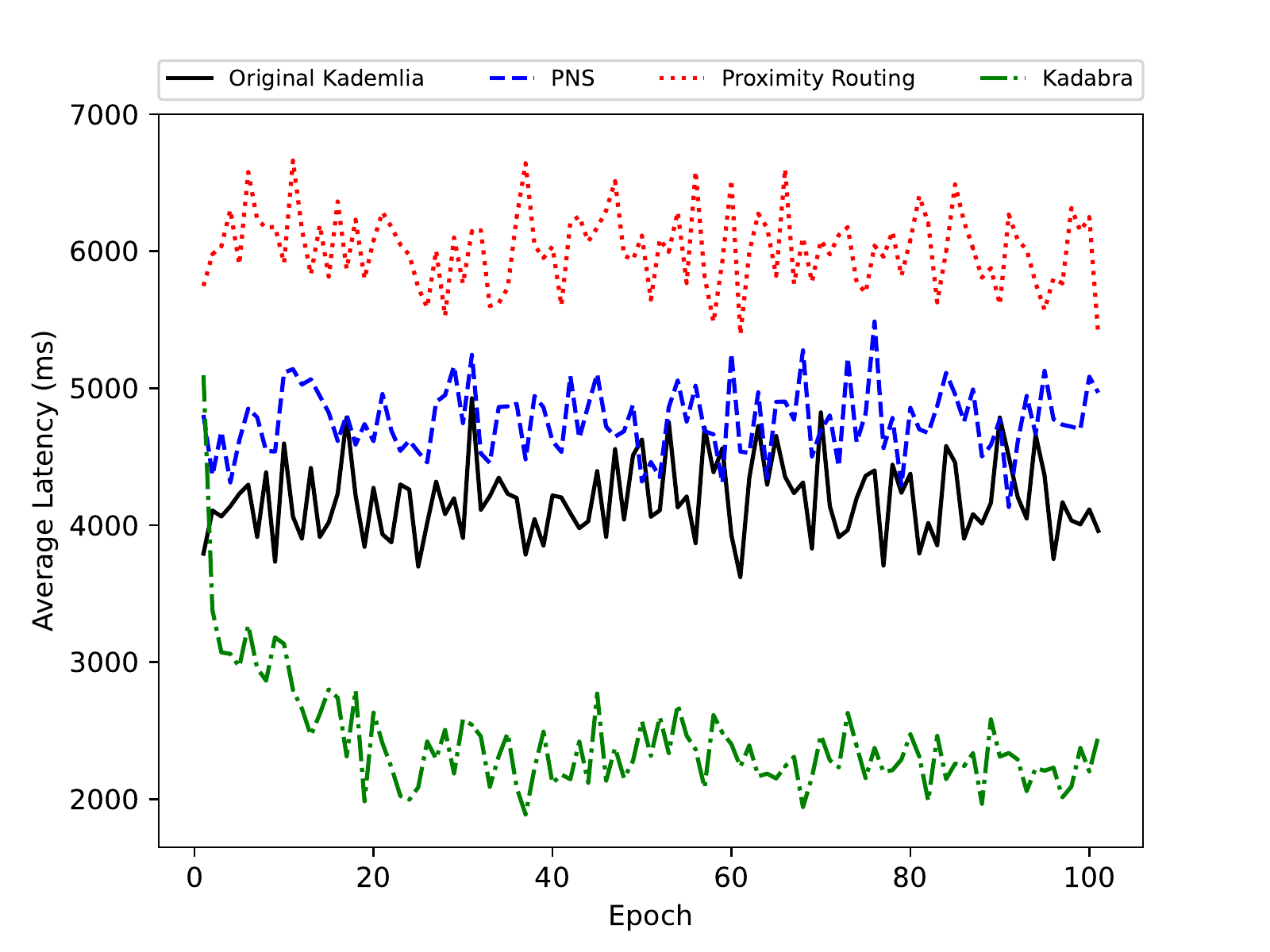}
    \caption{We performed iterative routing on nodes in the real world with KBR application under uniform demand. The result suggests that $\ourAlgorithm$ works effectively under iterative routing.}
    \label{fig:ndrwiterative}
\end{figure}

\section{Iterative routing}
\label{apx:iterativerouting}

There are two options when it comes to routing in Kademlia: recursive routing and iterative routing. In recursive routing, the source node contacts the first hop node, and the first hop node then contacts the second hop node. In iterative routing, the first hop node returns the second hop node information to the source node, and the source node contacts the second hop node by itself. Although most current Kademlia implementations use recursive routing, vanilla Kademlia uses iterative routing. In our evaluations we have mainly discussed recursive routing. 
In this section, we consider iterative routing under uniform traffic demand when nodes are located on the real world. The result is shown in Fig.~\ref{fig:ndrwiterative}, where \ourAlgorithm~outperforms the baselines. 

\begin{figure}[!tbp]
  \centering
  \subfloat[]{\includegraphics[width=0.5\textwidth]{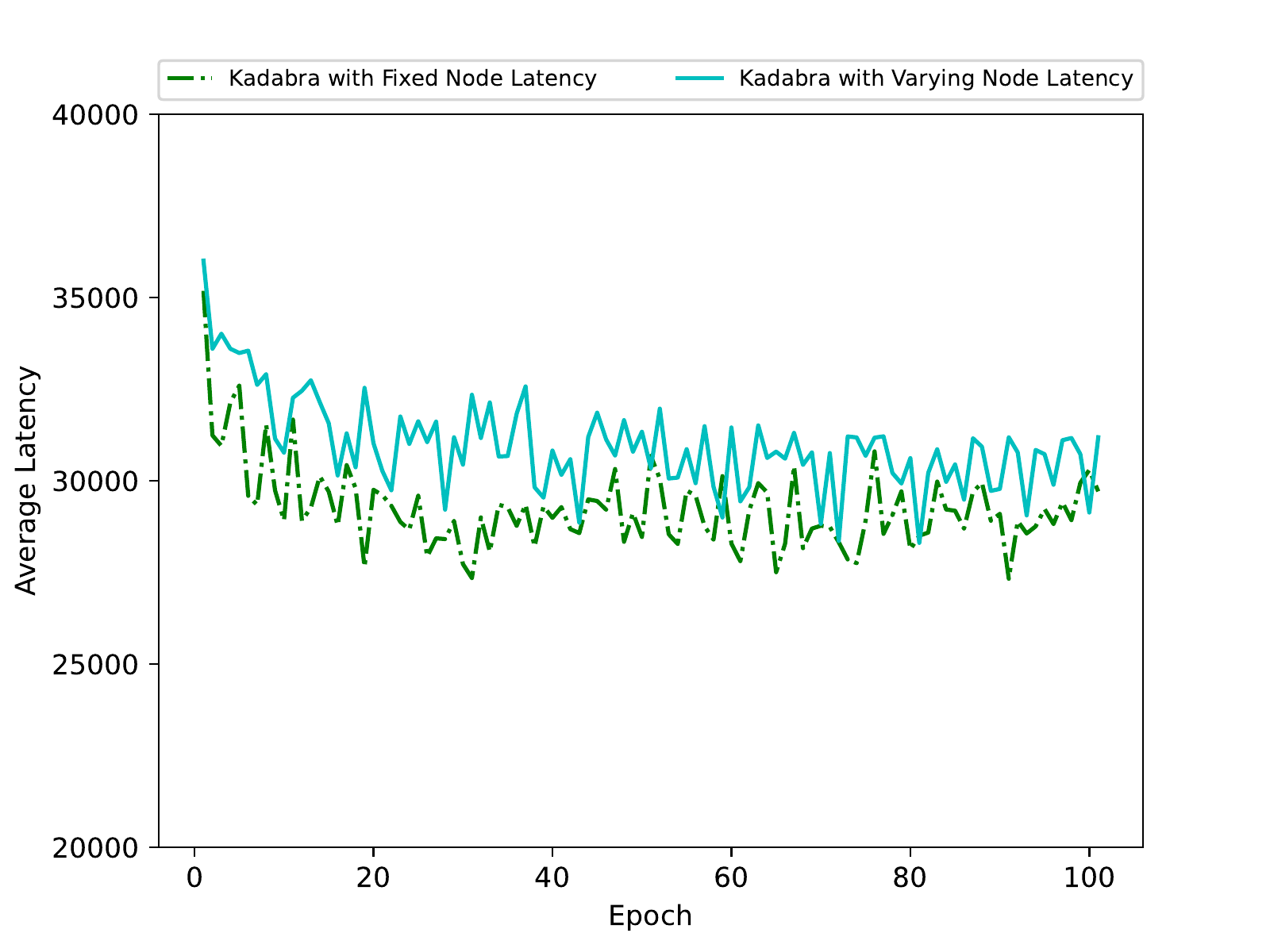}\label{fig:ednsc1randnl}}
  \hfill
  \subfloat[]{\includegraphics[width=0.5\textwidth]{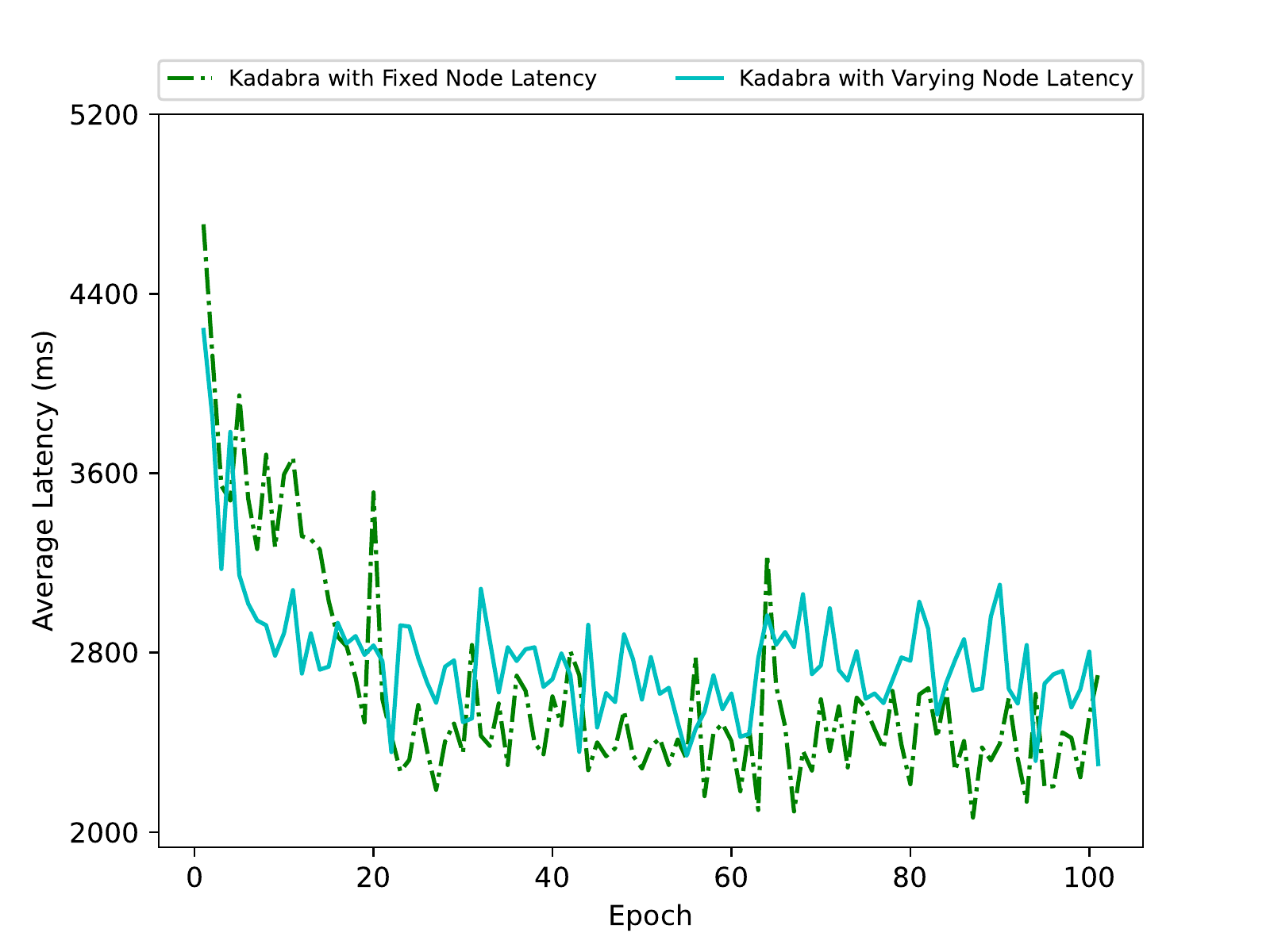}\label{fig:ndrwc1randnl}}
  \caption{Average latency during each epoch for queries
routed through the 1st k-bucket of an arbitrary node. KBR application with nodes that have noisy node latencies: (a) Nodes in a square. (b) Nodes in the real world.}
\end{figure}

\section{Network instability}
\label{apx:iterativerouting}

Node latency at the same node may vary for different queries. In our evaluations we have used fixed node latencies ($\delta$) for \ourAlgorithm~nodes. To capture potential network instabilities, in this section, we evaluate \ourAlgorithm~with varying node latencies. For each query, we add a random noise (random number from \([-0.05\times\delta_u, +0.05\times\delta_u]\)) to the fixed node latency \(\delta_u\) at a node \(u\). We consider the KBR application under uniform demand for both nodes in a square and nodes in the real world. The result in Fig.~\ref{fig:ednsc1randnl} and \ref{fig:ndrwc1randnl} indicates that the impact of the varying node latency on \ourAlgorithm~is insignificant. Specifically, for nodes in a square with varying node latencies, the performance of \ourAlgorithm~(average latency for the last 10 epochs) decreases by at most 5\%. For nodes in the real world with varying node latencies, the performance of \ourAlgorithm~decreases by at most 8\%.

Thus, we conclude that \ourAlgorithm~is a robust algorithm that can adapt to a wide variety of situations.

\end{document}